\documentclass[journal]{IEEEtran}\IEEEoverridecommandlockouts  

\ifCLASSINFOpdf

\else

\fi

\usepackage{xcolor,graphicx}       
\usepackage{subfigure}
\usepackage{indentfirst}

\usepackage{algorithm}
\usepackage{algorithmic}        
\usepackage{multirow}
\usepackage{array}

\usepackage{comment}            
\usepackage{epsfig}             
\usepackage{balance}
\usepackage{float}
\usepackage{caption}            

\usepackage{amsmath} \usepackage{amssymb}

\usepackage{cite} 
\usepackage{url}

\usepackage{multirow}
\usepackage{booktabs}
\usepackage{longtable}

\usepackage[colorlinks,bookmarksopen,bookmarksnumbered,citecolor=black,urlcolor=black]{hyperref}

\renewcommand{\eqref}[1]{Equation (\ref{#1})}

\usepackage{hyperref}
\hypersetup{
	colorlinks=true,
	allcolors=black
}

\usepackage{verbatim}
\usepackage{steinmetz}
\usepackage{bigstrut,multirow,rotating,booktabs}
\usepackage{bbding}
\usepackage{makecell}
\usepackage{graphicx}
\usepackage{threeparttable}
\usepackage{pifont}
\usepackage{longtable}
\usepackage{expl3}
\usepackage{amssymb}
\usepackage{pifont}
\usepackage{wasysym}
\usepackage{caption}
\usepackage[bottom]{footmisc}

\begin{document}
	
\title{6G Communication New Paradigm: The Integration of Unmanned Aerial Vehicles and Intelligent Reflecting Surfaces}

\author{Zhaolong Ning, Tengfeng Li, Yu Wu, Xiaojie Wang, Qingqing Wu, \\Fei Richard Yu,~\IEEEmembership{Fellow, IEEE}, and Song Guo,~\IEEEmembership{Fellow, IEEE}
	
	\thanks{\hspace{-2em} Zhaolong Ning, Tengfeng Li, and Xiaojie Wang (corresponding author) are with the School of Communications and Information Engineering, Chongqing University of Posts and Telecommunications, Chongqing 400065, China. E-mail: z.ning@ieee.org; s220101066@stu.cqupt.edu.cn; xiaojie.kara.wang@ieee.org.
	
		Yu Wu is with the School of Cyber Security and Information Law, Chongqing University of Posts and Telecommunications, Chongqing 400065, China. E-mail: wuy@cqupt.edu.cn.

		Qingqing Wu is with the Department of Electronic Engineering Shanghai Jiao Tong University Shanghai, Shanghai 200240, China. E-mail: qingqingwu@sjtu.edu.cn.
		
		F. Richard Yu is with the School of Information Technology, Carleton University, Ottawa, ON K1S 5B6, Canada. E-mail: richard.yu@carleton.ca.
	
		Song Guo is with the Department of Computer Science and Engineering, The Hong Kong University of Science and Technology, Kowloon, Hong Kong, China. E-mail: songguo@cse.ust.hk.
		}
}

\maketitle

\begin{abstract}
	 With the continuous development of Intelligent Reflecting Surfaces (IRSs) and Unmanned Aerial Vehicles (UAVs), their combination has become foundational technologies to complement the terrestrial network by providing communication enhancement services for large-scale users. This article provides a comprehensive overview of IRS-assisted UAV communications for 6th-Generation (6G) networks. First, the applications supported by IRS-assisted UAV communications for 6G networks are introduced, and key issues originated from applications supported by IRSs and UAVs for 6G networks are summarized and analyzed. Then, prototypes and main technologies related to the integration of IRSs and UAVs are introduced. Driven by applications and technologies of IRS-assisted UAV communications, existing solutions in the realms of energy-constrained communications, secure communications, and enhanced communications are summarized, and corresponding empirical lessons are provided. Finally, we discuss some research challenges and open issues in IRS-assisted UAV communications, offering directions for the future development.
\end{abstract}


\hypertarget{section1}{\section{Introduction}}
\indent With the continuous evolution of wireless networks, significant progress has been made in 5th-Generation (5G) networks, which are gradually being commercialized in certain regions. To meet the growing demands of networks, researchers are shifting their focus towards 6th-Generation (6G) wireless networks. Compared to 5G, 6G offers significant improvements in terms of rates, capacities, latency, and reliability \cite{8782879}. Furthermore, 6G networks are expected to support billions of interconnected devices, catering to different requirements of applications such as smart homes and Intelligent Transportation Systems (ITSs) \cite{9903905}. Realizing large-scale network connectivity of an Internet of Things (IoT) system has become a challenging issue. 6G networks are promising to cope with this requirement, aiming to achieve seamless coverage of not only terrestrial networks but also aerial and maritime domains \cite{10239285}. What's more, for new 6G bands such as millimeter Wave (mmWave) and TeraHertz (THz), serious path loss incurred over long distances can result in poor communications.

\indent Fortunately, the vision associated with 6G networks is promising to be realized, since Unmanned Aerial Vehicle (UAV) communication technologies continue to become mature. UAVs can be flexibly deployed in areas with dense network equipment, thus relieving the pressure of large-scale access networks on terrestrial networks, especially in scenarios where the density of network devices changes over time, such as vehicular communications \cite{9681714}. Main lobes of traditional ground-based Base Station (BS) antennas face downwards \cite{9681624}, which necessitates the construction of BSs to achieve wide-scale network coverage. By changing the flight altitude, it is possible to cover large ground areas when deploying UAVs as network relay stations or airborne BSs. Moreover, UAVs can be deployed in areas such as oceans and airspace where ground-based BS signals are difficult to reach, providing robust technical support for seamless network coverage. Furthermore, relying on their flexibility, UAVs can easily establish Line-of-Sight (LoS) links with other communication devices, effectively mitigating the high path loss caused by high-frequency communications.

\indent However, UAV-assisted communications still face development bottlenecks. First, UAVs have limited onboard energy, making it challenging to perform long-term communications and complex computational tasks \cite{9870557}. Second, the reliability of UAV communications is difficult to guarantee, especially in adverse weather conditions \cite{9599592}. It is worth noting that the aforementioned bottlenecks in UAV communications can be mitigated through Intelligent Reflecting Surface (IRS)-based technologies.

\subsection{Overview of IRS-Assisted UAV Communications for 6G Networks}

\begin{figure}[htb]
	\vspace{-0.2cm}
	\begin{center}
		\includegraphics[width=0.40\textwidth]{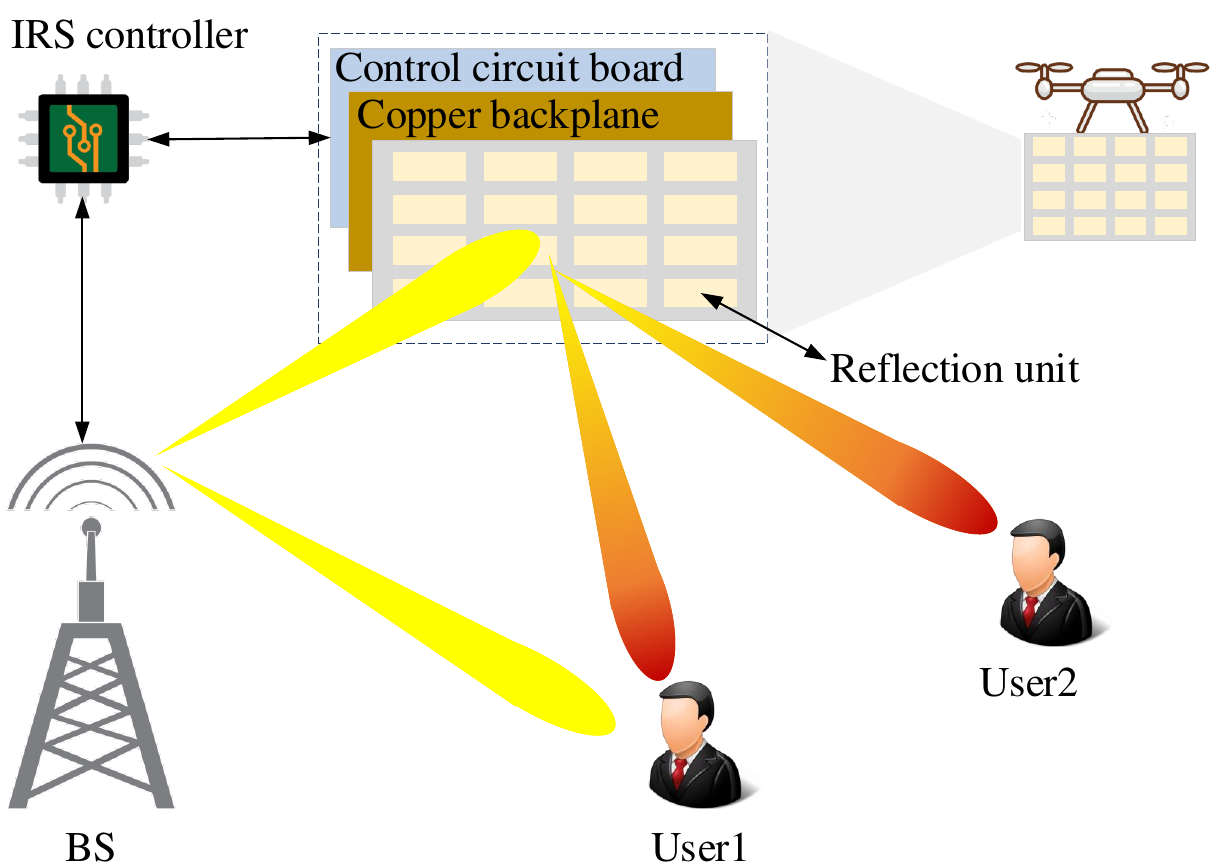}\\
		\hypertarget{Fig. 1}{\caption{A schematic diagram of IRS structure.}}
	\end{center}
	\vspace{-0.8cm}
\end{figure}

\indent As shown in \hyperlink{Fig. 1}{Fig. 1}, the IRS employs a three-layer architectural approach, where the innermost layer consists of the control circuit board, which excites the reflection unit to adjust the incident signal through a controller \cite{9326394}. The middle layer is made up of a copper backplane designed to minimize information and energy leakage. The outermost layer consists of a large number of passive reflection units, each of which can be dynamically controlled in a software-defined manner. By manipulating the phase and amplitude of incident signals, IRS achieves precise control and adjustment of the signal \cite{10061643}. 

\indent A UAV carrying IRS to assist ground user communications is illustrated in \hyperlink{Fig. 1}{Fig. 1}. A BS first generates IRS control messages and then sends them and communication messages to the aerial IRS based on special transmission protocols \cite{9453804,9198125}. The IRS controller receives control signals and transmits the IRS tuning strategies to the control circuit board, which then excites reflection units to intentionally adjust the incident signal. In this way, the BS can establish high-quality virtual LoS links with users, which can be coherently combined with the LoS link and other Non-LoS (NLoS) links to achieve signal enhancement and interference suppression. Generally speaking, IRS is passive and does not have signal processing capabilities. Therefore, the control strategies for IRS is typically determined at the BS through the cascaded BS-IRS-user channel. Fortunately, some reflection units can be replaced with dedicated sensing devices to perceive the wireless environment and assist in formulating the reflection strategies \cite{9326394}. Besides, the UAV can dynamically change flight trajectories to improve communication conditions.

\indent Based on the fact that the power consumption of IRS is much lower than that of Radio Frequency (RF) source and UAV flight power consumption \cite{9804220}, most studies typically assume that the IRS is passive, meaning it does not require energy for signal reflection. This is also the basic assumption for the scenarios considered in this article. In fact, maintaining the normal operation of IRS requires very low energy consumption, primarily for the IRS configuration receiver and control circuitry \cite{9140329}. Experiments show that the total power consumption of a large IRS, consisting of 1720 reflecting elements, is only 0.280 watts \cite{9279253}. Unlike most studies that overlook IRS power consumption, authors in \cite{10375242} considered IRS energy consumption in the scenario of aerial IRS-assisted ground user communications, where the power consumption of IRS control circuitry mainly depends on the number of reflection units and phase resolution, while the receiver power consumption is typically assumed constant and ignored.

\indent Currently, the practicality of IRS-assisted communications is thoroughly studied. 
Authors in \cite{zhang2018space} propose a metasurface that modulates electromagnetic waves in both spatial and frequency domains, validated by an experimental prototype. Similarly, authors in \cite{liu2022programmable} develop a programmable diffraction Deep Neural Network (DNN) for encoding metasurface arrays, providing practical evidence for IRS beamforming. However, research on IRS-assisted UAV communications remains in the theoretical stage. Athors in \cite{9386233} analyze the Bit Error Rate (BER) and ergodic capacity of IRS-assisted UAV-IoT networks. The analysis results indicate that introducing IRS into UAV networks can reduce BER by five orders of magnitude when the number of reflection elements exceeds 16. It is also shown that the ergodic capacity of IRS-assisted UAV communications is ten times higher than that of traditional UAV communications. For specific scenarios, authors in \cite{10.1145/3479239.3485700} propose a system-level simulation framework for Vehicle-to-Everything (V2X) networks assisted by hybrid aerial and ground IRSs to reduce reducing path loss and expand coverage in V2X networks. Based on the above analysis, we primarily offers a comprehensive overview of IRS-assisted UAV communications from theoretical perspective, providing a solid theoretical foundation for the practical design in the future.

\subsection{Features of IRS-Assisted UAV Communications}

\indent Unlike IRS-assisted terrestrial communications and traditional UAV communications, IRS-assisted UAV communications integrate the respective advantages of IRSs and UAVs to serve various scenarios within 6G networks in an efficient way. 

\indent IRS-assisted terrestrial communications aim to enhance signal quality by placing IRSs on the ground or building facades. The IRS is able to dynamically adjust the phase and amplitude of the incident signal, reflecting the signal in a different way from active relays to improve the propagation path of the signal \cite{9326394}. It can effectively extend signal coverage, especially in scenes where the direct LoS path is obstructed, such as urban, canyon or indoor environments. It can also improve the Energy Efficiency (EE) ratio and system capacity of communications through precise signal tuning. However, the performance of IRS-assisted terrestrial communications is limited by the location and number of IRSs, as well as their adaptability to environments.

\indent Traditional UAV communications rely on a direct wireless connection between UAVs and ground users. It can provide a high degree of mobility and flexibility, allowing UAVs to be quickly deployed in specific areas to support temporary or emergency communication needs \cite{8641424}. UAVs can take advantage of their height to establish LoS communication links with relative ease. However, there are still many challenges, including the battery life limitations of UAVs, the existence of obstacles, and the unstable connection in complex environments.

\indent IRS-assisted UAV communications represent a kind of highly innovative communications that can provide communication flexibility and efficiency by combining the passive signal enhancement capabilities of IRSs and the dynamic deployment capabilities of UAVs. Specifically, through strategic placement of IRSs, UAVs can establish communication links with devices without the need of flying closely, thus saving propulsion energy to a certain extent \cite{9690481}. By widening and flattening the three-dimensional beams, the coverage range of UAVs can be expanded with the assistance of IRSs \cite{9351782}. Additionally, this combination enables flexible implementation of airborne passive relays, which can effectively mitigate severe congestion and attenuation effects in THz and mmWave frequency bands \cite{9599592}. It is worth noting that the mobility of UAVs enables full-angle reflection, providing new degrees of freedom for IRS design and deployment. This combination overcomes the obstacles posed by terrain and buildings, and achieves real-time optimization of the signal propagation path, which is difficult to achieve in traditional IRS-assisted ground communications or UAV communications. Overall, IRS-assisted UAV communications provide a highly adaptable and scalable solution to address the limitations of traditional communication methods, and has significant research value.

\begin{figure*}[htb]
	\vspace{-1.2cm}
	\begin{center}
		\includegraphics[width=0.85\textwidth]{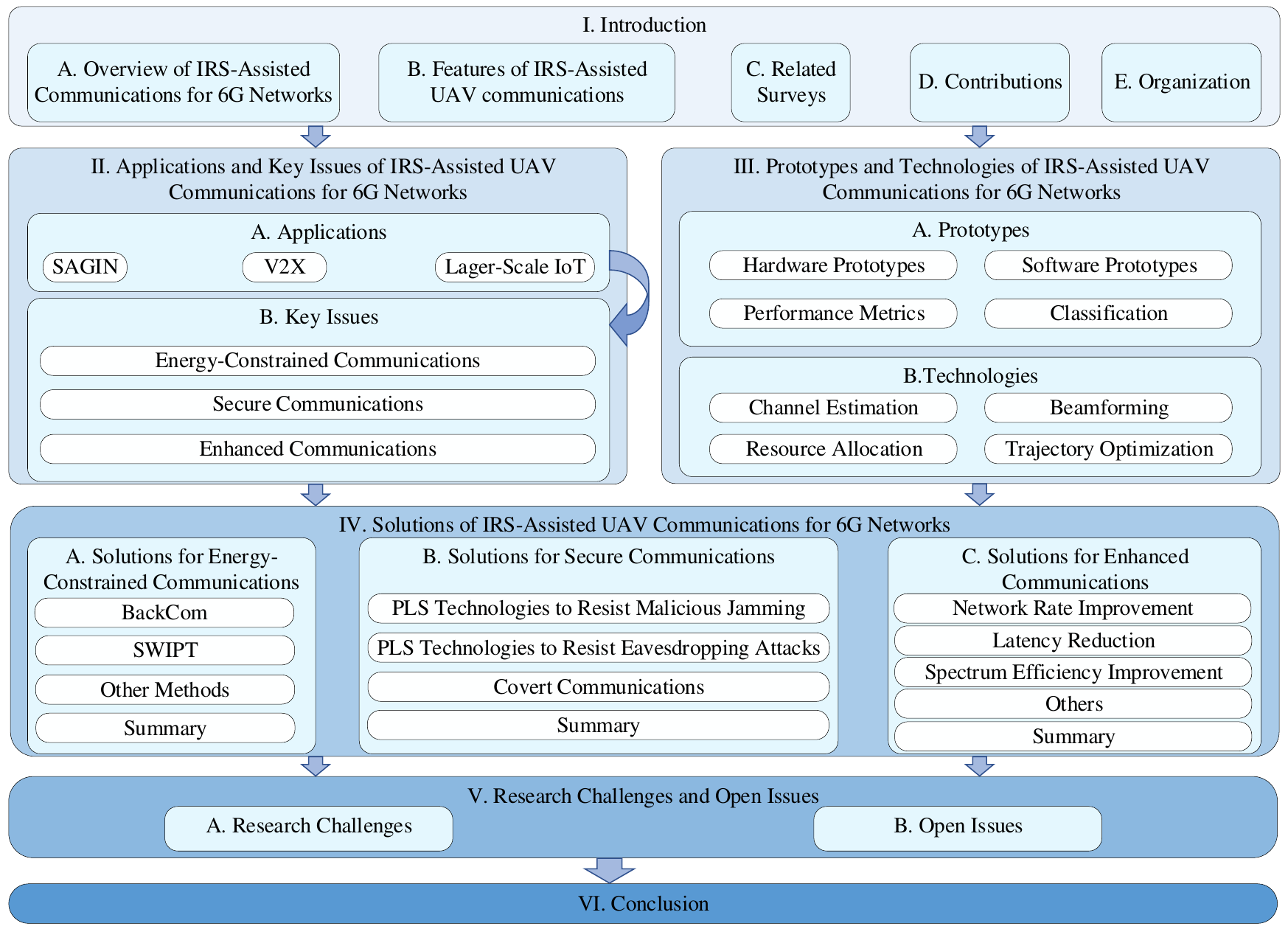}\\
		\hypertarget{Fig. 2}{\caption{The structure of this survey.}}
		\setlength{\parskip}{2cm plus4mm minus3mm}
	\end{center}
	\vspace{-0.5cm}
\end{figure*}

\subsection{Related Surveys}

\indent Some surveys have discussed IRSs and UAVs, and can be divided into three categories: specific implementation of IRSs or UAVs in wireless networks \cite{9722893,DAJER202287,SADIA2023,8411465,10.1145/3604933,8918497,8675384}, practical applications based on IRSs or UAVs \cite{9424177,9122596,9806434,9968053,9358097,9681624,9768113,9779853,7876852}, as well as the combination of IRSs and UAVs \cite{9599592,9690481,10061643}.

\indent For the specific implementation of IRSs or UAVs, authors in \cite{9722893} summarize the channel estimation and practical beamforming methods for imperfect IRS. Authors in \cite{DAJER202287} investigate the channel design of IRS-assisted wireless networks, while authors in \cite{SADIA2023} investigate network modeling based on the integration of IRSs and NOMA. Considering the spatiotemporal variability of UAV wireless channels, authors in \cite{8411465} discuss the wireless channel modeling of UAVs in detail. Authors in \cite{10.1145/3604933} summarize key technologies, issues and solutions in UAV networks with the assistance of Machine Learning (ML) and Mobile Edge Computing (MEC). Meanwhile, authors in \cite{8918497} discuss problems faced by UAV communications in future wireless networks and summarize feasible solutions. In addition, the standardization progress of UAV cellular communications is discussed in \cite{8675384}. 

\indent For applications supported by IRSs, authors in \cite{9424177,9122596} discuss the potential of IRS in wireless networks. Authors in \cite{9806434} summarize the current research on IRS-enabled ITSs in detail. Furthermore, visible light communications are considered as a crucial component of future communication networks, and its communication tutorial in combination with IRS is discussed in \cite{9968053}. For the applications of UAVs, authors in \cite{9358097} focus on UAV-assisted Aerial Radio Access Networks (ARANs), while authors in \cite{9681624,9768113} discuss UAV-assisted cellular networks from system design and industry perspectives, respectively. In addition, authors in \cite{9779853,7876852} introduce the application of both UAV-assisted data acquisition systems and ITSs.

\indent Authors in \cite{9599592,9690481,10061643} illustrate the combination of IRSs and UAVs for wireless networks. Specifically, benefits, current progress and development prospects of IRS-assisted UAV communications are discussed. It is worth noting that studies in \cite{9690481,9599592,10061643} all highlight the benefits of combining IRS and UAV through simulation tests, while there is still a lack of in-depth analysis of the relationship between corresponding scenarios and issues in the context of IRS-assisted UAV communications, as well as a comprehensive summary of effective solutions to mitigate these issues.

\hypertarget{table 1}{\begin{table*}[htbp]}
	\centering
	\vspace{-1.0cm}
	\fontsize{9}{12}
	\selectfont
	\begin{threeparttable}
		\caption{Comparisons of features and contributions among related surveys.}
		\begin{tabular}{|m{1.6cm}<{\centering}|m{0.8cm}<{\centering}|m{0.3cm}<{\centering}|m{0.3cm}<{\centering}|m{0.3cm}<{\centering}|m{0.3cm}<{\centering}|m{0.3cm}<{\centering}|m{11cm}<{\centering}|}
			\hline
			\multirow{2}[20]{*}{Categories} & \multirow{2}[20]{*}{Ref.} & \multicolumn{5}{c|}{Scopes} & \multirow{2}[20]{*}{Contributions} \bigstrut\\
			\cline{3-7}          &       &  \multirow{2}[15]{*}{\makebox[0.05cm][c]{IRS}}&    \multirow{2}[15]{*}{\makebox[0.05cm][c]{UAV}}    &   \multicolumn{3}{c|}{6G networks }   &  \bigstrut\\
			\cline{5-7} & & & &\vspace{0.2cm} \multirow{1}[-1]{*}{\makebox[0.05cm][c]{\begin{sideways}
					V2X\end{sideways}}} \vspace{0.2cm}&\vspace{0.2cm} \multirow{1}[-1]{*}{\begin{sideways}SAGIN\end{sideways}} \vspace{0.2cm}&\vspace{0.2cm} \multirow{1}[-1]{*}{\makebox[0.05cm][c]{\begin{sideways}IoT\end{sideways}}}\vspace{0.2cm}& \bigstrut\\
		
			\hline
			\multirow{7}[15]{*}{\makebox[0.05cm][c]{\makecell{Specific im-\\plementation \\of  IRSs \\or UAVs}}} & \multirow{1}[1]{*}{\cite{9722893}} & $\surd$ & $\times$ & $\times$  & $\times$ & $\times$ &  The design of channel estimation and passive beamforming for IRS is reviewed. \bigstrut\\
			\cline{2-8} &
			\multirow{1}{*}{\cite{DAJER202287}} & $\surd$ & $\times$ & $\times$ & $\times$ &$\times$ & ML-based solutions, channel and hardware design for IRS are discussed. \bigstrut\\
			\cline{2-8} &
			\multirow{1}{*}{\cite{SADIA2023}} & $\surd$ & $\times$ & $\times$ &$\times$ &$\times$ &The network modeling based on the integration of IRSs and NOMA is investigated. \bigstrut\\
			\cline{2-8} &
			\multirow{1}[1]{*}{\cite{8411465}} & $\times$ & $\surd$ & $\times$    & $\times$&$\times$ &Measurement schemes and channel characterization for UAV channels are reviewed. \bigstrut\\
			\cline{2-8} &
			\multirow{1}[1]{*}{\cite{10.1145/3604933}} & $\times$ & $\surd$ & $\times$ &$\times$ &$\times$ &Key technologies, issues, and solutions in UAV networks with the assistance of ML and MEC are summarized. \bigstrut\\ 
			\cline{2-8} &
			\multirow{1}[1]{*}{\cite{8918497}} & $\times$ & $\surd$ & $\times$   &$\times$ &$\times$ &The benefits of combining UAV and wireless networks are discussed. 
			\bigstrut\\
			\cline{2-8} &
			\multirow{1}[1]{*}{\cite{8675384}} & $\times$ & $\surd$ & $\times$   & $\times$&$\times$ &The standardization process of UAV communications and the testbed are reviewed. \bigstrut\\
			
			\hline
			\multirow{9}[45]{*}{\makebox[0.05cm][c]{\makecell{Applications \\based on \\IRSs or UAVs}}} & \multirow{1}[1]{*}{\cite{9424177}} & $\surd$ & $\times$ & $\times$ & $\times$&$\times$ &Principles, performance evaluation and enabling technologies for IRS-assisted wireless networks are presented. \bigstrut\\
			\cline{2-8}  &
			\multirow{1}[1]{*}{\cite{9122596}} & $\surd$ & $\times$ & $\times$  & $\times$&$\times$ &The applications of IRSs and performance enhancements in wireless communications are reviewed.  \bigstrut\\
			\cline{2-8} &
			\multirow{1}[1]{*}{\cite{9806434}} & $\surd$ & $\times$ & $\surd$ &$\times$ &$\times$ &The research on IRS-assisted ITSs is summarized for 6G networks. \bigstrut\\
			\cline{2-8}  &
			\multirow{1}[1]{*}{\cite{9968053}} & $\surd$ & $\times$ & $\times$ & $\times$ &$\times$ &A tutorial on IRS-based indoor visible light communications is investigated. \bigstrut\\
			\cline{2-8}  &
			\multirow{1}[1]{*}{\cite{9358097}} & $\times$ & $\surd$ & $\times$ & $\surd$&$\times$ &UAV-assisted ARANs for 6G networks are investigated. \bigstrut\\
			\cline{2-8} &
			\multirow{1}[1]{*}{\cite{9681624}} & $\times$ & $\surd$ & $\times$ &$\surd$ &$\times$ &The major barriers, design considerations, potential solutions, and the ability of cellular networks to support UAV communications are reviewed. \bigstrut\\
			\cline{2-8} &
			\multirow{1}[1]{*}{\cite{9768113}} & $\times$ & $\surd$ & $\times$  & $\surd$&$\times$ &The use cases, requirements, enabling technologies and unresolved issues of UAVs from 5G to 6G networks are reviewed. \bigstrut\\
			\cline{2-8} &
			\multirow{1}[1]{*}{\cite{9779853}} & $\times$ & $\surd$ & $\times$   & $\times$&$\surd$ &The development status and future trends of UAV-assisted data acquisition technologies are reviewed. \bigstrut\\
			\cline{2-8} &
			\multirow{1}[1]{*}{\cite{7876852}} & $\times$ & $\surd$ & $\surd$   &$\times$ &$\times$ &The application potential and challenges of UAV-based ITSs are reviewed. \bigstrut\\
			\hline

			\multirow{4}[15]{*}{\makebox[0.05cm][c]{\makecell{Combination \\of IRSs \\and UAVs}} }& \multirow{1}[1]{*}{\cite{9599592}} & $\surd$ & $\surd$ & $\times$ &$\times$ &$\times$ &The combination of UAVs and IRSs for air-ground network performance enhancement is discussed through two case studies. \bigstrut\\
			\cline{2-8} &
			\multirow{1}[1]{*}{\cite{9690481}} & $\surd$ & $\surd$ & $\times$  &$\times$ &$\times$ & The methods of integrating IRSs with UAVs to serve air-ground integrated networks are discussed. \bigstrut\\
			\cline{2-8} &
			\multirow{1}[1]{*}{\cite{10061643}} & $\surd$ & $\surd$ & $\times$ &$\times$ &$\times$ &The advantages of combining UAVs and active IRSs are demonstrated through two cases. \bigstrut\\
			\cline{2-8} & 
			\makecell{This\\ article} & $\surd$ & $\surd$ & $\surd$ &$\surd$ &$\surd$ &\makebox[0.2cm]{\makecell{The common problems, key technologies, application scenarios, solutions and open \\issues faced by IRS-assisted UAV communications for 6G networks are summarized.}} \bigstrut\\	
			\hline
		\end{tabular}%
		\begin{tablenotes}
			\centering
			\item The symbol ``$\surd$" represents the article satisfies the property, and ``$\times$" represents not. 
		\end{tablenotes}
	\end{threeparttable}
	
	\label{tab:addlabel}%
	\vspace{-0mm}
\end{table*}%

\indent In summary, we provide a comparison of the above related surveys in \hyperlink{table 1}{Table \uppercase\expandafter{\romannumeral1}}. It is obvious that the use of IRSs or UAVs for 6G networks has been discussed from different perspectives. However, the advantages of the combination of IRSs and UAVs bringing to 6G networks and the applications they support are not comprehensively summarized, and the common issues that exist in these applications are not analyzed in depth. Furthermore, there is still a lack of comprehensive investigation on how technologies and solutions related to IRS-assisted UAV communications can be effectively utilized to mitigate communication issues for 6G networks.

\subsection{Contributions}
\indent For 6G networks, the integration of IRS and UAV communications can complement non-terrestrial networks and drive a comprehensive development of future wireless communications and their applications. \textbf{\textit{To the best of our knowledge, we are the first to provide a survey on IRS-assisted UAV communications for 6G networks by exploring corresponding application scenarios, common key issues, system prototypes, technological support, specific implementations, and research prospects.}} Specifically, the contributions of this article can be summarized as follows:

\begin{itemize}
	\item We introduce three typical applications of IRS-assisted UAV communications for 6G networks, and analyze and summarize three common issues related to these applications.
	
	\item We summarize prototypes of IRS-assisted UAV communications from multiple perspectives, and provide a detailed description of the related technologies for IRS-assisted UAV communications.
	
	\item Based on key issues faced by IRS-assisted UAV communications and applications supported for 6G networks, we summarize existing solutions and provide corresponding lessons learned.
	
	\item We present challenges and potential research directions for IRS-assisted UAV communications, which can guide future research and exploration for 6G networks.
\end{itemize}

\subsection{Organization}

\indent \hyperlink{Fig. 2}{Fig. 2} illustrates the structure of this article. In \hyperlink{Section2}{Section \uppercase\expandafter{\romannumeral 2}}, we provide a detailed introduction to application scenarios of IRS-assisted UAV communications for 6G networks and analyze the common issues in these scenarios. In \hyperlink{Section3}{Section \uppercase\expandafter{\romannumeral 3}}, prototypes and technologies of IRS-assisted UAV communications are described in detail. In \hyperlink{Section4}{Section \uppercase\expandafter{\romannumeral 4}}, we summarize existing solutions to the specific issues presented in \hyperlink{Section2}{Section \uppercase\expandafter{\romannumeral 2}}, including energy-constrained communications, secure communications, and enhanced communications. Then, some future challenges and open issues are described in \hyperlink{Section5}{Section \uppercase\expandafter{\romannumeral 5}}. Finally,  we conclude this article in \hyperlink{Section6}{Section \uppercase\expandafter{\romannumeral 6}}. A list of abbreviations used in this article is given in \hyperlink{table 10}{Table \uppercase\expandafter{\romannumeral 2}}.


	\hypertarget{table 10}{\begin{table*}[htbp]}
		\vspace{-1cm}
		\centering
		\fontsize{8}{10}
		\selectfont
		\caption{List of abbreviations.}
		\begin{tabular}{|p{1.7cm}|p{6.45cm}||p{1.7cm}|p{6.55cm}|}
			\hline
			Abbreviation & Description & Abbreviation & Description \\
			\hline
			5G/6G & 5-Generation/6-Generation & MEC & Mobile Edge Computing \\
			AO    & Alternating Optimization & ML & Machine Learning \\
			ARANs & Aerial Radio Access Networks & mMTC & massive Machine Type Communication \\
			BackCom & Backscatter Communication & mmWave & millimeter Wave \\
			BCD   & Block Coordinate Descent & NOMA & Non-Orthogonal Multiple Access \\
			BD    & Backscatter Device & OFDMA & Orthogonal Frequency Division Multiple Access \\
			BER   & Bit Error Rate & PLS & Physical Layer Security \\
			BS    & Base Station & PSO & Particle Swarm Optimization \\
			CNN   & Convolutional Neural Network & RF & Radio Frequency \\
			CoMP  & Coordinated MultiPoint & RL & Reinforcement Learning \\
			CR    & Cognitive Radio & RNN & Recurrent Neural Network \\
			CS    & Compressive Sensing & RSU & RoadSide Unit \\
			CSI   & Channel State Information & SAGIN & Space-Air-Ground Integrated Networks \\
			DL    & Deep Learning & SCA & Successive Convex Approximation \\
			DNN   & Deep Neural Network & SDR & SemiDefinite Relaxation \\
			DRL   & Deep Reinforcement Learning & SIC & Successive Interference Cancellation \\
			EE    & Energy Efficiency & SR & Symbiotic Radio \\
			eMBB  & enhanced Mobile BroadBand & SWIPT & Simultaneous Wireless Information and Power Transfer \\
			FBC   & Finite Blocklength Coding & TDMA & Time-Division Multiple Access \\
			GU/PU/SU & Ground User/Primary User/Secondary User & THz & TeraHertz \\
			HAP   & High Altitude Platform & UAV & Unmanned Aerial Vehicle \\
			IoD-Sim & Internet of Drones Simulator & URLLC & Ultra-Reliable and Low Latency Communication \\
			IoT   & Internet of Things & V2I/V2N & Vehicle-to-Infrastructure/Vehicle-to-Network \\
			IRS   & Intelligent Reflective Surface & V2P/V2V & Vehicle-to-Pedestrian/Vehicle-to-Vehicle \\
			ITS   & Intelligent Transportation System & V2X & Vehicle-to-Everything \\
			LoS/NLoS & Line-of-Sight/Non-LoS & ZSPs & Zone Service Providers \\
			LSTM  & Long Short-Term Memory &       &  \\
			\hline
		\end{tabular}%
		\vspace{-0.4cm}
		\label{tab:addlabel}%
	\end{table*}%

\hypertarget{Section2}{\section{Applications and Key Issues of IRS-Assisted UAV Communications for 6G Networks}}

\indent For 6G networks, the rise of IRS-assisted UAV communication technologies not only marks a major step forward in the field of wireless communications, but also provides novel solutions to address the increasingly complex communication requirements. Thus, we first explore the wide range of applications of IRS-assisted UAV communications in three typical 6G networks, including Space-Air-Ground Integrated Networks (SAGINs), V2X, and large-scale IoT. 
Although IRS-assisted UAV communications provide a new way to realize these applications, there are still some issues and challenges in system implementation. Therefore, starting from application scenarios, we sprovide insights into the key issues and challenges in realizing energy-constrained, secure, and enhanced communications for 6G networks based on IRS-assisted UAV communications. The existence of these issues points to the direction of development of IRS-assisted UAV communications.

\subsection{Applications of IRS-Assisted UAV Communications for 6G Networks}

\indent Since IRS-assisted UAV communication technologies are maturing, they are widely used in a variety of application scenarios for 6G networks. So we take three typical application scenarios for 6G networks, i.e., SAGIN, V2X communications, and large-scale IoT, as examples to illustrate the benefits brought by the integration of IRSs and UAVs. This categorization is mainly based on three service requirements for 5G and beyond networks, i.e., enhanced Mobile BroadBand (eMBB), Ultra-Reliable and Low Latency Communication (URLLC), and massive Machine Type Communication (mMTC) \cite{10354514}. Specifically, eMBB provides high data rates and reliability for SAGIN, while V2X requires URLLC for low-latency and reliable communication services, and mMTC supports low-cost communication connectivity for massive IoT devices.

\subsubsection{\textbf{IRS-Assisted UAV Communications for SAGINs}}

\begin{figure}[t]
	\begin{center}
		\includegraphics[width=0.5\textwidth]{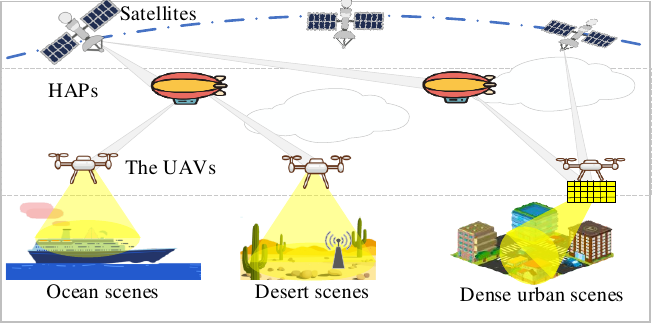}\\
		\hypertarget{Fig. 3}{\caption{A schematic of IRS-assisted UAV communications for SAGINs.}}
		\vspace{-8mm}
		\setlength{\parskip}{2cm plus4mm minus3mm}
	\end{center}
\end{figure}

\indent SAGIN, as a network architecture concept, is considered an important part of 6G network visions. The SAGIN refers to the integration of ground networks with aerial and space networks, providing global coverage and supporting communications for heterogeneous networks \cite{9520380}. This seamless coverage network provided by SAGINs plays a significant role in remote areas, maritime communications, post-disaster communications, and other fields. The SAGIN can be divided into three layers: space, air, and ground, which fully integrate communication resources of the three layers to leverage the advantages of heterogeneous networks \cite{10288083}. The air layer mainly refers to High Altitude Platforms (HAPs) composed of UAVs, balloons, airships, and other equipment, which can optimize and coordinate terrestrial and satellite resources for efficient communications and data transmission. However, unlike ground networks, limited computational capabilities and battery capacities of HAP equipments \cite{9520380}, as well as the impact of the wireless propagation environment on the propagation of communications among layers, can severely hinder the development of non-terrestrial networks. Therefore, the implementation of non-terrestrial networks is more challenging than that of ground networks.

\indent  In the three-layer architecture of SAGIN, as shown in \hyperlink{Fig. 3}{Fig. 3}, the space layer consists of various satellites to provide global communication, positioning, and surveillance capabilities. The ground layer includes ground stations, radar stations, and other communication and sensing devices, primarily involved in collecting, processing, and distributing data from the air and space layers. The air layer is made up of UAVs, IRSs, and airships, which mainly play the role of short-range data collection and communication relay. Specifically, the applications of IRS-assisted UAV communications in SAGIN are as follows:

\begin{itemize}
	\item Diversified link selection: Coordinating network resources of each network layer to realize efficient and reliable communication is the main design goal of SAGINs. HAPs usually operate in the air at 17-25km from the ground, and the weather has a negligible effect on them \cite{9822386}. Consequently, the links from satellites to HAPs should be maintained reliably \cite{9822386}. However, realizing reliable communications between HAP layers and the ground layers is challenging due to cloud cover and atmospheric turbulence. One potential solution is to use IRS-assisted UAVs as aerial relays to provide diversified link options when the communication quality between the air layer and the ground layer is weak, i.e., selecting IRSs on the UAVs for relaying while reducing the high signal attenuation caused by direct transmission. This is often attributed to the fact that IRSs can reflect the incident signal from the HAPs and ensure that the reflected signals are directed to the ground users. Moreover, UAVs can choose a suitable location to minimize the high signal attenuation caused by obstructions such as clouds.
	\item Multi-dimensional network coverage: Multi-dimensional network coverage is essential for applications like disaster response, remote sensing, and emergency communications in SAGINs. Traditional terrestrial networks struggle to provide effective global coverage, especially in inaccessible regions such as polar areas, deserts, and oceans. IRS-assisted UAV communications address these limitations by offering flexible, multi-dimensional coverage. IRSs can dynamically adjust the amplitude and phase shift of incident signals to reshape the propagation environment, enhancing signal reach and quality, especially for places that are difficult to reach signals directly \cite{9771729}. Furthermore, deploying IRSs on UAVs can extend coverage, providing reliable communication in remote regions. During emergencies, this approach can also help overcome signal blockages caused by infrastructure damage \cite{10330152}.

\end{itemize}

\subsubsection{\textbf{IRS-Assisted UAV Communications for V2X}}

\indent V2X refers to communication and interaction among vehicles and various entities in the surrounding environment. It mainly includes Vehicle-to-Vehicle (V2V), Vehicle-to-Infrastructure (V2I), Vehicle-to-Pedestrian (V2P), and Vehicle-to-Network (V2N) \cite{9681714}. It is a technology based on the Internet of vehicles and ITSs, aiming to improve vehicle safety, efficiency, and convenience. The implementation of V2X communications is required to handle real-time movement of vehicles and huge network demand \cite{9144463,9714139}. For the real-time mobility of vehicles, system design needs to be focused on problems including network topology transformation, and wireless access switching. Concurrently, high-quality links, imperceptible latency, and secure transmission are also demanded by V2X communications.

\begin{figure}
	\begin{center}
		\includegraphics[width=0.5\textwidth]{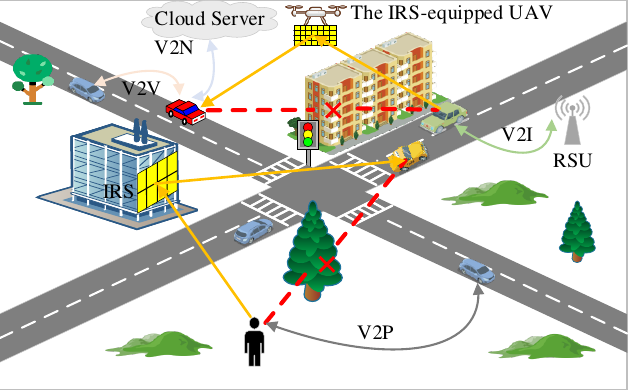}\\
		\hypertarget{Fig. 4}{\caption{A schematic of IRS-assisted UAV communications for V2X.}}
		\vspace{-5mm}
		\setlength{\parskip}{2cm plus4mm minus3mm}
		\vspace{-0.4cm}
	\end{center}
\end{figure}

\indent V2X represents a collection of various vehicle-related communications, primarily applied in urban areas to overcome communication barriers caused by terrain and buildings. In V2X, as shown in \hyperlink{Fig. 4}{Fig. 4}, IRSs can be deployed on buildings to provide communication services with LoS links for nearby users and devices. Furthermore, IRS-assisted UAV communications can be flexibly scheduled according to dynamic traffic communication requirements. Specifically, the applications of IRS-assisted UAV communications in V2X are summarized as follows:

\begin{itemize}
	\item Adaptability to dynamic V2X demands: In V2X communications, traffic flows and communication demands often vary dynamically over time and across regions, necessitating that V2X systems should adapt to these changes dynamically. However, traditional terrestrial networks typically rely on BSs for communication, where a few number of BSs can lead to network congestion during periods of high traffic flow, and conversely, an excess of BSs may result in resource wastage during idle periods. IRS-assisted UAV communications allow for flexible adjustments in response to various traffic flows and communication demands, for instance, by altering the UAVs' flight trajectories and adjusting the IRSs' reflection strategies to achieve demand-driven network deployment \cite{9681714}. Moreover, in emergency situations where RoadSide Units (RSUs) are damaged and non-operational, IRS-assisted UAV communications can also provide temporary communication services.
	\item Enhancing the quality and reliability of V2X communications: In urban environments, multipath propagation caused by signal reflection and refraction can interfere with signal reception. The intelligent control capability of IRS on incident signals reduces this interference by selectively enhancing the signals on favorable paths and attenuating the strength of interfering signals \cite{9144463}, which is particularly important for dynamic V2X communication environments. Meanwhile, the rational deployment of IRSs enables users to establish LoS links with ground and aerial BSs, significantly reducing signal attenuation caused by obstructions. Furthermore, UAVs equipped with IRSs can also be flexibly deployed, further reducing the risk of signal obstruction and improving the reliability of V2X communication links.
	\item Communication security and privacy protection: V2X-supported emerging applications, such as autonomous driving and ITS, impose high requirements on communication security and data privacy, aiming primarily at protecting users' life, property, and privacy \cite{9714139}. In dense urban environments, signal blockage caused by buildings or large vehicles often increases the risk of communication interference and privacy leakage. Therefore, by utilizing the beamforming technology of IRS to adjust the phase shift and amplitude of signals, a balance is struck between enhancing the quality of legitimate links and reducing that of illegitimate links, thereby maximizing the security of communication and minimizing the risk of privacy leakage. Furthermore, planning UAV trajectories to distance them from attackers can further mitigate the risk of communication attacks \cite{9416239}.
\end{itemize}

\subsubsection{\textbf{IRS-Assisted UAV Communications for Large-Scale IoT}}

\indent Large-scale IoT is a 6G vision that is gradually being proposed as networks evolve. IoT is a system that connects a wide range of intelligent digital devices with sensing and computing capabilities, widely used in urban construction, smart agriculture, healthcare, and home automation \cite{9685217}. However, with the continuous development of the IoT network, the shortcomings of traditional networks are gradually being exposed. First, the large number of IoT connections increases the network burden. Numerous network connections exacerbate spectrum scarcity, and the wide distribution of devices requires to overcome the path loss associated with long-distance transmission \cite{9804495}. Second, the shortage of energy for IoT devices limits the development of IoT. Finally, the security of IoT networks is crucial, since interconnectivity of IoT devices and their physical environment also introduce new attack surfaces \cite{8629941}. 

\begin{figure}[t]
	\begin{center}
		\includegraphics[width=0.5\textwidth]{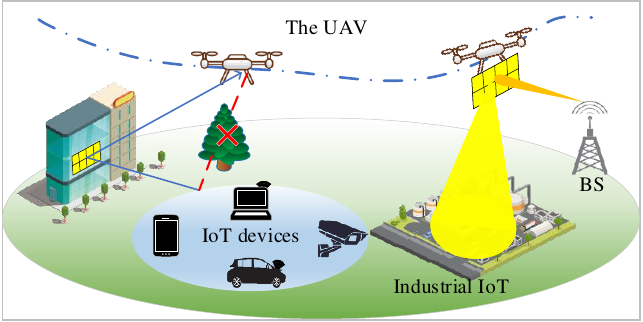}\\
		\hypertarget{Fig. 5}{\caption{A schematic of IRS-assisted UAV communications for large-scale IoT.}}
		\vspace{-5mm}
		\setlength{\parskip}{2cm plus4mm minus3mm}
		\vspace{-0.4cm}
	\end{center}
\end{figure}

\indent As illustrated in \hyperlink{Fig. 5}{Fig. 5}, UAVs are able to collect IoT device data in areas not covered by the cellular network. Deploying IRSs at suitable locations can mitigate signal attenuation caused by obstructions such as trees. In addition, aerial IRSs can also act as passive relays to enhance the network coverage quality of industrial IoT. Overall, IRS-assisted UAV communications help IoT networks to realize large-scale network accesses, high-quality transmission, prolonged IoT device lifetime and secure communications, which is described as follows:

\begin{itemize}
	\item Large-scale network accesses: Large-scale network access for IoT devices faces challenges with spectrum shortage and multi-user interference. IRS-assisted UAV communications offer solutions from multiple dimensions. First, multiple-access techniques such as NOMA are effective ways to enhance communication capacity. However, the performance of NOMA often depends on the differences among user channels. By designing the deployment location and reflection coefficient of IRSs, the channel differences among different users can be enlarged. The channel differences are even more pronounced in scenarios where the IRS is deployed dynamically by leveraging the mobility of UAV. Moreover, IRSs can reconfigure channel conditions, and thus the decoding order of users can be changed according to the demands of quality of service \cite{9454446}. Second, compared to active relays, IRSs offer advantages of low cost, low power consumption, and localized service range. This allows them to be densely deployed in IoT communication areas to increase communication capacity without causing environmental noise amplification or requiring complex interference management \cite{9326394}.

	\item High-quality transmission: IoT devices are usually widely distributed on the ground, which can cause huge deployment costs if cellular networks are utilized to collect IoT data by deploying BSs \cite{9685217}. IRSs and UAVs can provide network coverage for a designated area in a cost-effective way. Meanwhile, with the beamforming of IRSs and the mobility of UAVs, reliable LoS links can be established with ground nodes to improve transmission quality.

	\item Prolonged IoT device lifetime: Lifetime of IoT devices is crucial for the sustainability and efficiency of IoT, having a direct impact on their maintenance costs and operational reliability. IRSs and UAVs can prolong the lifetime of IoT devices from two perspectives: First, the combination of IRSs and UAVs can enable IoT devices to achieve low-power communications by specific optimisations including UAV trajectories, IRS deployment locations and beamforming. Second, they can improve the performance of Simultaneous Wireless Information and Power Transfer (SWIPT). IRS-based passive relaying and beamforming can reduce the energy loss caused by long-distance energy transmission, while UAVs can fly near the ground nodes to reduce the energy transmission distance, which extends the lifetime of IoT devices from a sustainable perspective.
	
	\item Secure communications: The security of IoT networks can be enhanced from the physical layer through the use of IRS beamforming and design of trajectories for UAVs to reconfigure the wireless environments. In addition, UAVs can generate artificial noise to assist in the joint optimization of IRS beamforming design, effectively countering illegal eavesdropping.
	
\end{itemize}

\indent In summary, through the analysis of three typical applications: SAGIN, V2X, and large-scale IoT, it can be confirmed that IRS-assisted UAV communications can provide diversified services for various applications supported by 6G networks. Indeed, based on these three applications, multiple 6G network-driven application scenarios are also expected to realize. For example, SAGIN can provide comprehensive connectivity services for smart cities. Based on the low latency and high reliability of communications offered by V2X, ITSs and autonomous driving for 6G networks can be achieved. The characteristic of massive network access in large-scale IoT can contribute to the realization of smart healthcare, by connecting hospital, home, wearable medical devices and sensor data for real-time health monitoring and analysis. In addition, IRS and UAV can play a pivotal role in future fields such as virtual reality \cite{10379001} and environmental monitoring \cite{9580624}.

\hypertarget{Section2.B}{\subsection{Key Issues of IRS-Assisted UAV Communications for 6G Networks}}

\indent Although IRS-assisted UAV communications offer a new perspective for realizing the aforementioned applications, the utilization of this emerging method to achieve these applications still faces numerous challenges. Through an in-depth analysis of these applications and IRS-assisted UAV communication systems, we summarize key issues of these applications, and analyze corresponding challenges as follows:

\subsubsection{\textbf{Energy-Constrained Communications}}

\indent In IRS-assisted UAV communications, energy limitations are obvious, which challenge the process of communications. For IRS-assisted UAV communication systems, the limited on-board energy of UAVs severely impacts the service time of UAV networks\cite{9870557}. Specifically, the onboard energy of UAVs is not only used to maintain their propulsion, but also for signal relaying or transmission \cite{9690481}, while the former typically consumes more energy than the latter. In order to efficiently utilize the limited service time of UAVs supported by limited energy, a trade-off needs to be made between the service time of UAVs and other performance metrics such as throughput \cite{10108047}, which has become a major bottleneck limiting the development of IRS-assisted UAV communications.

\indent For applications supported by IRS-assisted UAV communications, the limited battery capacity of devices affects their normal operation. In general, most IoT devices supported by IRS-assisted UAV communications are powered by on-board batteries. Effective energy replenishment and low-power transmission methods are crucial to ensure the long-term stability of IoT devices \cite{9894720,9756208}. SWIPT is one effective means of energy supplementation. However, the significant energy loss associated with long-distance transmission limits the use of SWIPT. Furthermore, network devices supported by IRS-assisted UAV communications in remote areas face challenges for low-power transmission due to the complexity of wireless environments.

\indent There are multiple challenges to addressing energy-constrained communications in IRS-assisted UAV communications and the application scenarios they support, which are listed below:

\begin{itemize}
	\item Complex wireless environments: UAVs usually work in complex and dynamic wireless environments, such as high-speed mobile scenarios based on V2X communications and high-altitude scenarios based on SAGIN. This poses various challenges on signal and energy transmission of UAVs. Therefore, the design of IRSs and UAVs must be appropriately adapted with the dynamic wireless environments. First, UAVs need to rationally avoid ground obstacles and balance the flight energy consumption and communication quality. Second, the IRS needs to adjust its reflection matrix instantly to assist UAV communications with the changing environments, which increases the difficulty of IRS beamforming and beam tracking. In addition, efficient communication relies on accurate environment modeling and prediction, which is challenging to realize due to the passive nature of IRSs and highly dynamic nature of the wireless environment caused by UAV motion \cite{9743298}. Especially for air-to-ground communication scenarios, UAV jitter should be carefully treated \cite{10061643,9681714}.
	
	\item Multiple factors to balance: Energy consumption can be caused by multiple factors, including UAV flight trajectories, resource allocation strategies, environmental impact factors, IRS design, transmission beams, and the number of served users. Since these factors need to be comprehensively considered, it is significantly difficult to solve the formulated optimization problems.
	
	\item Difficulties in technology fusion: To effectively alleviate the problem of limited energy, it is often necessary to integrate other technologies (such as SWIPT and Backscatter Communication (BackCom)) into the framework of IRS-assisted UAV communications, which brings new problems. For example, when using SWIPT technology, energy collection and information transmission are conflicted with each other \cite{9894720}. In addition, in the BackCom scenario, the trade off between energy collection and signal reflection is also challenging \cite{10066841}.
\end{itemize}

\subsubsection{\textbf{Secure Communications}}

\indent They are critical to the applications supported by IRS-assisted UAV communications. For SAGINs, secure communications are the cornerstone for enabling cross-layer network collaboration, optimizing resource allocation, and securing the transmission of critical information. For V2X, secure communications are closely related to the security of users' lives, properties and privacy. For large-scale IoT, secure communications are the key to protect the IoT data from unauthorized access or tampering, ensuring overall network security and user trust. In the context of IRS-assisted UAV communications, security concerns primarily involve two aspects: ensuring the integrity of transmission data and maintaining the confidentiality of communication contents. The former aims to prevent data tampering, damage or loss, while the latter focuses on preventing information leakage. Based on different security concerns, common security threats in IRS-assisted UAV communications can be classified into two categories: malicious jamming and eavesdropping attacks.

\indent Malicious jamming aims to disrupt, compromise, or undermine the performance and reliability of communication systems, which is one of the most easily achievable attacks in IRS-assisted UAV communications. It does not require much information about the target user, but rather interferes with and floods data towards the target user with malicious jamming devices \cite{10139787}.

\indent Eavesdropping attacks are able to illegally acquire network data, leading to information leakage. They can be active or passive \cite{9696283}. In the passive eavesdropping attacks, the eavesdroppers do not take any actions other than potentially obtaining the transmission information, thus not blocking the legitimate users' information reception. Active eavesdroppers use malicious jamming devices to send jamming signals to the target channel, which are typically disguised as indistinguishable natural noise. Therefore, the sending end usually increases the transmission power to maintain system transmission performance, which increases the probability of successful eavesdropping \cite{9940551}.

\indent In IRS-assisted UAV communications, although other attacks, such as time delay attack \cite{10115020}, still exist, this paper only focuses on jamming and eavesdropping attacks. The specific reasons are as follows: First, for IRS-assisted UAV communication systems, jamming and eavesdropping are easier to implement and more prevalent compared to other attack methods. Second, they can be effectively mitigated through system-level design using Physical Layer Security (PLS) technologies. A discussion of other UAV or IRS related security attacks can be found in \cite{10261240,10121733}.

\indent There are two reasons that make it challenging to address security issues in IRS-assisted UAV communications:

\begin{itemize}
	\item Randomness of attacker's illegal access: Since UAVs move along optimized trajectories within designated areas, both legitimate and illegitimate users randomly access the network based on their communication requirements. This dynamic random access mechanism makes it difficult to distinguish illegal attackers. In addition, obtaining the attacker's accurate Channel State Information (CSI) and location information is the basis for effective defense against attacks. However, the dynamic random access also makes the above information difficult to obtain \cite{9771762,9538830}. In fact, part of the attacker's CSI and location information can be obtained through sensing devices mounted on UAVs and IRSs, but how to robustly design IRS-assisted UAV communications and its supported applications with partial information is still challenging.
	
	\item Limited energy and computational capabilities: UAVs and IRSs typically rely on built-in batteries with limited capacities, constraining their ability to perform complex data processing over extended periods. Moreover, UAVs and other devices generally have limited computational power, rendering them incapable of executing complex encryption algorithms and real-time security protocols, thereby leaving them vulnerable to attackers with substantial computing resources. Attackers can exploit this computational advantage to launch sophisticated eavesdropping or jamming attacks, while the constrained computational resources and energy capacity make it challenging for IRS-assisted UAV communication systems to effectively identify and defend against these attacks \cite{9527176}. Therefore, designing secure communication strategies while efficiently utilizing limited computational capabilities and energy becomes significant.
\end{itemize}

\subsubsection{\textbf{Enhanced Communications}}
\indent For 6G networks, enhanced communications are needed in the fields of SAGIN, V2X, and large-scale IoT, requiring the achievement of ultra-low latency, high reliability, broad coverage, and massive connectivity \cite{8869705}. Specifically, enhanced communications need to support seamless data transmission globally, real-time response of ITSs, and efficient connection of massive IoT devices, while focusing on the EE and cost control of network deployment. Although IRS-assisted UAV communications provide possibilities for enhanced communications, there are still multiple challenges to achieve fast network rates, large network coverage, short network latency, high spectrum utilization and reliability. These challenges include difficulties of interference management, spectrum resource shortage, and nonconvexity of multivariate coupling.

\begin{itemize}
	\item Difficulties of interference management: The mobility of UAVs results in UAV communication systems vulnerable to unpredictable signal interferences, including interference from broadcast signals, signal congestion caused by simultaneous communications among multiple users, and unintentional interference from other devices. Unlike jamming attacks, this unpredictable interference does not lead to information leakage, but it seriously affects the signal transmission quality. Although IRSs can avoid some signal interference by beamforming, coordinating beamforming becomes a complex task for dense-user scenarios. In particular, this interference is exacerbated when the IRSs are deployed on UAVs. In addition, it is possible to jointly optimize the beamforming of IRSs, power allocation, and trajectory design of UAVs to reduce the impact of communication interference. However, taking multiple factors into account makes the optimization problem extremely challenging to resolve \cite{9749020,9912224}.
	
	\item Shortage of spectrum resources: The spectrum resource shortage arises along with the proliferation of network devices. In IRS-assisted UAV communications, only a small amount of spectrum resources can be used. Although some articles \cite{9745104,9789841,9367288,9528924} have proposed to use frequency bands such as mmWave and THz for communications to alleviate the spectrum shortage, the impact of path loss on the system caused by ultra-high frequency cannot be ignored. In addition, multiple access technologies can also effectively improve spectrum utilization, whereas interference management among users and cells is necessary, which increases the difficulty of beamforming of IRSs and joint design with other technologies.
	
	\item Nonconvexity of multivariate coupling: In order to improve the performance of IRS-assisted UAV communication systems, the joint design of multiple variables is usually required, including UAV trajectories, IRS phases, BS transmission power and bandwidths. However, variables, such as UAV trajectories and beamforming of IRSs, the decoding order of Non-Orthogonal Multiple Access (NOMA) and association results between IRSs and users are often coupled \cite{9893192,9454446}, resulting in difficulties of problem solving with the characteristic of non-convexity. Therefore, how to design reasonable algorithms for the non-convex problems is the key to enhance communications.
\end{itemize}

\indent In summary, the main challenge to realize enhanced communications in IRS-assisted UAV communications arises from comprehensive consideration of the aforementioned issues in the system design. Furthermore, realistic conditions, such as non-ideal channel conditions and UAV jitter, need to be taken into account.

\hypertarget{Section3}{\section{Prototypes and Technologies of IRS-Assisted UAV Communications for 6G Networks}}

\indent The combination of IRSs and UAVs fully utilizes the advantages of each and eases the pressure on ground communications in an innovative way. In this section we first present the system prototypes for IRS-assisted UAV communications from four perspectives. Then we present in detail the technologies related to IRS-assisted UAV communications for 6G networks.

\subsection{Prototypes of IRS-Assisted UAV Communications for 6G Networks}
\indent To comprehensively analyze IRS-assisted UAV communications, we introduce the prototypes from the perspectives of hardware, software, performance metrics, and prototype classification.

\subsubsection{\textbf{Hardware Prototypes for IRS-Assisted UAV Communications}}
\indent In general, hardware prototypes of UAVs typically contain propulsion module, control module, sensing module, and communication module \cite{10.1007/s10846-021-01383-5,9836083}. Propulsion module mainly provides the propulsion of UAV, enabling it to fly and maneuver. Control module mainly manages the flight attitude, position and flight path of the UAV to ensure it is stable and controllable, and Pixhawk is a common flight controller. Sensing modules provide the data required for flight control and mission execution by sensing the environment and its own state. Communication model realizes data transmissions between the UAV and the BS or other devices, supporting remote control and data return. As the technology continues to evolve, UAV communication prototypes have been studied. Typically, authors in \cite{9836083} construct a modular autonomous UAV platform designed for a variety of indoor and outdoor applications, simplifying the transition from simulation experiments to actual deployment, and providing detailed mechanical design, electrical configuration, and dynamic modeling to support diverse experiments and research.

\indent Although the reflecting unit of an IRS can be composed of a variety of materials, such as diodes, liquid crystals and graphene meta-atoms, the IRS hardware prototype typically contains reflecting unit array module, control circuit module, power management module, and communication module \cite{10596064}. The reflecting unit array module adjusts the electromagnetic response of reflecting units to incident signals by means of control information generated in the control circuit module. Control circuit module usually consists of microcontroller or field-programmable gate array for executing control algorithms and transmitting control commands to reflecting units \cite{zhang2018space}. Although reflecting units do not need to consume energy, their control circuits usually need to consume a weak amount of power, which is usually accomplished by power management modules \cite{9140329}. In addition, the communication module needs to receive control signals from UAVs or BSs to the IRS.

\indent Currently, research on IRS-assisted UAV communications is in the transitional phase from theoretical study to practical application. To maximize the flexibility of IRS, authors in \cite{9794781} propose a typical aerial IRS control architecture. This architecture is divided into two main parts: ground control layer and UAV control layer. The former includes a processing center responsible for analyzing environmental data and assigning communication tasks. Considering the limited energy capacity of IRS and UAV in handling complex computational tasks, the ground control layer also calculates UAV trajectories and IRS phase shift control information. The latter consists of an onboard antenna coupled with the UAV, managing the battery and communication gateway. The integrated UAV control module and IRS control module can receive control information from the ground control unit to adjust UAV trajectories and IRS reflection coefficients, thereby enhancing communication quality.

\subsubsection{\textbf{Software Prototypes for IRS-Assisted UAV Communications}}
\indent In IRS-assisted UAV communications, software prototypes are used to simulate the proposed schemes in different environments and conditions. It is mainly used to evaluate the system rationality and algorithm performance. For IRS software prototypes, authors in \cite{9852389} utilize Coffee Grinder Simulator, a system-level simulation platform for evaluating the deployment of IRS in 5G networks, to quantify the expected enhancement of IRS in terms of coverage probability and network performance by simulating a variety of urban environments in the medium and high frequency bands. For UAV software prototypes, authors in \cite{park2020devising} propose a distributed UAV network simulation coordinator, which significantly improves the accuracy and efficiency of multi-UAV network scenario simulation by running flight and network simulations simultaneously and dealing with the time synchronization between them. However, the software prototype in \cite{park2020devising} lacks the features and local details of UAVs, wireless communications and network protocol simulation. For this reason, authors in \cite{9893879} propose an open source simulation platform called Internet of Drones Simulator (IoD-Sim), which can be extended on network simulator 3 to support a multi-level simulation architecture, and thus UAV mission design, trajectory planning, hardware configuration, and communication techniques are simulated and verified with high accuracy.

\hypertarget{Fig. 13}{\begin{figure}[t]}
	\begin{center}
		\includegraphics[width=0.48\textwidth]{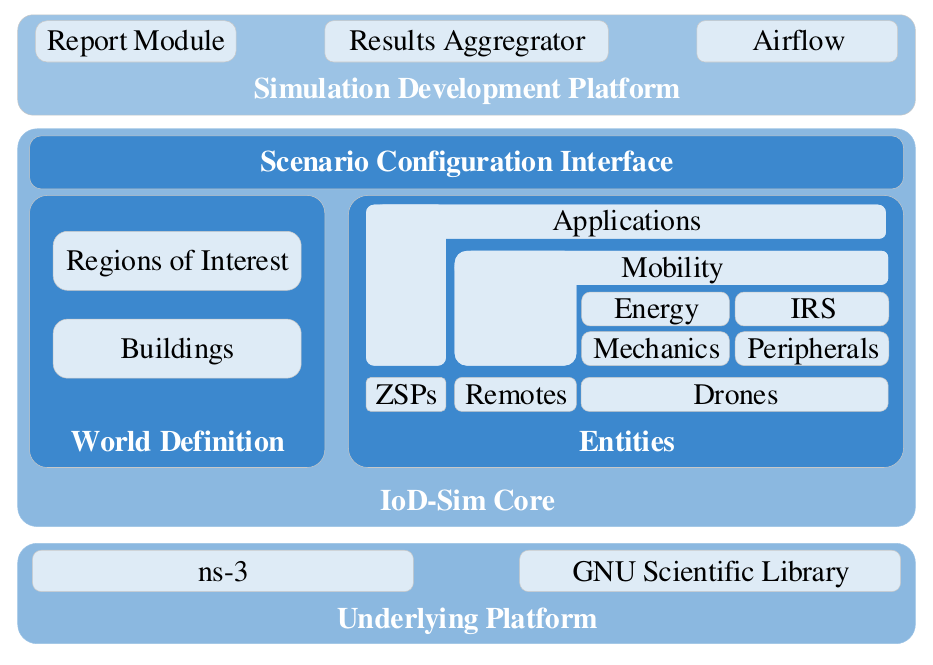}\\
		\caption{A schematic of IoD-Sim-based IRS-assisted UAV communication simulation architecture}
		\vspace{-5mm}
		\setlength{\parskip}{2cm plus4mm minus3mm}
		\vspace{-0.4 cm}
	\end{center}
\end{figure}

\indent In order to further study the advantages of IRSs combined with UAVs in wireless communications, authors in \cite{10356747} propose a typical IRS-assisted UAV communication software simulation prototype by adding an IRS simulation module on the basis of IoD-Sim. As shown in \hyperlink{Fig. 13}{Fig. 6}, the platform is divided into three layers: the underlying platform, the IoD-Sim core and the simulation development platform. Specifically, the underlying platform adopts network simulator-3 for network simulation and GNU scientific library for mathematical operations. IoD-Sim core introduces IoD entities, 3D world models and remote cloud services. In particular, the UAV is mechanically modeled by considering physical and power consumption characteristics. Zone Service Providers (ZSPs) are introduced to provide services beyond traditional coverage. The top-level simulation development platform provides scenario configuration through a user-friendly graphical interface or JSON. The software prototype supports a range of applications and peripherals, including IRS modules, which assist in the performance evaluation of IRS-assisted UAV communications.

\subsubsection{\textbf{Performance Metrics for IRS-Assisted UAV Communications}}
\indent When evaluating IRS-assisted UAV communication systems, performance metrics not only provide clear objectives for the design, but also provide quantitative criteria for system evaluation and optimization. Based on the three issues considered in this article, the performance metrics can be broadly summarized as: EE, secure communication rates, and network rates. 
\begin{itemize}
	\item EE: It is defined as the ratio of system network rates to energy consumption and is often used in energy-constrained communication scenarios to help evaluate system performance. 
	Authors in \cite{10296049} analyze the relationship between the number of IRS reflecting units and EE in the presence of eavesdropping scenarios by modeling the actual air-to-ground channel and phase estimation errors, and their results validate the superiority of IRS-assisted UAV communications. Unlike the eavesdropping scenario in \cite{10296049}, authors in \cite{10330152} analyze the EE of IRS-assisted UAV communications in an emergency communication scenario. Based on the newly developed fading-shadowing model, authors provide insights into the impact of the number of reflecting elements and UAV deployment altitude on system EE.

	\item Secure Communication Rates: It usually refers to the ability to maintain secure communications in the presence of electromagnetic jamming or eavesdropping. For eavesdropping scenarios, secure rates refer to the difference between the users' acceptance rates and the eavesdropping rate. For jamming scenarios, secure communication rates are expressed as the network rates that maintains stable communications. 
	The analysis of secure communication rates for IRS-assisted UAV communications in jamming and eavesdropping scenarios is conducted in \cite{9538830} and \cite{10049533}, respectively. Authors in \cite{9538830} utilize stochastic geometry to model eavesdroppers in a distributed manner and characterize the impact of the number of IRS reflecting units and UAV positioning on secure communication rates in the presence of channel estimation errors. For jamming scenario, authors in \cite{10049533} derive the overall bit error rate for two IRS deployment scenarios. The simulation results indicate that jamming to the UAV causes more significant performance degradation compared to jamming at the receiving end, and deploying the IRS near the transmitter can maximally suppress jamming.
	
	\item Network Rates: It usually refers to the amount of data that can be transmitted per unit of time, and is defined in most studies by Shannon's formula. It is commonly used in enhanced communication scenarios to measure the communication efficiency of systems. For example, authors in \cite{9395180} derive the network rate of UAV communications assisted by flying IRS under imperfect phase compensation.
\end{itemize}

\subsubsection{\textbf{Classification for IRS-assisted UAV communications}}
\indent In IRS-assisted UAV communications, to fully leverage the advantages of IRSs and UAVs, different prototypes are usually required for various application scenarios. Below, we classify and discuss the prototypes of IRS-assisted UAV communications from two perspectives: service objects and service types.

\indent Prototypes Based on Different Service Objects: IRS-assisted UAV communications can be divided into two main prototypes from the perspective of service objects: ground-based IRS-assisted UAV communications and UAV-carried IRS-assisted ground communications. 
In the former case, IRSs are deployed on the ground to assist aerial UAV communications. The aerial UAVs typically serve as aerial BSs to provide communication services for ground users \cite{9293155}. They can also act as edge servers, collecting and processing computational tasks generated by ground devices \cite{9804341,9771971}. Meanwhile, ground-based IRSs usually act as passive relays, providing virtual LoS links for ground users and UAVs. Although this method can enhance the coverage and quality of UAV signals, it has high deployment and maintenance costs. As a result,  it is suitable for dense urban environments or other complex terrains, especially when direct LoS links between UAVs and ground users are not achievable.

\indent UAV-carried IRS-assisted ground communications treat the IRS and the UAV as a whole aerial relay unit, where the UAV carries the IRS to an appropriate position, and the IRS relays signals in the air \cite{9785612}. This prototype boasts high flexibility, since strategic planning of the UAV's flight path reduces terrain obstructions. However, the UAV's battery life limits the duration of continuous communications, and instability during flight can significantly impair the qualities of communications. Therefore, this prototype is suitable for scenarios requiring rapid deployment, such as disaster response, temporary events, and V2X communications.
	
\indent Prototypes Based on Different Service Types: According to different functions and types, IRS-assisted UAV communications can be divided into: passive IRS-assisted UAV communications, active IRS-assisted UAV communications, and hybrid IRS-assisted UAV communications. Without loss of generality, IRS is usually passive, and its phase shift and amplitude adjustments for incident signals do not need to consume energy, only need to consume a subtle amount of energy to control the reflection matrix and receive the IRS control signal. Consequently, the energy consumption of the IRS is usually negligible for both dynamic and static IRS scenes. Correspondingly, this prototype is usually suitable for remote areas where energy cannot be replenished in time to support applications such as environmental monitoring.

\indent Active and passive IRS have the same structure, with the only difference being that the passive load impedances in passive IRS is replaced by active ones in active IRS. So in addition to the passive IRS power consumption, active IRS requires extra power to support the active load impedances. Studies show that the power consumption of active IRS reflection units can be reduced to the microwatt level \cite{9377648}. Therefore, without loss of generality, active IRS is also considered passive. Due to its ability to adjust both the phase and amplitude of incident signals, active IRS-assisted UAV communications are typically used to mitigate the ``double fading" effect caused by passive IRS to further enhance communication qualities \cite{10061643}. However, this prototype, while amplifying the signal, also amplifies noise, increasing the complexity of system interference management, especially in the dynamic scenario with the active IRS on the UAV \cite{10214219}. As a result, this prototype is suitable for long-distance and low power communication scenarios.

\hypertarget{Fig. 6}{\begin{figure}[t]}
	\begin{center}
		\includegraphics[width=0.4\textwidth]{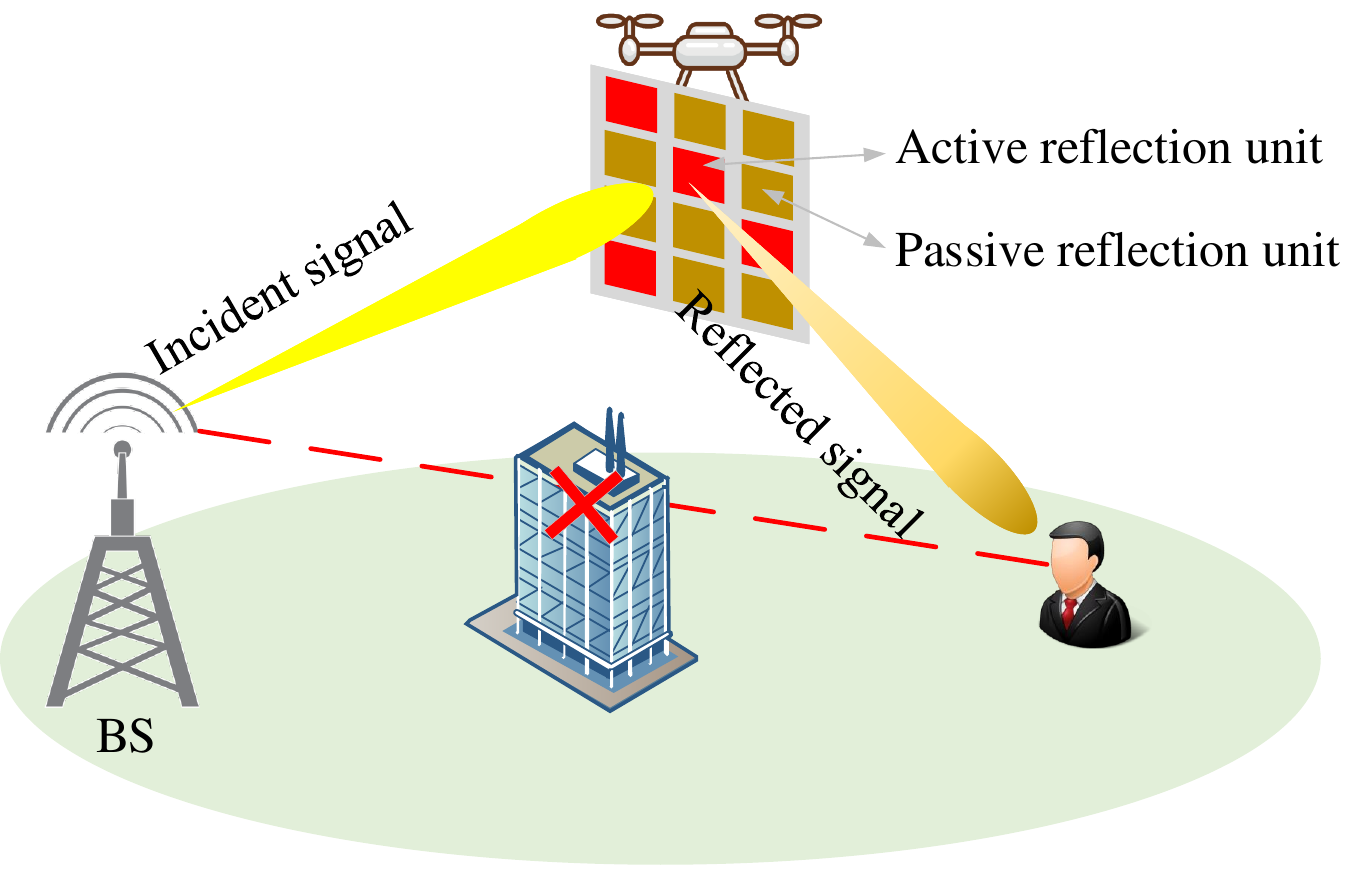}\\
		\caption{A schematic of hybird IRS-assisted UAV communications.}
		\vspace{-5mm}
		\setlength{\parskip}{2cm plus4mm minus3mm}
		\vspace{-0.4 cm}
	\end{center}
\end{figure}

\indent In order to overcome the respective drawbacks of active and passive IRS, a hybrid IRS architecture is proposed. As shown in \hyperlink{Fig. 6}{Fig. 7}, each reflection unit of the hybrid IRS is able to adaptively switch between active and passive reflection modes according to the communication requirements. In this way, by reasonably controlling the reflection unit model, not only the ``double fading" caused by passive IRS can be effectively overcome, but also the noise amplification caused by active IRS can be effectively restrained, so as to enhance the beamforming gain. Based on this feature, the hybrid IRS can further enhance the performance of UAV communications. For example, it can be deployed on ground buildings to enhance the EE of UAV communication systems \cite{10472415}. Hybrid IRS-assisted UAV communications can also improve communication security \cite{10198901}. However, it requires flexible adjustment of reflection patterns of reflection units, which undoubtedly increases the difficulty of system design. Therefore, hybrid IRS-assisted UAV communications are suitable for scenarios with strict requirements on communication qualities.

\subsection{Technologies of IRS-Assisted UAV Communications for 6G Networks}
\indent In order to enhance the performance of IRS-assisted UAV communications to support applications such as SAGIN, V2X, and large-scale IoT, researchers typically integrate multiple technologies to enhance the system performance, such as channel estimation, beamforming, resource allocation, and trajectory optimization. In the following, we provide a detailed description of these technologies, which are the basic for solutions in \hyperlink{Section4}{ Section \uppercase\expandafter{\romannumeral4}}. 

\subsubsection{\textbf{Channel Estimation for IRS-Assisted UAV Communications}}

\indent It refers to the estimation and inference of channel characteristics based on the received signal in wireless communication systems, to correctly process and decode signals at the receiving end. In general, channel estimation relies on the selection of appropriate channel models and estimation algorithms to accurately and efficiently obtain the complete CSI. In other words, it intends to obtain a complete and accurate channel state based on the limited CSI. In IRS-assisted UAV communications, the challenge of channel estimation comes from two aspects: First, for UAVs, channel estimation needs to have a strong adaptive nature. Because UAVs are with mobilities and cause wireless channel switching, UAV locations and environmental characteristics play a crucial role in determining the primary channel qualities \cite{8411465}. Second, for IRSs, the number of reflection units is proportional to that of channel coefficients \cite{9771718}, which may cause a huge estimation overhead. In addition, the IRS lacks signal processing capabilities and traditional channel estimation methods are not fully applicable \cite{9559873}.

\indent For IRS-assisted UAV communications, accurate channel estimation is the foundation of system design. In order to improve the performance of channel estimation in IRS-assisted UAV communications, many methods are proposed, including Compressive Sensing (CS)-based channel estimation, cascade-based channel estimation, anchor-assisted channel estimation, Deep Learning (DL)-based channel estimation, and tensor decomposition-based channel estimation. Each method is introduced in the following content.

\paragraph{CS-based channel estimation}
\indent It is commonly utilized for IRSs and UAVs, and can be integrated with other channel estimation techniques for performance improvement \cite{9967965}. The basic idea is to leverage the high sparsity of the wireless channel by extracting a sparse representation from a limited set of measurement data, to recover the complete CSI. A few pilot signals are utilized to obtain relatively accurate CSI, which greatly reduces the training overhead of channel estimation while effectively estimating channel parameters \cite{9521836,9513592}. 

\paragraph{Cascade-based channel estimation}
\indent It is suitable for scenarios with multi-level channels and multiple users. This method exploits the correlation property of channel cascading, i.e., the cascaded channel coefficients are scaled versions of the superimposed low-dimensional CSI of the specific channel \cite{9373363,9771718}, thus significantly reducing the training overhead of channel estimation. For multi-user scenarios, the channel of any single user is also a low-dimensional scaled version of other channels, thus allowing effective estimation of all channels with a low overhead on the BS \cite{9373363}.

\paragraph{Anchor-assisted channel estimation}
\indent It is mainly applicable to scenarios with a large number of fully passive IRS reflective elements and users. The method first allows anchor nodes to be deployed near the IRS to obtain partial CSI through anchor-assisted training and feedback. The information is then used to efficiently estimate the cascaded channels between the BS and the IRS with additional training by users \cite{9603291}. Thus, with the assistance of anchor nodes, this method can reduce the huge training overhead incurred by signal transmission and reception, due to the increasing number of users and IRS reflection units.

\paragraph{DL-based channel estimation}	
\indent DL-based channel estimation is suitable for highly complex dynamic channel environments with high-dimensional spatial features and nonlinear channel characteristics. Its main idea is to use DL methods, such as neural networks, to analyze and process received signals and obtain CSI. Recurrent Neural Networks (RNNs) and Convolutional Neural Networks (CNNs) are commonly used methods. RNNs can capture the temporal dependencies in sequential data, allowing them to estimate the current channel based on previous channel samples, and is therefore commonly used for continuous-flight UAV channel estimation \cite{8847452}. Long Short-Term Memory (LSTM) is a special RNN algorithm that overcomes the gradient vanishing and gradient explosion problems of traditional RNNs when dealing with long sequences. Through the gating mechanism and the memory unit, LSTM can capture the long-term dependency of the channel, so as to track the channel and realize dynamic channel estimation \cite{9743298}. 

\indent CNNs can autonomously learn feature representations suitable for channel characteristics. With their deep structures, weight sharing, and parallel computing capabilities, CNNs can reduce the algorithm complexity of channel estimation for a large number of IRS reflection units, while ensuring estimation accuracy \cite{9505267,9622178}. In addition, the offset learning in DL can simulate dynamic channel states \cite{9622178}, and deep residual learning provides a method for recovering channel coefficients from the noise-based pilot observation data \cite{9505267}.

\paragraph{Tensor decomposition-based channel estimation}	
\indent It is mainly applied in large-scale antenna scenarios with multiple inputs and multiple outputs. The tensor-based channel estimation describes the spatio-temporal characteristics of the channel by constructing a channel tensor model, and then uses mathematical methods such as tensor decomposition to process and optimize the channel tensor to estimate CSI. This method can capture the high-order relationships and spatial correlations of the channel, thus improving the accuracy and efficiency of channel estimation. In addition, tensor-based channel estimation can also exploit channel sparsity to further improve its performance \cite{9745538}.

\subsubsection{\textbf{Beamforming for IRS-Assisted UAV Communications}}

\indent Beamforming refers to adjusting the phase shift and amplitude of signals by controlling the electromagnetic response of the corresponding devices (e.g., IRSs and transmitting antennas) to form different radiation patterns on the signals \cite{665}. In essence, beamforming is a spatial filtering method, which is initially used for specific directional radiation or energy acquisition. In IRS-assisted UAV communications, beamforming is the key to reconfigure the wireless propagation environment and reduce inter-user or inter-cell signal interference\cite{10075500}. It also plays a significant role in ensuring system security, since it can suppress interference signals while enhancing the desired signal \cite{9810528,9656117}. However, when the IRS assists many users or cells at the same time, the effectiveness of beamforming in mitigating interference decreases as the number of users or cells increases. Fortunately, this problem can be effectively addressed by integrating other technologies and increasing the number of IRSs \cite{9893192}. Besides, when IRSs are deployed on UAVs, the interference suppression effect is dramatically improved by coordinating UAV trajectories and beamforming design \cite{9810528}.

\indent To reduce the energy consumption of UAVs and extend communication duration, current research defines that the control information for IRS should be generated by BS \cite{9453804,9198125}. Specifically, the ground BS collects CSI from the UAV, IRS, and other communication nodes, then formulates optimal IRS beamforming and UAV trajectory control strategies, and provides the control information as feedback. However, direct feedback of control information and data transmission can inevitably cause interference. Therefore, authors in \cite{9453804} propose an adaptive IRS-assisted transmission protocol, which alternates channel estimation, transmission strategies, and data transmission within a frame to reduce interference and adapt to the dynamic wireless environment. Similarly, authors in \cite{9198125} propose a dual time-scale transmission protocol and jointly utilize instantaneous and statistical CSI to reduce beamforming overhead. It is noteworthy that the method of generating and feeding back IRS control information by the ground BS can lead to some latency in IRS beamforming, but this can be negligible in short time frame structures.

\indent In IRS-assisted UAV communications, beam tracking design during the beamforming process is needed to adapt the beams to the UAV's movement. Therefore, beamforming can be further categorized into static beamforming and beam tracking (dynamic beamforming).

\paragraph{Static beamforming}
\indent Its critical design challenge is the trade-off between the overhead of acquiring instantaneous CSI and the performance of beamforming. In fact, the effectiveness of beamforming largely depends on the accuracy of CSI, where better beamforming results are achieved with more accurate CSI. However, there is currently no mature method to obtain precise instantaneous CSI for IRSs \cite{9198125,9325920,9348156}. Most beamforming designs for IRSs are based on statistical/mixed CSI \cite{9479733}, using statistical CSI of IRSs as a substitute for part of the instantaneous IRS to balance the channel estimation overhead and beamforming performance. It is worth noting that for scenarios with unavailable instantaneous CSI of IRSs, beamforming can be realized by techniques such as beam training, channel tracking, and heuristic algorithms \cite{9722893}. The beamforming for hybrid CSI and non-explicit CSI is described below.

\begin{itemize}
	\item Beamforming with hybrid CSI: It is a beamforming approach by reasonably weighing and combining statistical and instantaneous CSI. Statistical CSI changes slowly and only statistical characteristics of the channel are needed to know, and thus it is easier to obtain compared with instantaneous CSI. Therefore, this method has the advantage of low overhead. A common channel estimation method for hybrid CSI is the dual time-scale based hybrid CSI beamforming \cite{9198125}. Specifically, the phase shift of the passive IRS is first optimized by statistical CSI, and then the transmit beamforming of the access point is optimized to cater to the instantaneous CSI of the user's effective fading channel.
	
	\item Beamforming without explicit CSI: This is a beamforming method that does not require any instantaneous CSI. Beam training is one of the common methods, which tries to achieve the best result for the system by selecting the best beam from a predefined beam set. However, beam training tends to incur a significant overhead, and authors in \cite{9325920} propose hierarchical and random training beamforming methods to further reduce the beam training overhead. DL-based beamforming is another beamforming method without instantaneous CSI, which learns the mapping relationship between channel characteristics and beamforming from training data to achieve intelligent beamforming \cite{9348156}. In addition, Reinforcement Learning (RL) can also be used for beamforming design. In this method, the system is described as a learning agent, which optimizes the global behavioral strategy to select the optimal beamforming parameters based on the current state and the received reward \cite{10139787}.
\end{itemize}

\paragraph{Beam tracking}
\indent For dynamic beamforming, fast-changing channels and hardware/resource limitations hinder the implementation of beam tracking. In addition, the time overhead of beam tracking implementation should be taken seriously in scenarios with real-time requirements. To solve the above challenges, filter-based beam tracking and learning-based beam tracking are proposed.

\begin{itemize}
	\item Filter-based beam tracking: Its main idea is to continuously adjust the coefficients of the filter in real-time based on feedback information, to adapt to channel variations and achieve beam tracking. Due to the simplicity of the filter update process and clear target orientation, it can enhance signals in specific target directions. Therefore, it is suitable for scenarios with relatively dynamic environments and time-critical requirements. It is worth noting that the utilization of single-pulse signals in designing filter-based beam tracking methods is promising to address the high nonlinearity problem of codebook-based beam tracking models and reduce the overhead of beam scanning \cite{9374101}. Moreover, the distributed beam tracking approach with multi-anchor node collaboration is suitable for scenarios where environmental factors have a significant impact on beam tracking \cite{9869296}. 
	
	\item Learning-based beam tracking: It utilizes ML technologies such as DL, to train models and learn beam tracking strategies. This approach exhibits strong adaptability to environmental changes and has powerful generalization capabilities, but may not be suitable for scenarios with limited computational resources. Among them, Q-learning can be used for beamforming design based on current and past observation data, striking a balance between data acquisition and beam tracking costs \cite{9390407}. Additionally, LSTM can leverage the temporal correlation of beams to model the channel, which reduces the training overhead of beam tracking to some extent \cite{10120938}.
\end{itemize}

\indent In general, the selection of beamforming methods is different according to different scenarios. In order to obtain a good beamforming effect, the combination of multiple methods is a good choice \cite{9813590}. The ability of IRS to reconstruct the environment through beamforming depends greatly on the CSI acquisition accuracy. For scenarios where it is difficult to acquire accurate CSI, such as when IRSs are mounted on UAVs, it is a good choice to utilize the S-process to deal with the uncertainty of CSI for robust beamforming design \cite{10288199}. Beamforming is also needed to jointly consider with channel estimation to ensure the coordination between beamforming performance and channel CSI acquisition overhead.

\subsubsection{\textbf{Resource Allocation for IRS-Assisted UAV Communications}}

\indent For IRS-assisted UAV communications, the limited transmission bandwidth, UAV energy, and transmission power restrict the system performance. To fully utilize the limited network resources, e.g., power \cite{9417539,9749020,9293155}, bandwidth
\cite{9293155}, IRS reflection units \cite{9293155,9944150}, and computing
resources \cite{9789841,9771971}, efficient resource allocation strategies should be designed. By uniformly managing and allocating resources in the communication system, various performance metrics can be improved, including system delay \cite{9789841}, system energy consumption \cite{9417539}, network rates \cite{9893192}, and spectrum utilization \cite{9454446}.

\indent It is common to follow certain principles for resource allocation. The multi-level water-filling principle aims to maximize the overall system performance, e.g., total network rates, by allocating communication resources to each channel, and then gradually reducing the allocated resources for each channel based on the quality of wireless channel conditions \cite{9293155}. However, this allocation principle requires the accurate acquisition of user channel information, which may increase the system overhead and require complex control algorithms to ensure appropriate resource allocation. In the system design, resource allocation is not strictly based on the multi-level water-filling principle; instead, practical design requirements are taken into consideration. For example, authors in \cite{9293155} consider per-user heterogeneous quality-of-service requirements. Authors in \cite{9804220} consider constraints on individual data rate requirements and the maximum tolerable outage probability. Another common resource allocation strategy is the priority-based allocation, where users with higher priority or specific needs are allocated with more resources. This allocation principle is widely used in the spectrum allocation process of Cognitive Radio (CR) systems, ensuring spectrum resources for Primary Users (PUs) first and then Secondary Users (SUs) \cite{9400768}.

\indent In IRS-assisted UAV communication systems, resource allocation is usually formulated as optimization problems, and the corresponding algorithms include game theory, Deep Reinforcement Learning (DRL), heuristic, and approximation algorithms, which are described separately below.

\paragraph{Game theory-based resource allocation}
\indent The basic idea of this method is to establish a game model that considers both competitions and cooperations among users, making resource allocation decisions based on game strategies to balance the interests of different users and achieve optimization goals \cite{9804495,9625737}. The resource allocation model based on game theory is able to adaptively adjust the strategy to fit the dynamic behavior of participants and changes in the environment, which improves the flexibility and adaptability of the system. At the same time, the method can realize decentralized decision making of resources and reduce the complexity of centralized control, which is suitable for multi-IRS-assisted multi-UAV communication scenarios \cite{9625737}. However, the method usually requires sufficient information interaction among participants, which is difficult to realize in dynamic scenarios. In addition, in some game models, the Nash equilibrium may be slow or even unable to reach.

\paragraph{DRL-based resource allocation}
\indent In IRS-assisted UAV communications, resource allocation based on DRL represents an intelligent method that dynamically optimizes wireless resources through autonomous learning. DRL learns resource allocation strategies through interaction with environments and possesses robust learning capabilities, making it suitable for highly complex dynamic scenarios induced by UAV mobility \cite{9685217,9804495}. However, DRL models typically require substantial data and computational resources for training, and their learning algorithms are of high complexity. Therefore, leveraging techniques such as meta-learning to reduce training overhead, or designing efficient learning algorithms and reward mechanisms, is crucial for dynamic IRS-assisted UAV communication scenarios.

\paragraph{Heuristic algorithm-based resource allocation}
\indent It can heuristically obtain solutions for resource allocation problems, but without guaranteeing the effectiveness of obtained solutions. It simulates the human heuristic thinking process by introducing a series of heuristic rules, strategies, and methods to search the solution space. Heuristic algorithms simplify the decision-making process, and can find a preferable solution in a short time. Heuristic rules can also be adjusted according to the optimization problem and the algorithm performance can be optimized by combining domain knowledge \cite{9944150,10285605}. Therefore, it is suitable for dynamic IRS-assisted UAV communication scenarios that require fast response and real-time decision making. However, the quality of resource allocation depends on the heuristic rules, and the optimal solution may not be found. Moreover, due to the limitation of heuristic rules, heuristic algorithm-based resource allocation has poor generalization ability. Common heuristic algorithms include simulated annealing algorithms, genetic algorithms, and Particle Swarm Optimization (PSO) algorithms \cite{10003080}.

\paragraph{Approximation algorithm-based resource allocation}	
\indent Its basic idea is to find a near optimal solution or a solution that satisfies specific conditions within an acceptable time frame \cite{9749020}. Since the method is based on strict mathematical theory, the solution of resource allocation obtained by approximation algorithm is near optimal and has theoretical performance guarantee \cite{9454446}. Therefore, to obtain strict theoretical performance upper bounds, approximation algorithms are the most commonly used approach for resource allocation problems in IRS-assisted communications. It provides efficient solutions with stability and reliability, and is suitable for IRS-assisted UAV communication scenarios with limited computational resources and high requirements on solution quality. However, in some cases, approximation algorithms may require strict mathematical constructions, which limits their flexibility and applicability. Additionally, approximation algorithms necessitate a profound comprehension of the problem's structure, with implementation and optimization often being complex processes. Common approximation algorithms include greedy algorithms, SemiDefinite Relaxation (SDR) algorithms \cite{9810528}, and Successive Convex Approximation (SCA) algorithms \cite{9804220,9454446}.

\subsubsection{\textbf{Trajectory Optimization for IRS-Assisted UAV Communications}}

\indent In IRS-assisted UAV communications, trajectory optimization refers to the specific design of the UAV's movement trajectories to improve the system performance. The UAV's trajectories and positions can greatly affect performance metrics such as communication delay, coverage ranges, power consumption, and system throughput. It is worth noting that the height of the UAV is critical for LoS link establishment. Although authors in \cite{9367288} discuss performance improvement brought by trajectory optimization, they simplify the UAV's three-dimensional trajectories and positions to a two-dimensional plane, ignoring the impact of height on the system. In addition, IRSs can enhance the flexibility of UAV trajectories. For instance, in scenarios with multiple Ground Users (GUs), UAVs no longer need to alter their original trajectories to maintain the minimum distance from all users. Instead, by intelligently deploying IRSs near GUs, the system can satisfy users' connectivity requirements without consuming excessive time and energy.

\indent Similar to resource allocation, trajectory optimization is often formulated as an optimization problem in IRS-assisted UAV communication systems. The process of trajectory optimization is complicated. First, optimization variables are diverse, including positions, velocities, and flight angles of UAVs, while there are mutual constraints and interactions among these variables. Second, when multi-user or wide-service-range scenarios are involved, the trajectory optimization becomes much complicated and the computational complexity of the solution is very high. Therefore, efficient trajectory optimization algorithms are necessary. Generally, some algorithm, such as greedy algorithm, PSO, DL, and RL can be used for trajectory design, which are described in detail below.

\paragraph{Trajectory optimization based on greedy algorithms}
\indent It tries to construct the UAV trajectory based on the current optimal choice (such as the shortest flying distance, strongest signal, and minimum interference) to achieve a local optimum \cite{9804341}. The greedy algorithm has low complexity and is commonly used for simple and real-time UAV communication scenarios. However, its main drawbacks are lack of backtracking abilities, tendency to reach local optima, and sensitivity to initial conditions.

\paragraph{Trajectory optimization based on PSO}	
\indent It is a swarm intelligence algorithm that optimizes the UAV trajectories by simulating the movement of a particle swarm in the search space. Here, each particle represents a potential solution, and each particle can simultaneously update its velocity and position to facilitate the search for the optimal solution, thus approaching the global optimization \cite{10088448}. PSO exhibits commendable global search capabilities, effectively preventing convergence to local optima. However, the performance of the algorithm heavily relies on the parameter settings, with improper configurations potentially leading to diminished effectiveness. Additionally, PSO tends to converge slowly in the context of high dimensional and complex problems. Therefore, trajectory optimization based on PSO are suited for scenarios where the problem model is intricate and the requirements for real-time performance are not exceedingly stringent.

\paragraph{Trajectory optimization based on imitation learning}
\indent It is a trajectory optimization approach that learns and optimizes its own flight path by imitating the trajectories of experts or other UAVs. This method exhibits pronounced effectiveness in intricate environmental flights and multi-UAV collaborative tasks \cite{9552547}. However, its effect is profoundly contingent upon high quality training datasets, with potential limitations in generalization when encountered with novel scenarios.

\paragraph{Trajectory optimization based on DL}	
\indent This method utilizes DL models to learn the optimal UAV flight trajectories. By inputting the state information of the UAV and the communication environment, the DL model can learn the mapping relationship between the system performance and the state, and output the optimal trajectory \cite{9802633}. Trajectory optimization based on DL exhibits strong applicability and high accuracy, and is used to solve complex nonlinear trajectory optimization problems in large-scale communication environments. However, the demand for large amounts of tagged data restricts its effectiveness for trajectory optimization.

\paragraph{Trajectory optimization based on RL}	
\indent By establishing an interaction model between the intelligent agent and the environment, the agent learns the optimal action strategies through continuous trials to maximize cumulative rewards such as EE, which is the ratio of transmission rates to system energy consumption \cite{9685217}. The greatest advantage of this method is robust environmental adaptability without precise prior data \cite{9817819}. Consequently, it is widely used in UAV trajectory optimization scenarios with multiple objectives and complex dynamic environments. However, the elevated computational complexity demands additional computational resources and time, particularly for trajectory optimization of multiple users in complex environments.

\indent In practical system design of IRS-assisted UAV communications, the selection of the above technologies usually depends on system requirements and constraints. It should be noted that the above technologies are interactive with each other, and a single optimization has limited performance enhancement for systems. Thus, it is necessary to select multiple technologies simultaneously for joint design. However, this joint design increases the system complexity. This also drives researchers to investigate efficient algorithms for IRS-assisted UAV communications, which is analyzed in \hyperlink{Section4}{Section \uppercase\expandafter{\romannumeral4}}.

\hypertarget{Section4}{\section{Solutions of IRS-Assisted UAV Communications for 6G Networks}}

\indent In this section, we provide a detailed description of solutions to the related issues summarized in \hyperlink{Section2.B}{Section \uppercase\expandafter{\romannumeral2}.B}, including energy-constrained communications, secure communications, and enhanced communications.

\subsection{Solutions for Energy-Constrained Communications}

\indent In IRS-assisted UAV communications, energy-constrained communications can be solved not only from the perspective of reducing system energy consumption, but also from a sustainable perspective based on wireless power transfer. The following are examples of SWIPT, BackCom and other methods to illustrate solutions for energy-constrained communications in detail. \hyperlink{table 3}{Table \uppercase\expandafter{\romannumeral3}} provides a corresponding summary of these solutions.\footnote{Although the articles in \hyperlink{table 3}{Table \uppercase\expandafter{\romannumeral2}} do not provide an extensive description of the channel estimation process, the choice of channel models partially reflects the channel estimation. The same with \hyperlink{table 4}{Tables \uppercase\expandafter{\romannumeral3}} and \hyperlink{table 5}{\uppercase\expandafter{\romannumeral4}}.}

\renewcommand{\arraystretch}{1.1} 
\hypertarget{table 3}{\begin{table*}[htbp]}
	\centering
	\vspace{-1.0cm}
	\fontsize{9}{11}\selectfont 
	\begin{threeparttable}
		\caption{Summary of solutions for energy-constrained communications.}
		\begin{tabular}{|m{1.5cm}<{\centering}|m{0.5cm}<{\centering}|m{8.2cm}<{\centering}|m{0.4cm}<{\centering}|m{0.4cm}<{\centering}|m{0.4cm}<{\centering}|m{0.4cm}<{\centering}|m{0.4cm}<{\centering}|m{0.4cm}<{\centering}|m{0.4cm}<{\centering}|m{0.4cm}<{\centering}|}
			\hline
			\multirow{3}[40]{*}{Categories} & \multirow{3}[40]{*}{Ref.} & \multirow{3}[40]{*}{Description} & \multicolumn{6}{c|}{Technologies} & \multicolumn{2}{c|}{\makecell{Involved\\ issues}} \bigstrut\\
			\cline{4-11} & & & \multirow{2}[40]{*}{\begin{sideways}Channel estimation  \end{sideways}} & \multicolumn{2}{c|}{\makecell{Beamfor-\\ming}} & \multicolumn{2}{c|}{ \makecell{Resource \\ allocation}} & \multirow{2}[40]{*}{\begin{sideways}Trajectory 
					optimization\end{sideways}} & \multirow{2}[40]{*}{\begin{sideways}Low power communications \end{sideways}} & \multirow{2}[40]{*}{\begin{sideways}Sustainable communications\end{sideways}} \bigstrut\\
			\cline{5-8}          &       &       &        & \vspace{0.2cm} \begin{sideways}Active beamforming\end{sideways} \vspace{0.2cm}& \begin{sideways}Passive beamforming\end{sideways} & \begin{sideways}Transmission power \end{sideways} & \begin{sideways}Computing resources\end{sideways} &       &       &  \bigstrut\\
			
			\hline
			 \multirow{2}[3]{*}{BackCom} & \multirow{1}{*}{\makebox[-0.1cm]{\cite{10066841}}} & An AO algorithm based on  fractional program, semidefinite program, and RL to maximize total reception rate of all users. &  $\surd$ & $\surd$ & $\surd$ & $\times$ & $\times$ & $\surd$ & $\surd$ & $\times$ \bigstrut\\
			 \cline{2-11} & 
			 \multirow{1}[0]{*}{\makebox[-0.1cm]{\cite{10044705}}} & An AO algorithm based on semidefinite program, SCA, and RL to maximize the broadcast secrecy rate. & $\surd$ & $\surd$ & $\surd$ & $\times$ & $\times$ & $\surd$ & $\surd$ & $\times$ \bigstrut\\
			 
			 \hline 
			 \multirow{5}[30]{*}{SWIPT} & \multirow{1}{*}{\makebox[-0.1cm]{\cite{9771577}}} & An optimization algorithm based on Lagrangian dual to maximize the average harvested energy. & $\surd$ & $\times$ & $\surd$ & $\surd$ & $\times$ & $\surd$ & $\times$ & $\surd$ \bigstrut\\
			 \cline{2-11} &
			 \multirow{1}{*}{\makebox[-0.1cm]{\cite{9756208}}} & An iterative algorithm based on SCA and BCD to maximize the minimum average achievable rate. & $\surd$ & $\times$ & $\surd$ & $\surd$ & $\times$ & $\surd$ & $\times$ & $\surd$ \bigstrut\\
			 \cline{2-11} &
			 \multirow{1}[2]{*}{\makebox[-0.1cm][c]{\cite{9849020}}} & An AO algorithm based on SCA, penalty function method, and difference-convex programming to maximize achievable sum-rate for all users. & $\surd$ & $\times$ & $\surd$ & $\surd$ & $\times$ & $\surd$ & $\times$ & $\surd$ \bigstrut\\
			 \cline{2-11} &
			 \multirow{1}{*}{\makebox[-0.1cm]{\cite{9894720}}} & An AO algorithm based on convex programming and SCA to minimize the maximum energy consumption. & $\surd$ & $\times$ & $\surd$ & $\surd$ & $\times$ & $\surd$ & $\times$ & $\surd$ \bigstrut\\
			 \cline{2-11} &
			 \multirow{1}{*}{\makebox[-0.1cm]{\cite{9479733}}} &  A double iteration algorithm to maximize average achievable rate over time slots. & $\surd$ & $\surd$ & $\surd$ & $\surd$ & $\times$ & $\surd$ & $\times$ & $\surd$ \bigstrut\\
			 	
			 \hline
			 \multirow{9}[40]{*}{\makecell{Other sys-\\tem  EE\\ optimisation}} & \multirow{1}{*}{\makebox[-0.1cm]{\cite{9771971}}} & An AO algorithm to minimize UAV’s total flying time. & $\surd$ & $\times$ & $\surd$ & $\times$ & $\surd$ & $\surd$ & $\surd$ & $\times$ \bigstrut\\
			 \cline{2-11} & 
			 \multirow{1}{*}{\makebox[-0.1cm]{\cite{9870557}}} & A two-phase approach to improve the global EE of the system. & $\surd$ & $\times$ & $\surd$ & $\times$ & $\times$ & $\surd$ & $\surd$ & $\times$  \bigstrut\\
			 \cline{2-11} &
			 \multirow{1}{*}{\makebox[-0.1cm]{\cite{9804220}}} & An AO algorithm and DNN to minimize average system energy consumption. & $\surd$ & $\times$ & $\surd$ & $\surd$ & $\times$ & $\surd$ & $\surd$ & $\times$ \bigstrut\\
			 \cline{2-11} &
			 \multirow{1}{*}{\makebox[-0.1cm]{\cite{10075533}}} & A DL based algorithm to minimize energy consumption. & $\surd$ & $\surd$ & $\surd$ & $\times$ & $\times$ & $\surd$ & $\surd$ & $\times$ \bigstrut\\
			 \cline{2-11} &
			 \multirow{1}{*}{\makebox[-0.1cm]{\cite{9417539}}} & An AO algorithm based on SDR to maximize the EE. & $\surd$ & $\surd$ & $\surd$ & $\surd$ & $\times$ & $\times$ & $\surd$ & $\times$ \bigstrut\\
			 \cline{2-11} &
			 \multirow{1}{*}{\makebox[-0.1cm]{\cite{9906843}}} & An AO algorithm to maximize the EE. & $\surd$ & $\surd$ & $\surd$ & $\times$ & $\times$ & $\times$ & $\surd$ & $\times$ \bigstrut\\	
			 \cline{2-11} &
			 \multirow{1}{*}{\makebox[-0.1cm]{\cite{9866052}}} & An iterative algorithm based on SCA and Dinkelbach’s method to maximize the spectrum efficiency and the EE.   & $\surd$ & $\surd$ & $\surd$ & $\times$ & $\times$ & $\surd$ & $\surd$ & $\times$ \bigstrut\\
			 \cline{2-11} &
			 \multirow{1}{*}{\makebox[-0.1cm]{\cite{9526285}}} & An approach based on deep Q-network and SCA to minimizie total transmission power. & $\surd$ & $\surd$ & $\surd$ & $\times$ & $\times$ & $\surd$ & $\surd$ & $\times$ \bigstrut\\
			 \hline
		\end{tabular}%
		\begin{tablenotes}
			\centering
			\item The symbol ``$\surd$" represents the article satisfies the property, and ``$\times$" represents not.
		\end{tablenotes}
		\vspace{-0.2cm}
	\end{threeparttable}
	\label{tab:addlabel}%
\end{table*}%

\subsubsection{\textbf{BackCom for Energy-Constrained Communications}}

\indent BackCom is a reflection-based wireless communication technology. Devices using BackCom can not only transmit information by designing the impedance matching state in the antenna, but also obtain energy from the Radio Frequency (RF) signal to maintain normal operation, realizing a green communication paradigm \cite{9789440}. 
The simplest single-base BackCom system consists of a Backscatter Device (BD) and a reader, where the reader includes a power beacon and a backscatter receiver. During communication, the RF source generates an RF signal to activate the tag, and the backscatter transmitter loads the sent information into the RF signal and reflects the modulated signal to the backscatter receiver \cite{8368232}.

\hypertarget{Fig. 7}{\begin{figure}[t]}
	\begin{center}
		\includegraphics[width=0.5\textwidth]{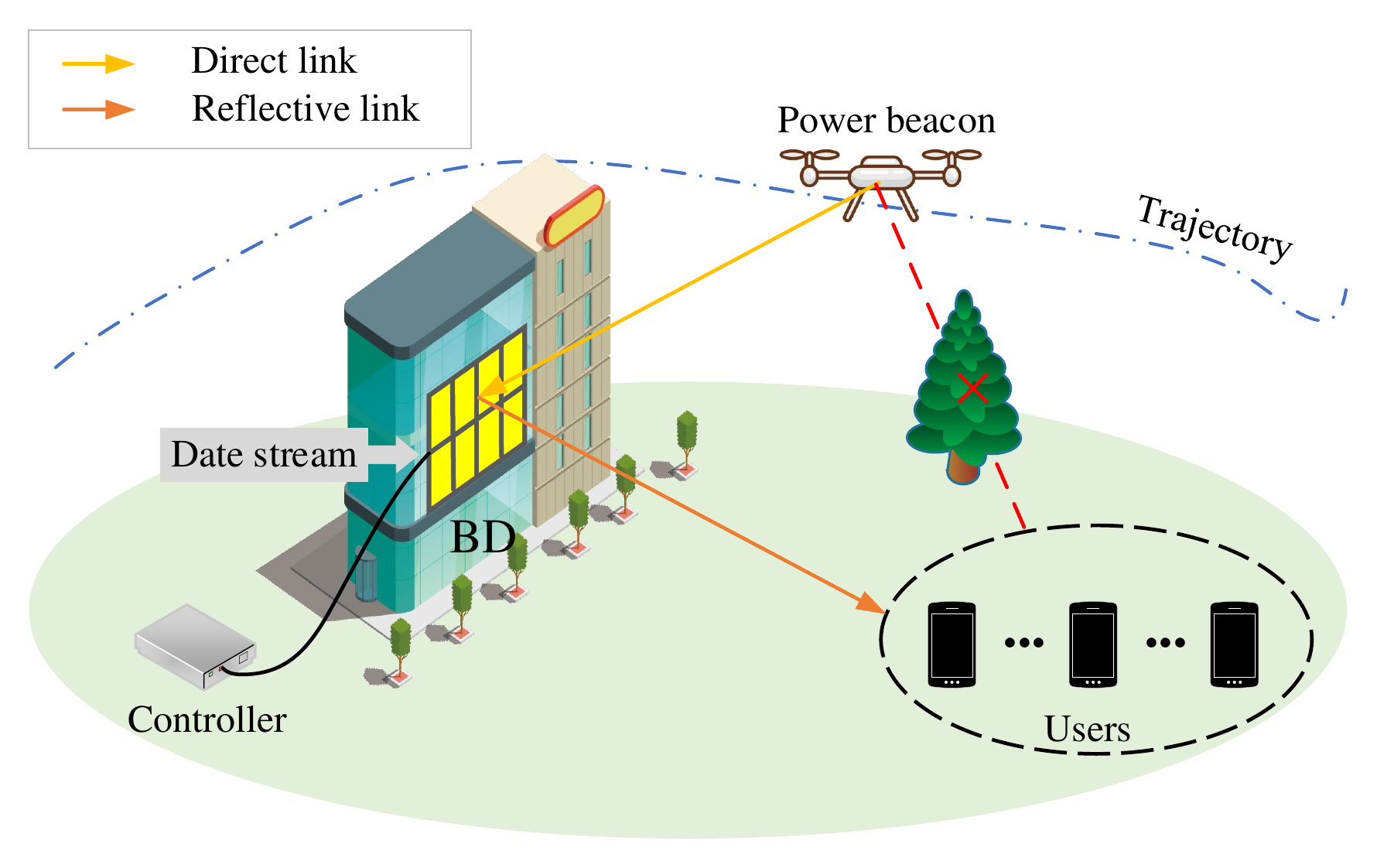}\\
		\caption{A schematic of BackCom in IRS-assisted UAV communications.}
		\vspace{-5mm}
		\setlength{\parskip}{2cm plus4mm minus3mm}
	\end{center}
\end{figure}

\indent In IRS-assisted UAV communications, using BackCom can reduce system energy consumption while satisfying system requirements for communication performance. \hyperlink{Fig. 7}{Fig. 8} illustrates the communication principle of BackCom. In the BackCom system, the UAV acts as airborne power beacon, providing RF signals to specific areas of the IRS through carefully designed UAV trajectories. The IRS, acting as the BD, leverages beamforming to enhance the reverse scattering effect while ensuring low energy consumption, and it transmits information to users via RF signals. In \cite{10066841}, authors formulate a problem to maximize the reception sum rate of all users under constraints of UAV transmission power and trajectory, as well as IRS reflection coefficients. Due to the existence of multiple deeply coupled variables, the problem is decomposed into three sub-problems using Block Coordinate Descent (BCD), and an Alternating Optimization (AO) algorithm is proposed to iteratively solve each subproblem. Compared to traditional BackCom, BackCom in an IRS-assisted UAV communications exhibits cost-effectiveness and EE. Unlike \cite{10066841}, authors in \cite{10044705} study the BackCom communication system in the presence of multiple illegal eavesdroppers, considering both system energy consumption and security.

\indent \textbf{Lesson 1}: While IRSs can replace traditional BDs for low-power communications, the deployment location and number of IRSs can have a large impact on system performance and power consumption. Additionally, BackCom needs to consume the energy of the UAV to activate the reflected signal. To further reduce the transmission energy consumption, the approach of jointly modulating environmental signals and RF signals is also worthy of further research.

\subsubsection{\textbf{SWIPT for Energy-Constrained Communications}}
\indent SWIPT is an evolution of wireless power transfer that utilizes the properties of wireless signals to couple energy into signals for simultaneous transmission. SWIPT works by transmitting a combined wireless signal that carries information and energy, and allows receiving devices or intermediate devices to simultaneously decode/relay information and derive energy from the same signal \cite{8476597}. \hyperlink{Fig. 8}{Fig. 9} illustrates the basic working principle of SWIPT-based IRS-assisted UAV communications. Herein, IRSs act as intermediate devices to relay wireless signals. Users obtain the energy and information from the received signal by power splitting \cite{9531372}. It is worth noting that the power splitting ratio has a direct impact on the efficiency of user energy harvesting and information transmission. In addition, the mobility of UAVs can be utilized to add flexibility to IRS deployment.

\hypertarget{Fig. 8}{\begin{figure}[t]}
	\begin{center}
		\includegraphics[width=0.5\textwidth]{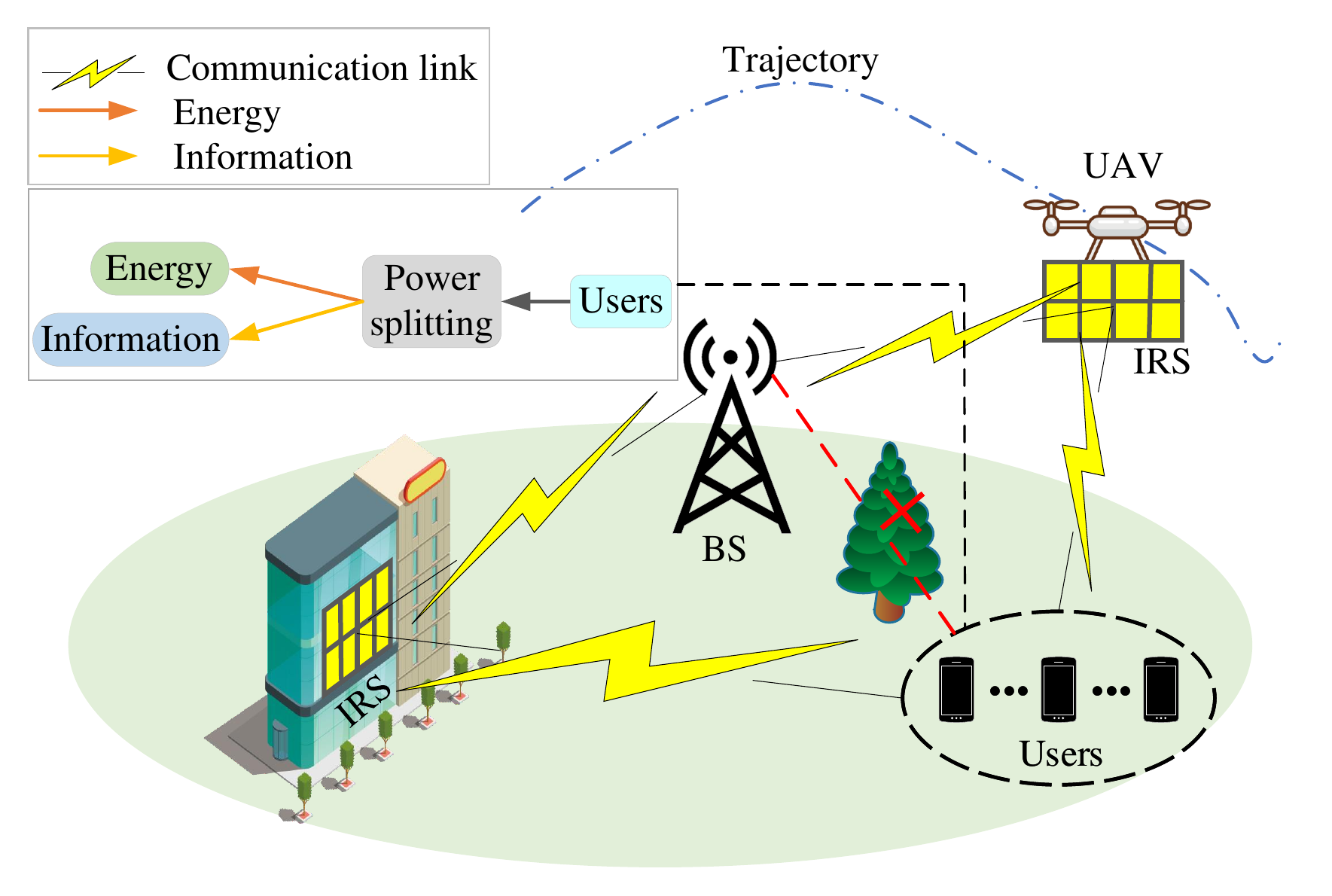}\\
		\caption{A schematic of SWIPT in IRS-assisted UAV communications.}
		\vspace{-5mm}
		\setlength{\parskip}{2cm plus4mm minus3mm}
		\vspace{-0.4 cm}
	\end{center}
\end{figure}

\indent SWIPT can be also widely used in IRS-assisted UAV communication scenarios with energy-constrained devices. Authors in \cite{9771577} investigate an IRS and UAV-assisted SWIPT system to maximize the average harvested energy of users, and formulate a joint optimization problem of UAV trajectories, power allocation ratio, and IRS phase shift. A method based on Lagrangian dual algorithm is proposed to solve the formulation problem, aiming to overcome the drawback of traditional BCD-SCA methods, which are sensitive to initial parameters.

\indent Authors in \cite{9756208,9849020} investigate multi-user SWIPT scenarios, different from single-user scenario mentioned in \cite{9771577}. In order to satisfy communication requirements of multiple users, authors in \cite{9756208} propose a Time-Division Multiple Access (TDMA) scheduling protocol when the UAV flies along an optimized trajectory. It divides the transmission time into different time slots to ensure that each IoT device is allocated with a specific slot for energy and information transmission. Unfortunately, authors in \cite{9756208} use a simple linear SWIPT model, which results in energy saturation at high power levels and difficulty in energy harvesting at low power levels. Instead, authors in \cite{9849020} propose a non-linear SWIPT framework with IRS-assisted UAV communications for energy-limited IoT devices. This article uses Successive Interference Cancellation (SIC) to eliminate interference among different users when adopting the NOMA scheme. Specifically, the channel gains of all users are first estimated and sorted, and then the user with stronger channel power helps the user with weaker channel power to decode the signal, and finally, its own signal is decoded.

\indent Authors in \cite{9894720} study the three-dimensional trajectory of the UAV, which is different from studies that assume the UAV moves at a fixed altitude in \cite{9771577,9756208,9849020}. An onboard IRS is utilized to enhance the uplink signal in a non-linear SWIPT system. Considering the impact of UAV's three-dimensional trajectory on system performance, authors optimize UAV trajectories, IRS phase shift, and user scheduling to minimize energy consumption for all users.

\indent Authors in \cite{9479733} compare the performance of perfect CSI and statistical CSI for IRS-assisted UAV communications in SWIPT systems, which is contrary to the assumption that the system can obtain perfect CSI in \cite{9849020,9894720}. Authors optimize the power allocation ratio, transmission beamforming, UAV trajectories, and IRS phase shift to maximize the average achievable reception rate during UAV flight slots. Simulation results show that the performance on statistical CSI is worse than that on perfect CSI, but statistical CSI is more suitable for practical applications.

\indent \textbf{Lesson 2}: 
The aforementioned research on utilizing SWIPT technology in IRS-assisted UAV communication scenarios is focused solely on system transmission rates and energy consumption, while ignoring the fair distribution of communication resources, such as energy, among multiple users. Moreover, in IRS-assisted UAV communications and its supported applications, each device's requirements for energy and communication are constantly changing. Therefore, it is essential to address the conflict between system fairness and performance.

\subsubsection{\textbf{Other Methods for Energy-Constrained Communications}}

\indent In addition to SWIPT and BackCom, optimizing specific performance metrics such as EE is another way to address energy-constrained communication issues. Besides, it's worth noting that authors in \cite{9453799} illustrate that solar energy can alleviate the energy issues of UAVs and their supporting communication devices. Due to the complexity of hardware design, uneconomical energy consumption caused by the weight of solar panels, and a heavy reliance on weather conditions, the use of solar energy is limited. Since the placement of IRSs affects UAV flight trajectories, which is also closely related to UAV propulsion energy consumption, the improvement of EE from the perspective of IRS placement is introduced in the following content.

\indent Ground IRS placement on buildings is considered in \cite{9771971,9870557,9804220}. Authors in \cite{9771971} construct a problem of jointly optimising UAV trajectories, IRS phase shifts, scheduling of ground nodes and computational resource allocation to minimize the total flight time of the UAV, thus overcoming the energy-constrained issue of IRSs and UAV-based MEC systems. Unlike the linear discharge of the UAV battery in \cite{9771971}, the nonlinear discharge is discussed in \cite{9870557}. Authors first propose an algorithm to estimate the flight time of the airborne UAV, and then optimize the IRS phase shift matrix and deployment position. Finally, the UAV flight trajectory is optimized based on the estimated UAV flight time and optimized IRS deployment position to improve the global system EE. In particular, this system considers practical constraints such as discrete phase compensation of the IRS and phase-amplitude relationships. 

\indent Authors in \cite{9804220} approximate effective channel gain by DNN models based on imperfect CSI, which is different from \cite{9870557}, and conduct research on the average system energy consumption minimization problem to jointly optimize UAV trajectories, IRS phase shifts, and resource allocation strategies by an AO approach. Similar to \cite{9804220}, imperfect CSI and other system design constraints, such as UAV flight jitter, and practical hardware constraints are also taken into account in \cite{10075533}. In order to minimize energy consumption of the UAV, authors jointly optimize UAV trajectories, UAV active beamforming, and IRS passive beamforming. Specifically, to alleviate the challenges of obtaining the real-time varying CSI of IRS-UAV, UAV-user, and IRS-user links under UAV jitter and system hardware constraints, the authors propose a hybrid semi-unfolding DNN.

\indent In fact, placing IRSs on UAVs can achieve flexible deployment of IRSs while ensuring their support for communications \cite{9417539,9906843}. The flexibility of airborne IRS is reflected in the fact that it is easier to establish LoS links compared to ground-based IRS, while allowing flexible reflections from all angles. In order to maximize the EE of aerial IRS-assisted UAV communication systems, authors in \cite{9417539} optimize the user transmission power, BS active beamforming, and IRS passive beamforming under user minimum transmission rate and power constraints. It is worth noting that deploying aerial IRSs near users rather than BSs leads to a more pronounced enhancement in system EE. Polarisation technologies allow signals of the same frequency to be transmitted in different polarisation directions and are considered to be an effective method of achieving multi-mode transmission. In \cite{9906843}, authors introduce polarization technologies in an aerial IRS communication system, propose an energy-efficient framework for joint broadcast-unicast communication and jointly optimize passive beamforming of IRSs and active beamforming of BSs. However, to simplify the system model, authors in \cite{9417539,9906843} assume all CSI can be perfectly obtained.

\indent Authors in \cite{9866052} study trajectory optimization of UAVs equipped with IRSs, unlike the static deployment of UAVs equipped with IRSs \cite{9417539,9906843}. The goal of trajectory optimization is to minimize the propulsion energy consumption of UAVs while ensuring the system's service quality. Authors jointly optimize UAV trajectories, active and passive beamforming to achieve a balance between achievable reception rates and energy consumption, thereby improving the energy and spectral efficiency of the system. Furthermore, for trajectory optimization, authors use SCA and first-order Taylor expansion to reformulate the problem and adopt the Dinkelbach method to solve it. Similar to \cite{9866052}, authors in \cite{9526285} aim to minimize the transmission power of a communication system aided by multiple UAVs carrying IRSs under heterogeneous networks. Each aerial IRS has the capability to adjust its position and phase shift to serve users with poor channel conditions. To tackle the formulated highly non-convex problem, authors decompose it into two subproblems, and employ the dueling deep Q network and SCA to sequentially solve them.

\indent \textbf{Lesson 3}: Although IRS is a passive device, the control and economic costs of deploying IRSs cannot be ignored. Additionally, the computational energy consumption of algorithms can increase the system's energy consumption. Therefore, introducing lightweight and low-complexity algorithms into the system design is promising.

\subsubsection{\textbf{Solutions for Energy-Constrained Communications in Different Scenarios}} 

The aforementioned studies primarily discuss general solutions for energy-constrained communication in two scenarios: ground-based IRS-assisted UAV communications \cite{10066841,10044705,9771577,9849020,9771971,9870557,9804220,10075533} and aerial IRS-assisted communications \cite{9756208,9894720,9479733,9417539,9906843,9866052,9526285}. These solutions mainly focus on minimizing system energy consumption without specific scenario constraints. While most of the above solutions can be applied to SAGIN scenarios, V2X and large-scale IoT scenarios typically require specialized designs.

\indent In V2X communications, the main challenge for achieving energy-constrained communications lies in dynamic resource management in the context of vehicles' mobility. For example, authors in \cite{10285605} study the relationship between task offloading and computational capabilities of vehicular communications, optimizing the BS transmit beamforming and reflection matrix of IRS carried on UAV to minimize system energy consumption. Similarly, for task offloading in dense urban areas with electric vehicles, authors in \cite{10555361} deploy aerial IRS to expand offloading nodes and develop a game-based algorithm to solve the formulated multi-objective optimization problem by considering energy consumption. For energy constrained communications in large-scale IoT, the issue of resource balancing among multiple devices is needed to address. Authors in \cite{9804495} propose a novel IoT network model utilizing an auxiliary antenna IRS powered by a master UAV, allowing two UAVs to move freely and achieve wireless charging. The goal is to maximize network throughput through collaboration strategies based on multi-agent DRL.

\subsection{Solutions for Secure Communications}
\indent In IRS-assisted UAV communications, malicious jamming and eavesdropping attacks are the most common attacks, and can be also effectively defended against through system design and optimization without over-reliance on network protocols, encryption techniques, and identity authentication. Thus, we take these two security threats as examples and provide a detailed overview of existing solutions for secure communications. Common anti-jamming and eavesdropping methods are summarized in \hyperlink{table 4}{Table \uppercase\expandafter{\romannumeral4}}, including PLS technologies and covert communications.

\renewcommand{\arraystretch}{1.1}
\hypertarget{table 4}{\begin{table*}[htbp]}
	\centering
	\vspace{-1.0cm}
	\fontsize{9}{11}\selectfont 
	\begin{threeparttable}
		\caption{Summary of solutions for security communications.}
		\begin{tabular}{|m{1.6cm}<{\centering}|m{0.8cm}<{\centering}|m{8.0cm}<{\centering}|m{0.4cm}<{\centering}|m{0.4cm}<{\centering}|m{0.4cm}<{\centering}|m{0.4cm}<{\centering}|m{0.4cm}<{\centering}|m{0.4cm}<{\centering}|m{0.4cm}<{\centering}|m{0.4cm}<{\centering}|}
			\hline
			\multirow{3}[38]{*}{Categories} & \multirow{3}[38]{*}{Ref.} & \multirow{3}[38]{*}{Description} & \multicolumn{6}{c|}{Technologies} & \multicolumn{2}{c|}{\makecell{Involved \\issues}} \bigstrut\\
			\cline{4-11}     &     &       &  \multirow{2}[38]{*}{\begin{sideways}Channel estimation\end{sideways}}
			& \multicolumn{2}{c|}{\makecell{Beamfor-\\ming}} & \multicolumn{2}{c|}{ \makecell{Resource \\ allocation}} & \multirow{2}[38]{*}{\begin{sideways}Trajectory 
					optimization\end{sideways}} & \multirow{2}[38]{*}{\begin{sideways}Malicious jamming\end{sideways}} & \multirow{2}[38]{*}{\begin{sideways}Eavesdropping attacks\end{sideways}} \bigstrut\\
			\cline{5-8}          &       &       &       & \vspace{0.2cm} \begin{sideways}Active
				beamforming\end{sideways} \vspace{0.2cm}& \begin{sideways}Passive
				beamforming\end{sideways} & \begin{sideways}Transmission power\end{sideways} & \begin{sideways}IRS \end{sideways} &       &       &  \bigstrut\\
			
			\hline
			\multirow{11}[60]{*}{\makecell{PLS \\technologies }} &\multirow{1}{*}{\makebox[-0.1cm]{\cite{9810528}}} & An AO algorithm based on SCA and SDR to maximize the average transmission rate. & $\surd$ & $\times$ & $\surd$ & $\surd$ & $\times$ & $\surd$ & $\surd$ & $\times$ \bigstrut\\
			\cline{2-11} &
			\multirow{1}{*}{\makebox[-0.1cm]{\cite{9771762}}} & An iterative algorithm based on AO algorithm, SDR and SCA to maximize the EE.  & $\surd$ & $\times$ & $\surd$ & $\times$ & $\times$ & $\surd$ & $\surd$ & $\times$ \bigstrut\\
			\cline{2-11} &
			\multirow{1}{*}{\makebox[-0.1cm]{\cite{10017780}}} & A DRL-based defensive deception approach to realize efficient communications. & $\surd$ & $\times$ & $\surd$ & $\surd$ & $\times$ & $\times$ & $\surd$ & $\times$ \bigstrut\\
			\cline{2-11} &
			\multirow{1}{*}{\makebox[-0.1cm]{\cite{10139787}}} & An optimization algorithm based on distributed matching and Q-learning to maximize achievable communication rates. & $\surd$ & $\times$ & $\surd$ & $\times$ & $\surd$ & $\times$ & $\surd$ & $\times$ \bigstrut\\
			\cline{2-11} &
			\multirow{1}{*}{\makebox[-0.1cm]{\cite{10452297}}} & An optimization algorithm based on DRL to maximize EE. & $\surd$ & $\times$ & $\surd$ & $\surd$ & $\surd$ & $\surd$ & $\surd$ & $\times$ \bigstrut\\
			
			\cline{2-11} &
			\multirow{1}{*}{\makebox[-0.1cm]{\cite{9656117}}} & An iterative algorithm based on SCA to maximize the average secrecy rate. & $\surd$ & $\surd$ & $\surd$ & $\times$ & $\times$ & $\surd$ & $\times$ & $\surd$ \bigstrut\\
			\cline{2-11} &
			\multirow{1}{*}{\makebox[-0.1cm]{\cite{10070838}}} & An AO algorithm based on SCA and SDR to maximize the worst-case sum secrecy rate. & $\surd$ & $\surd$ & $\surd$ & $\times$ & $\times$ & $\surd$ & $\times$ & $\surd$ \bigstrut\\
			\cline{2-11} &
			\multirow{1}{*}{\makebox[-0.1cm]{\cite{9940551}}} & An AO algorithm based on BCD, SDR and SCA to maximize the average secrecy rate. & $\surd$ & $\surd$ & $\surd$ & $\surd$ & $\times$ & $\surd$ & $\times$ & $\surd$ \bigstrut\\
			\cline{2-11} &
			\multirow{1}{*}{\makebox[-0.1cm]{\cite{9528924}}} & An AO algorithm based on SDR to maximize the system secrecy rate. & $\surd$ & $\surd$ & $\surd$ & $\times$ & $\times$ & $\surd$ & $\times$ & $\surd$ \bigstrut\\
			\cline{2-11} &
			\multirow{1}{*}{\makebox[-0.1cm]{\cite{10044705}}} & An AO algorithm based on RL, SCA, and semidefinite program to maximize the broadcast secrecy rate. & $\surd$ & $\surd$ & $\surd$ & $\times$ & $\times$ & $\surd$ & $\times$ & $\surd$ \bigstrut\\
			\cline{2-11} &
			\multirow{1}{*}{\makebox[-0.1cm]{\cite{9434412}}} &  An optimization algorithm based on DL to maximize the total system secrecy rate. & $\surd$ & $\surd$ & $\surd$ & $\times$ & $\times$ & $\surd$ & $\times$ & $\surd$ \bigstrut\\
			
			\hline
			\multirow{6}[2]{*}{\makecell{Covert \\communica-\\tions}} & \multirow{1}{*}{\makebox[-0.1cm]{\cite{10111039}}} & An AO algorithm based on SCA to maximize the average transmission rate.  & $\surd$ & $\times$ & $\surd$ & $\times$ & $\times$ & $\surd$ & $\surd$ & $\surd$ \bigstrut\\
			\cline{2-11} &
			\multirow{1}{*}{\makebox[-0.1cm]{\cite{9943536}}} & An optimization methods based on closed-form solutions to maximize the covert transmission rate. & $\surd$ & $\times$ & $\surd$ & $\surd$ & $\times$ & $\surd$ & $\surd$ & $\surd$ \bigstrut\\
			\cline{2-11} &
			\multirow{1}{*}{\makebox[-0.1cm]{\cite{9745104}}} & A block SCA algorithm to minimize average EE. & $\surd$ & $\times$ & $\surd$ & $\surd$ & $\times$ & $\surd$ & $\surd$ & $\surd$ \bigstrut\\
			\cline{2-11} &
			\multirow{1}{*}{\makebox[-0.1cm]{\cite{10271264}}} & Mathematical analysis and derivation to maximize the worst-case transmission rate. & $\surd$ & $\times$ & $\times$ & $\times$ & $\times$ & $\times$ & $\surd$ & $\surd$ \bigstrut\\
			
			\hline
		\end{tabular}%
		\begin{tablenotes}
			\centering
			\item The symbol ``$\surd$" represents the article satisfies the property, and ``$\times$" represents not. 
		\end{tablenotes}
	\end{threeparttable}
	\label{tab:addlabel}%
\end{table*}%

\subsubsection{\textbf{PLS Technologies to Resist Malicious Jamming}}

\indent PLS technologies utilize the characteristics and corresponding technologies of the physical layer to protect communication channels and transmission media, enabling secure communications. In general, the PLS method focuses on the wireless environment and the physical characteristics of both IRSs and UAVs. As shown in \hyperlink{Fig. 9}{Fig. 10}, leveraging the maneuverability of UAVs, the trajectory of UAVs is set in advance, allowing them to move away from jamming sources and reduce the impact of jamming on communication performance \cite{9454372,9200570}. 

\hypertarget{Fig. 9}{\begin{figure}[t]}
	\begin{center}
		\vspace{-0.8 cm}
		\includegraphics[width=0.5\textwidth]{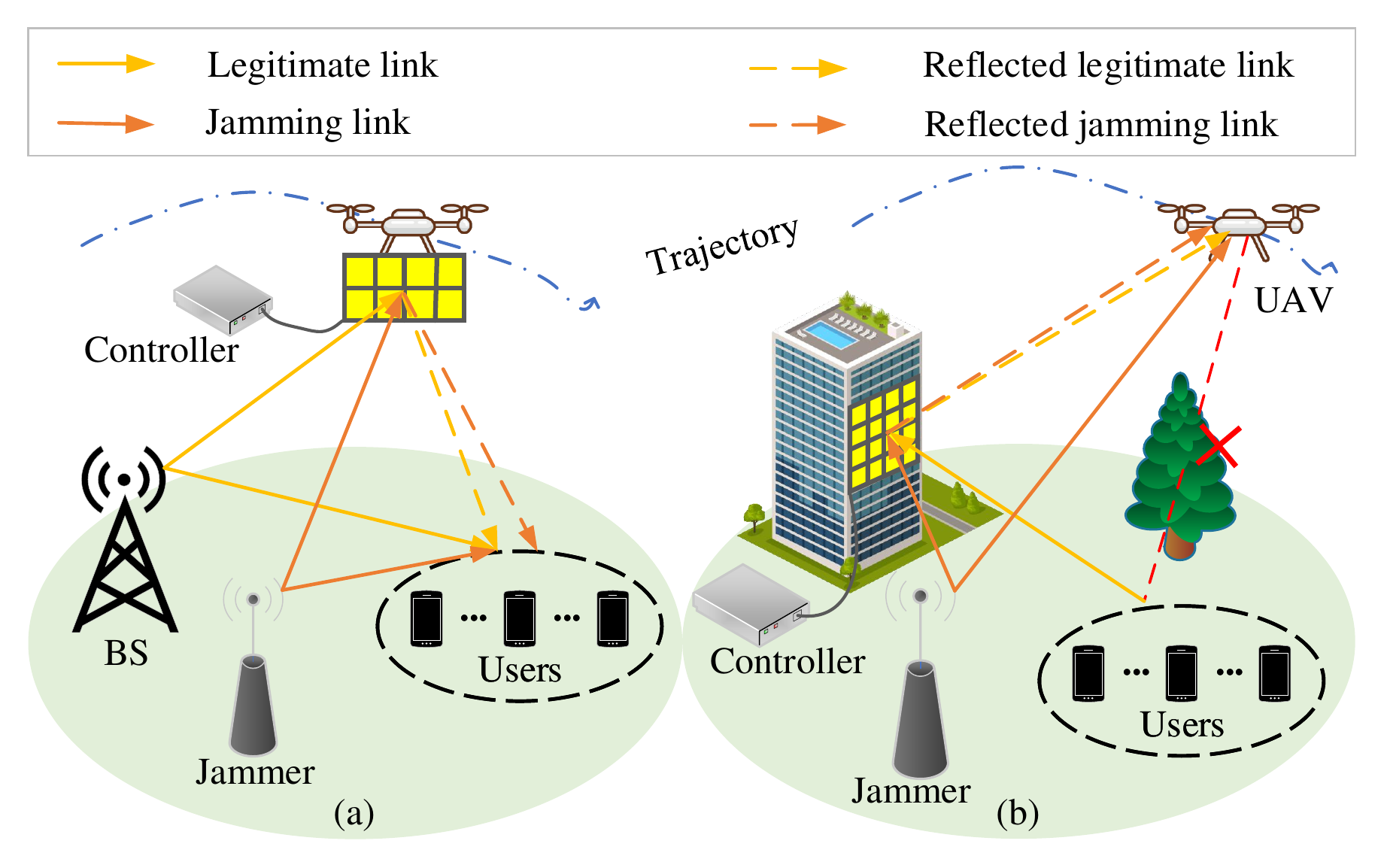}\\
		\caption{Schematics of malicious jamming in IRS-assisted UAV communications: (a) aerial IRS scenes; and (b) ground IRS scenes. }
		\vspace{-5mm}
		\setlength{\parskip}{2cm plus4mm minus3mm}
		\vspace{-0.4 cm}
	\end{center}
\end{figure}

\indent The use of passive beamforming by IRSs can effectively suppress jamming signals while controlling excessive energy consumption of the system. Authors assume two scenarios in \cite{9810528}: one with the IRS deployed in an area far from the jamming source, and the other with that near the jamming source. They optimize UAV trajectories, passive beamforming of IRSs, and transmission power of ground networks to maximize the UAV's average reception rates in the wireless environment where the signal jamming is present. The numerical results show that deploying the IRS near the jamming source rather than far from it can provide better performance for the system. 

\indent The accuracy of the CSI about the jamming channel directly affects the anti-jamming effect. Authors in \cite{9771762} analyze the EE of the system under imperfect CSI jamming, unlike the ideal channel in \cite{9810528}. They formulate a nonconvex problem of jointly optimizing UAV trajectories and IRS beamforming, and solve it by an iterative algorithm based on SDR and SCA. Although the system performance is reduced compared to the perfect CSI, analyzing jamming attacks under imperfect jamming CSI is more realistic. 

\indent Defensive deception strategies are also used to resist jamming attacks, by misleading and confusing attackers into believing the success of their attacks. Authors in \cite{10017780} propose a combination of defensive deception and ML to resist jamming attacks, in which a DRL-based power allocation scheme combined with passive beamforming by IRSs aims to obfuscate the attack surface and lure the jamming attack to the designated channel, with the purpose of achieving anti-jamming effects.

\indent The anti-jamming scenarios described above in \cite{9810528,9771762,10017780} are static scenes for IRSs. As \hyperlink{Fig. 9}{Fig. 10a} shows, in order to provide IRSs with the same maneuverability as UAVs, authors in \cite{10139787} propose to place IRSs on UAVs and investigate the anti-jamming scenario where the jamming source location is uncertain. To maximize the system's anti-jamming performance, author jointly optimize the selection of IRS and beamforming. In particular, authors propose a game-theory-based distributed matching selection algorithm to address the matching problem between IRS-equipped UAVs and multiple users. Furthermore, a Q-learning-based beamforming algorithm is proposed to mitigate the impact of CSI acquisition accuracy on the system's anti-jamming capability. 

\indent Unlike the study in \cite{10139787}, authors in \cite{10049533} study the system performance of IRS-assisted UAV communications in free-space optical systems under malicious UAV jamming. Compared with ground jamming, UAV-mounted jamming sources change positions when the UAV moves, making it difficult to detect and track the jamming source. The authors also derive closed-form expressions for the end-to-end average BER and average outage probability, emphasizing advantages of IRSs under aerial jamming resistance. In addition, authors in \cite{10452297} investigate the performance of airborne IRS-assisted anti-jamming communications in land-to-sea communication scenarios. An intelligent resource management method based on DRL is proposed to optimize the transmission power, UAV deployment position, and IRS beamforming.

\subsubsection{\textbf{PLS Technologies to Resist Eavesdropping Attacks}}

\hypertarget{Fig. 10}{\begin{figure}[t]}
	\begin{center}
		\vspace{-0.8 cm}
		\includegraphics[width=0.5\textwidth]{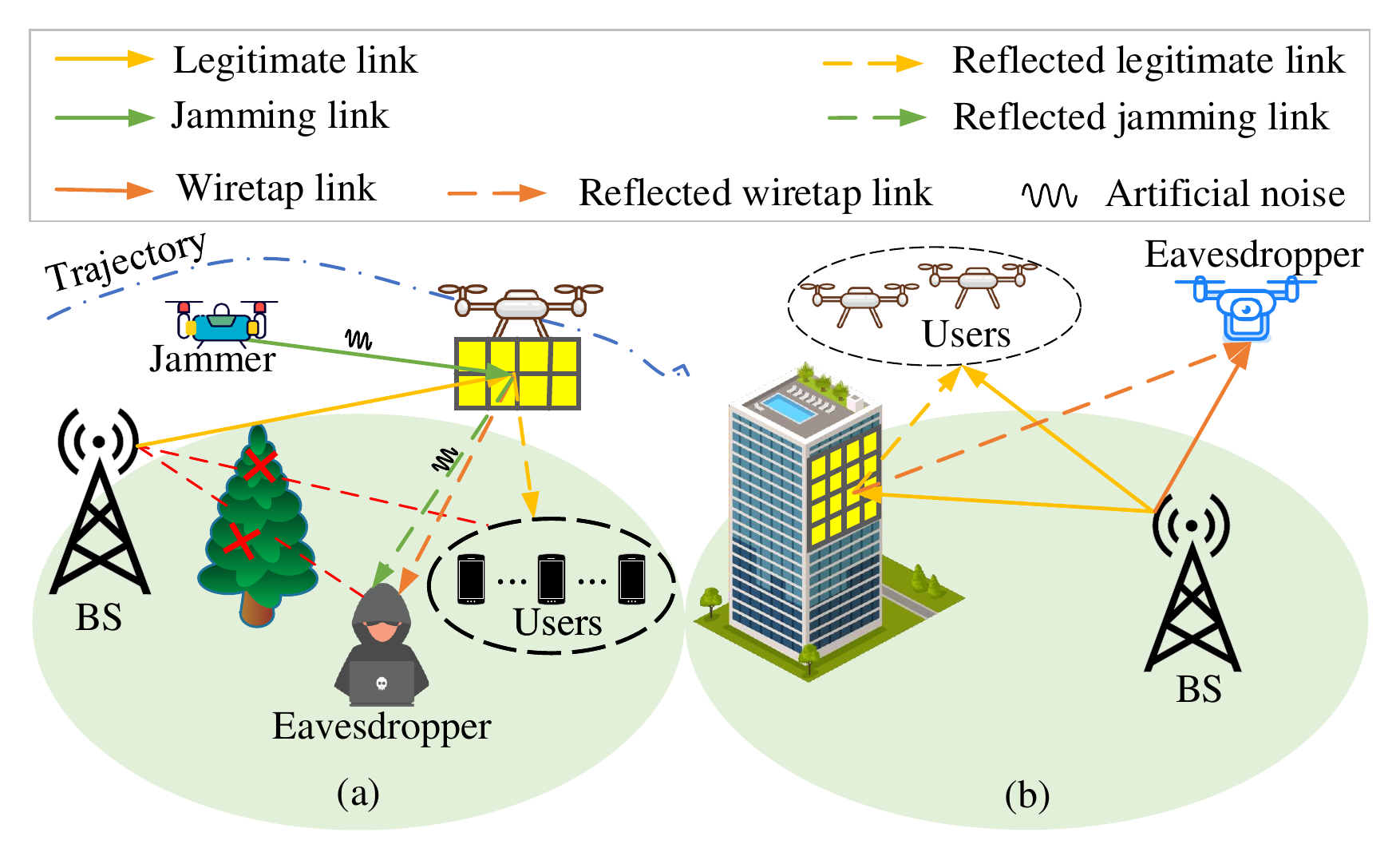}\\
		\caption{Schematics of eavesdropping attracks in IRS-assisted UAV communications: (a) ground eavesdropper scene; and (b) aerial eavesdropper scene. }
		\vspace{-5mm}
		\setlength{\parskip}{2cm plus4mm minus3mm}
		\vspace{-0.3 cm}
	\end{center}
\end{figure}

\indent Relying on its unique advantages at the physical layer, PLS technologies can also play a role in countering eavesdropping in UAV communication scenarios. However, its performance may still  be affected by the specific scene \cite{9656117,10070838,9940551,9681714,10044705} and device constraints \cite{9538830,9434412}. Additionally, the combination of PLS technologies and artificial noise can further enhance the anti-eavesdropping performance of UAV communications \cite{9528924}.

\indent Relevant PLS technologies of countering ground eavesdroppers are studied in \cite{9656117,10070838}. Authors in \cite{9656117} focus on the single-user and single-eavesdropper scenario, where the IRS is fixed on the ground to achieve the maximum secrecy rate of the system through passive beamforming of the IRS, active beamforming of the transmitted signal, and UAV trajectory optimization. Unlike the assumption in \cite{9656117} where the system can accurately obtain the eavesdropper's location information and CSI, authors in \cite{10070838} consider that the eavesdropper's location is unknown. Although the exact locations of eavesdroppers are not known, their approximate positions can be locked into a circular area by the UAV's aerial target detection. Hence, authors derive the worst-case secrecy rate of legitimate users and use this to formulate a problem for jointly optimising the phase shift of the IRS, the transmitted beamforming and the hovering position of the UAV. The formulated problem is solved by an AO algorithm based on SCA and SDR to maximize the worst-case secrecy rate of all system users.

\indent PLS technologies can also be used to solve the security problem caused by airborne eavesdroppers. As shown in \hyperlink{Fig. 10}{Fig. 11b}, aerial eavesdroppers based on UAVs can easily establish a LoS link with the ground BS, and these channels facilitate the reception of eavesdroppers' signals. Therefore, compared with ground eavesdroppers, airborne eavesdroppers pose more serious security threats to the network. Authors in \cite{9940551} study the security of a single airborne user in the presence of multiple airborne eavesdropper scenarios. Faced with a scenario where both the user and eavesdroppers are in the air, three-dimensional trajectory optimization of the UAV is more advantageous than two-dimensional trajectory optimization when adjusting distances among users, IRSs, eavesdroppers, and BSs. Unlike the use of a virtual antenna array constructed by a drone swarm \cite{9552611}, authors use the IRS to achieve energy-efficient secure communications. By jointly optimizing the transmission power, active and passive beamforming, and the UAV's three-dimensional trajectories, the system's secrecy rate is improved. 

\indent The combination of PLS technologies and artificial noise can achieve enhanced anti-eavesdropping performance of UAV communications. Artificial noise can introduce interference, preventing eavesdroppers from accurately capturing the original signal. The use of artificial noise and PLS technologies against eavesdropping attacks is investigated in \cite{9528924}. The authors consider a scenario in which a ground eavesdropper eavesdrops on airborne BS signals. By simultaneously introducing artificial noise and employing both active and passive beamforming, the eavesdropper's signal interception quality is disturbed. Furthermore, the authors optimize the deployment locations of UAVs and IRSs to maximize the system's secrecy rate.

\indent For the anti-eavesdropping communication scenario involving multiple UAVs, as shown in \hyperlink{Fig. 10}{Fig. 11a}, some UAVs can act as friendly jammers to interfere with eavesdropping, and other IRSs and UAVs can be used as airborne relays to compensate for the reduced received quality caused by artificial noise \cite{9681714}. Authors in \cite{10044705} also consider the broadcast secrecy rate of BackCom in scenarios with multiple eavesdroppers. The signals emitted by the UAV, serving as the carrier for IRS information transmission, can cause certain interference to both users and eavesdroppers. Therefore, they optimize the UAV's trajectory to balance the actual impact of UAV transmission signals on user interference, eavesdropper interference, and IRS information transmission. At the same time, passive beamforming for the IRS and active beamforming for the UAV are jointly optimized.

\indent Regrettably, in the aforementioned process of using PLS technologies to address eavesdropping attacks \cite{9656117,10070838,10044705,9940551}, perfect CSI is assumed, which is not consistent with actual scenarios, especially in situations where positions of the eavesdropper and jamming sources are unknown. This issue is discussed in \cite{9538830}, where authors assume perfect channel estimation between the BS and the IRS, as well as between the IRS and users, but only partially available CSI for the channel between the IRS and the eavesdropper. By deriving expressions of probability density function and moment generating function of the instantaneous secrecy rate, the system's secrecy rate is analyzed. Similar to \cite{9538830}, authors in \cite{9434412} investigate the secure transmission problem of IRS-assisted UAV mmWave communication in the presence of imperfect CSI. They jointly optimize active/passive beamforming and UAV trajectories to maximize the system secrecy rate. In particular, to overcome the challenges posed by high coupling of CSI and UAV trajectories, authors propose a DRL algorithm based on the two-deep deterministic policy gradient framework, which has a strong decoupling capability. Specifically, the first policy gradient is used for active and passive beamforming while the second policy gradient is used for trajectory optimisation of the UAV.

\indent \textbf{Lesson 4}: For malicious jamming and eavesdropping attacks, accurately obtaining the attacker's CSI is crucial. However, currently, acquiring accurate CSI of the attacker remains challenging. Fortunately, it is possible to obtain imperfect CSI of the attacker effectively, and its error can be modeled through specific random distributions \cite{9374975}. Therefore, designing efficient robust optimization algorithms to counteract attacks can greatly mitigate system security issues caused by the inaccurate acquisition of the attacker's CSI. Moreover, the measures taken by the system to counter attacks often introduce additional overhead, prompting researchers to strike a balance between system security and other performance indicators.

\subsubsection{\textbf{Covert Communications to Resist Malicious Jamming and Eavesdropping Attacks}}

\indent Another way to address the security threats of UAV communications is covert communications. \hyperlink{Fig. 11}{Fig. 12} illustrates the basic operating principle of IRS-assisted UAV covert communications. Alice (transmitter) sends signal to Bobs (legitimate users), and Whillie (warden) determines whether Alice is in a transmitting state by detected signal power $P_W$ from Alice and detection threshold $\varGamma $. Covert communications utilize randomization techniques such as artificial noise \cite{9745104,10271264} and power control \cite{9943536} to conceal transmission signals within environmental noise or artificial uncertainties, aiming to reduce the detectability of the transmission, hence they are also known as low probability of detection communications. More importantly, when covert communications are utilized for wireless information transmission, the signal should not be perceivable by eavesdroppers \cite{9382022}. 

\indent As a complement of PLS technologies to resist eavesdropping attacks and malicious jamming, covert communications are also extensively studied in \cite{9943536,10111039}. 
Authors in \cite{9943536,10111039} investigate the covert transmission rate of the aerial IRS-assisted covert communication system. They first determine the optimal detection threshold and derive the error detection probability for eavesdroppers. Then, they formulate an optimization problem with variables of UAV trajectories and IRS phase shifts to maximize the covert transmission rate, and solve it by deriving a closed-form solution and an AO algorithm. Different from \cite{10111039}, authors in \cite{9943536} consider the uncertainty of the eavesdropper's location and optimize the signal transmission power. 

\begin{figure}[t]
	\begin{center}
		\includegraphics[width=0.5\textwidth]{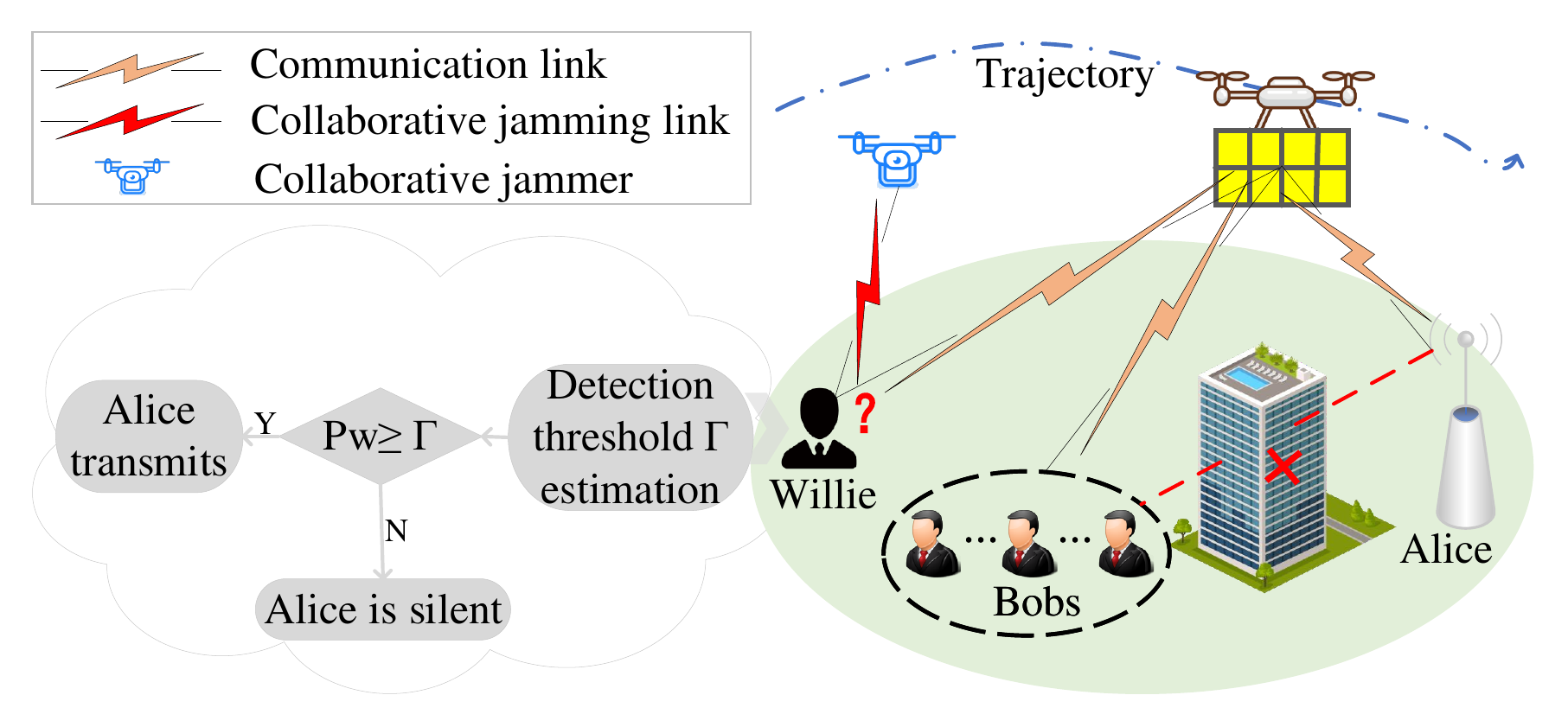}\\
		\hypertarget{Fig. 11}{\caption{A schematic of covert communications in IRS-assisted UAV communications.}}
		\vspace{-5mm}
		\setlength{\parskip}{2cm plus4mm minus3mm}
		\vspace{-0.4 cm}
	\end{center}
\end{figure}

\indent The combination of artificial noise and covert communications can further enhance the security of UAV communications. As shown in \hyperlink{Fig. 11}{Fig. 12}, artificial noise can interfere with illegal users to confuse them to determine whether a legitimate user is communicating. This approach is investigated in \cite{9745104,10271264}. Different from the above mentioned articles \cite{9943536,10111039}, where all UAVs serve as one role, authors in \cite{9745104} consider that UAVs undertake two distinct roles: one is used to carry the IRS for reliable data transmission, and the other acts as a collaborative jammer to enhance the secrecy of the transmission. A power-efficient multi-UAV covert communication scheme is designed for scenarios with multiple eavesdroppers in the THz frequency band. In particular, an optimization problem is constructed to both improve the throughput of covert communications and reduce energy consumption of UAVs, which is iteratively solved by the block SCA method. Unlike the scenario where UAVs have two different roles in \cite{9745104}, authors in \cite{10271264} set up the legitimate receiver with the full-duplex mode to receive signals while also generating jamming signals to further ensure the confidentiality of covert communications. Through theoretical analysis and derivation of the proposed system, the effectiveness of the scheme is proven.

\indent \textbf{Lesson 5}: Covert communications enhance data security and privacy, making unauthorized users difficult to detect or interfere with the transmission content. But extended communication durations increase the risk of communication exposure. Therefore, to enhance the performance of covert communications, multi-modal communication approaches that include both communication and silence modes are worth researching. Furthermore, in practical scenarios where eavesdropping and interference are encountered, covert communications can be used as a supplement and combined with other methods to achieve security outcomes.

\subsubsection{ \textbf{Summary for Secure Communications in Different Scenarios}} 
\indent The aforementioned studies mainly discuss general anti-jamming and anti-eavesdropping solutions in IRS-assisted UAV communications, such as defensive deception strategies \cite{10017780} and artificial noise \cite{9528924,9745104,10271264}, while also considering some non-ideal channel conditions \cite{9771762,9538830,9434412}. However, in practical application scenarios, customized designs are often required to meet specific constraint needs.

\indent In SAGIN, the broadcast nature of air-to-ground communications significantly increases the risk of jamming and eavesdropping. Although authors in \cite{9810528,9771762,10017780,9656117,9940551,9528924,9434412} provide general solutions for secure communications in air-to-ground scenarios, they do not consider the overall security performance of integrated communication systems involving satellites, UAVs, and ground terminals. To enhance jamming elimination in SAGIN scenarios, authors in \cite{10499205} propose an aerial IRS-assisted SAGIN system, integrating satellites, UAV-IRS, and ground terminals. The jamming management scheme adapts to different types of CSI (i.e., no CSI, instantaneous CSI, and delayed CSI), allowing optimal performance through spatial dimension compression, beamforming, and joint precoding strategies. Similarly, to improve the anti-eavesdropping performance of space-ground communication systems, authors in \cite{10089834} consider both double-IRS and single-IRS scenarios to maximize secrecy rate. Placing IRSs near the source of attack can significantly enhance security \cite{9810528}. However, the mobility of vehicle increases the difficulty of secure communications in V2X. To handle this issue, authors in \cite{9760044} study a rate-splitting multiple access scheme, analyze the outage performance in multi-jamming vehicle environments, and propose a solution to optimize power allocation coefficients. The results show that integrating IRS significantly enhances system performance even in the presence of direct links.

\subsection{Solutions for Enhanced Communications}
\indent Up to now, researchers have proposed many solutions for enhanced communications, which typically focus on performance metrics including network rates, network latency, coverage ranges, spectrum efficiency, and reliability. In the following, we introduce relevant solutions for enhanced communications based on these system metrics, and summarize them in \hyperlink{table 5}{Table \uppercase\expandafter{\romannumeral5}}.

\subsubsection{\textbf{Enhanced Communications for Network Rate Improvement}}

\indent The network rate is one of the crucial metrics in wireless transmission and is also the joint result of various factors. Specifically, interference management \cite{9749020,9367288,9293155,10638476,10411934}, multi-user sub-channel allocation\cite{9893192,9454446}, communication links \cite{9685217,9785612}, and hardware constraints \cite{9944150} all have impacts on the network rate.

\indent In IRS-assisted UAV communications, interference can be primarily categorized into intra-cell multi-user interference and inter-cell interference. Regarding the former, the feasibility of constructing multiple interference-free beams based on IRS is validated in \cite{9013433}, providing a theoretical foundation for subsequent IRS beamforming to mitigate intra-cell multi-user interference. Additionally, researchers in \cite{9749020,9367288,9293155} discover that it is generally necessary to jointly consider factors including transmission power allocation and UAV trajectories to maximize the beamforming gain.

\indent Authors in \cite{9749020,9367288,9293155} focus on intra-cell multi-user interference. Without loss of generality, only one set of reflection matrices is available at the IRS in each time slot. Therefore, IRS beamforming typically requires a balance between mitigating multi-user interference and enhancing signal strength. Authors in \cite{9749020} investigate the impact of IRS beamforming, transmission power allocation, and UAV trajectory on the total network rate in IRS-assisted UAV downlink communication networks. By virtue of beamforming of IRSs, signal interference of users is reduced, and the interference suppression effect is more pronounced in multi-user scenarios. Similarly, authors in \cite{9367288} formulate the problem of maximizing the minimum average achievable rate in a UAV scenario with THz communications. In this case, beamforming not only affects interference among multiple users but also plays a crucial role in reducing path loss at high frequencies. However, it is worth noting that in \cite{9749020} and \cite{9367288}, only the IRS phase shift is optimized, and the reflection coefficient is fixed. The research on joint consideration of IRS reflection amplitudes and phase shift is discussed in \cite{9293155}. To mitigate interference, authors employ Orthogonal Frequency Division Multiple Access (OFDMA) technology to maximize the transmission sum rate by jointly optimizing IRS beamforming, subchannel assignment, power allocation and UAV trajectories. The simulation results demonstrate that the usage of OFDMA technology and reasonable deployment of IRS can effectively reduce the interference among multiple users.

\indent The ``double fading" effect of passive IRS means it can only serve its localized areas \cite{9326394}, thereby avoiding unnecessary inter-cell interference. However, this does not completely eliminate the impact of inter-cell interference on communications. To address this, authors in \cite{10411934} investigate the inter-cell interference resistance of IRS-assisted multi-UAV multicast communications. Specifically, authors leverage Coordinated Multi-Point (CoMP) technology, which can convert interference signals into useful ones through coordinated beamforming at the transmitter, to mitigate inter-cell interference. Fixed IRS is also introduced to improve channel conditions for users at the cell edge while increasing the flexibility of UAV trajectory optimization. 

\indent Unlike the static IRS scenario in \cite{10411934}, authors in \cite{10638476} examine a dynamic UAV-assisted multi-cell and multi-user interference-resistant communication scheme with IRS onboard. Specifically, during the UAV's flight along optimized trajectories, authors jointly optimize active beamforming at ground BSs and passive beamforming at IRS to maximize the minimum user transmission rate. Simulation results show that users near the cell edge receive more IRS service time, thereby minimizing both inter-cell and intra-cell multi-user interference.

\indent NOMA technology is another common approach to mitigate inter-cell and intra-cell multi-user interference in IRS-assisted UAV communication scenarios to enhance network rates. It provides services for multiple users by dividing a subcarrier into different power levels, allowing users to share the same frequency and time resources while using different signal power and codewords to differentiate among users. To eliminate interference among different users, interference-cancellation technology such as SIC is often employed \cite{9893192,9454446}. In addition, the system performance brought by NOMA depends on the differences in wireless channels among different users, with greater channel differences resulting in better system performance. Traditionally, wireless channel gain has been thought to be entirely determined by the environment, which means that the differences in randomness can severely affect the performance of NOMA. Fortunately, the reconfiguration of wireless environments by IRSs can change the design paradigm of NOMA.

An IRS-assisted multi-UAV NOMA network is studied in \cite{9454446}. Specifically, UAVs use NOMA to serve ground users, while IRS is deployed at the cell edge to enhance communication links and reduce both inter-cell and intra-cell interference. To maximize the transmission rate, authors jointly optimize UAV positioning and transmission power, NOMA decoding order, and IRS beamforming, and propose an iterative algorithm based on BCD. Since the IRS service coverage is limited \cite{9326394}, authors in \cite{9893192} discuss the association problem between multiple IRSs and multiple users. To decouple user association and NOMA decoding order, they propose a sequence interference extension method, transforming the IRS association problem into a convex optimization problem for multi-user and multi-cell scenarios with severe interference.

\indent In an aerial IRS scenario, UAV trajectories affect the quality of LoS links \cite{9685217,9785612}. Authors in \cite{9785612} discuss the scenario of an airborne IRS assisting multiple GUs. In order to maximize the minimum average transmission rate, the authors optimize the trajectory of the UAV, the phase shift of the IRS, and the communication scheduling. Different from the aforementioned IRS-assisted UAV communication scenarios, where UAVs only play a role as BSs \cite{9749020,9367288,9293155,9893192,9454446} or active/passive relays \cite{9785612}. The UAVs play two important roles in \cite{9685217}, with the main UAV acting as an airborne RF transmitter and the auxiliary UAV acting as a passive relay to enhance the signal reception of GUs. To maximize the cumulative system throughput, authors jointly optimize trajectories of multiple UAVs and the transmission power of the main UAV, and propose an algorithm based multi-agent DRL to solve the optimization problem.

\renewcommand{\arraystretch}{1.1}
\hypertarget{table 5}{\begin{table*}[htbp]}
	\vspace{-1.25cm}
	\centering
	\fontsize{9}{11}\selectfont 
	\begin{threeparttable}
		\caption{Summary of solutions for enhanced communications.}
		\begin{tabular}{|m{1.2cm}<{\centering}|m{0.5cm}<{\centering}|m{7.0cm}<{\centering}|m{0.3cm}<{\centering}|m{0.3cm}<{\centering}|m{0.3cm}<{\centering}|m{0.3cm}<{\centering}|m{0.3cm}<{\centering}|m{0.3cm}<{\centering}|m{0.3cm}<{\centering}|m{0.3cm}<{\centering}|m{0.3cm}<{\centering}|m{0.3cm}<{\centering}|m{0.3cm}<{\centering}|}
			\hline
			\multirow{3}[65]{*}{\makecell{Metrics \\for \\perform-\\ance\\optimiza-\\tion}} & \multirow{3}[65]{*}{Ref.} & \multirow{3}[65]{*}{Description} & 
			\multicolumn{8}{c|}{Technologies}             & 
			\multicolumn{3}{c|}{Involved issues} \bigstrut\\
			\cline{4-14}          &       &  &
			\multirow{2}[70]{*}{\begin{sideways}Channel estimation\end{sideways}} &  
			\multicolumn{2}{c|}{\makecell{Beamf-\\orming}} & 
			\multicolumn{4}{c|}{\makecell{Resource allocation}} & 
			\multirow{2}[70]{*}{\begin{sideways}Trajectory 
					optimization\end{sideways}} & 
			\multicolumn{1}{c|}{\multirow{2}[70]{*}{\begin{sideways}Difficulties of interference management\end{sideways}}} & 
			\multicolumn{1}{c|}{\multirow{2}[70]{*}{\begin{sideways} 
						Shortage of spectrum 
						resources\end{sideways}}} &  
			\multicolumn{1}{c|}{\multirow{2}[70]{*}{\multirow{1}{*}{\begin{sideways}\makebox[8pt][c]{Nonconvexity of multivariate coupling} \end{sideways}}}} \bigstrut\\
			\cline{5-10}          &       &       &       &  
			\vspace{0.95cm} \begin{sideways} Active beamforming\end{sideways} 
			\vspace{0.95cm} & \begin{sideways}Passive
				beamforming\end{sideways} & \begin{sideways}Transmission 
				power\end{sideways} & \begin{sideways}IRS 
				\end{sideways} & \begin{sideways}Computing 
				resources\end{sideways} & 
			\begin{sideways}Frequency\end{sideways} &       &       &       
			&   \bigstrut\\
			
			\hline
			\multirow{8}[55]{*}{\makecell{Network \\rates}} & \multirow{1}{*}{\makebox[-0.1cm]{\cite{9749020}}} & An AO algorithm 
			based on BCD to maximize the total reception rate of users. & $\surd$ & $\times$ & $\surd$ & $\surd$ & $\times$ & $\times$ & $\times$ & $\surd$ & $\surd$ & $\times$  & $\surd$ \bigstrut\\
			\cline{2-14} &
			\multirow{1}{*}{\makebox[-0.1cm]{\cite{9367288}}} & An iteration 
			algorithm based on SCA to maximize the minimum average achievable rate of users. & $\surd$ & $\times$ & $\surd$ & $\surd$ & $\times$ & $\times$ & $\surd$ & $\surd$ & $\surd$ & $\times$  & $\surd$ \bigstrut\\
			\cline{2-14} &
			\multirow{1}{*}{\makebox[-0.1cm]{\cite{9293155}}} & A parametric approximation method and AO algorithm to maximize the total reception rate of users. & $\surd$ & $\times$ & 
			$\surd$ & $\surd$ & $\surd$ & $\times$ & $\surd$ & $\surd$ & 
			$\surd$ & $\times$  & $\surd$ \bigstrut\\
			\cline{2-14} &
			\multirow{1}{*}{\makebox[-0.1cm]{\cite{9785612}} }& An AO algorithm 
			based on SCA to maximize the minimum average transmission rate.  & $\surd$ & $\times$ & $\surd$ & $\times$ & $\times$ & $\times$ & $\times$ & $\surd$ & $\surd$ & 
			$\times$ & $\surd$ \bigstrut\\
			\cline{2-14} &
			\multirow{1}{*}{\makebox[-0.1cm]{\cite{9685217}}} & A multi-agent 
			DRL based algorithm to maximize the total throughput. & $\surd$ & $\times$ & $\times$ & $\surd$ & $\times$ & $\times$ & $\times$ & $\surd$ & $\times$ & $\times$  & $\surd$ \bigstrut\\
			\cline{2-14} &
			\multirow{1}{*}{\makebox[-0.1cm]{\cite{9944150}}} & A centralized 
			algorithm to  maximize the total reception rate of UAVs. & $\surd$ & $\times$ & $\surd$ & $\surd$ & $\surd$ & $\times$ & $\times$ & $\times$ & $\surd$ & $\times$ & $\surd$ \bigstrut\\
			\cline{2-14} &
			\multirow{1}{*}{\makebox[-0.1cm]{\cite{10411934}}} & A hybrid learning algorithms to maximize the sum of the minimum reception rates. & $\surd$ & $\surd$ & $\surd$ & $\times$ & $\times$ & $\times$ & $\times$ &$\surd$ & $\surd$ & $\times$ & $\times$ \bigstrut\\
			\cline{2-14} &
			\multirow{1}{*}{\makebox[-0.1cm]{\cite{10638476}}} & A BCD method to maximize the minimum  ergodic reception rate. & $\surd$ & $\surd$ & $\surd$ & $\times$ & $\surd$ & $\times$ & $\times$ &$\surd$ & $\surd$ & $\times$ & $\times$ \bigstrut\\
			\cline{2-14} &
			\multirow{1}{*}{\makebox[-0.1cm]{\cite{9893192}}} & A three-stage optimization algorithm to maximize the total reception rate of users. & $\surd$ & $\times$ & $\surd$ & $\times$ & $\surd$ &	$\times$ & $\times$ & $\surd$ & $\surd$ & $\surd$ & $\surd$ \bigstrut\\
			\cline{2-14} &
			\multirow{1}{*}{\makebox[-0.1cm]{\cite{9454446}}} & An iterative 
			algorithm based on BCD to maximize the total reception rate of users. & $\surd$ & $\times$ & $\surd$ & $\surd$ & $\times$ & $\times$ & $\times$ & $\surd$ & $\surd$ & $\surd$  & $\surd$ \bigstrut\\

			\hline
			\multirow{5}[55]{*}{\makecell{Latency}} & \multirow{1}{*}{\makebox[-0.1cm]{\cite{9013626}}} & A RL-based approach to maximize the downlink transmission capacity. & $\surd$ & $\times$ & $\surd$ & $\times$ & $\times$ & $\times$ & $\times$ & $\surd$ & $\surd$ & $\surd$  & $\surd$ \bigstrut\\
			\cline{2-14} &
			\multirow{1}{*}{\makebox[-0.1cm]{\cite{9348040}}} & An optimization algorithm based distributed RL to maximize the total reception rate of users. & $\surd$ & $\surd$ & $\surd$ & $\times$ & $\times$ & $\times$ & $\times$ & $\surd$ & $\surd$ & $\surd$  & $\surd$ \bigstrut\\
			\cline{2-14} &
			\multirow{1}{*}{\makebox[-0.1cm]{\cite{9781659}}} & An AO algorithm 
			based on SCA and double deep Q-network to reduce statistical delay 
			and error rates of users.  & $\surd$ & $\times$ & $\times$ & $\times$ & $\surd$ & $\times$ & $\times$ & $\surd$ & $\times$ & $\times$  & $\surd$ \bigstrut\\
			\cline{2-14} &
			\multirow{1}{*}{\makebox[-0.1cm]{\cite{9912224}}} & An AO algorithm 
			based on SCA and Dinkelbach’s transform to reduce statistical delay 
			and error rates of users.  & $\surd$ & $\surd$ & $\surd$ & $\surd$ & $\times$ & $\times$ & $\times$ & $\surd$ & $\surd$ & $\times$  & $\surd$ \bigstrut\\
			\cline{2-14} &
			\multirow{1}{*}{\makebox[-0.1cm][c]{\cite{9789841}}} & An iterative 
			algorithm based on  Hungarian algorithm and  whale optimization 
			algorithm to reduce total network latency. & $\surd$ & $\times$ & $\surd$ & $\times$ & $\times$ & $\surd$ & $\surd$ & $\times$ & $\surd$ & $\times$  & $\surd$ \bigstrut\\
			\cline{2-14} &
			\multirow{1}{*}{\makebox[-0.1cm][c]{\cite{9804341}}} & An optimization algorithm based on differential evolution, clustering algorithm, and greedy algorithm to minimize the total system cost. & $\surd$ & $\times$ & $\surd$ & $\times$ & $\times$ & $\times$ & $\times$ & $\surd$ & $\surd$ & $\surd$  & $\surd$ \bigstrut\\
			\hline	
			
			\multirow{2}[15]{*}{\makecell{Spectrum \\efficiency}} & 
			\multirow{1}{*}{\makebox[-0.1cm][c]{\cite{10108047}}} & An iterative optimization algorithm to maximize the throughput of SUs. & $\surd$ & $\times$ & $\surd$ & $\surd$ & $\times$ & $\times$ & $\times$ & $\surd$ & $\surd$ & $\surd$  & $\surd$ \bigstrut\\
			\cline{2-14} &
			\multirow{1}{*}{\makebox[-0.1cm][c]{\cite{9400768}}} & An algorithm based on relaxation and penalty to minimize weighted sum BER among all IRSs. & $\surd$ & $\times$ & $\surd$ & $\times$ & $\surd$ & $\times$ & $\times$ & $\surd$ & $\surd$ & $\surd$  & $\surd$ \bigstrut\\
			
			\hline
			\multirow{1}[2]{*}{Others} & \multirow{1}{*}{\makebox[-0.1cm][c]{\cite{9817819}}} & An AO algorithm based on binary integer linear programming and soft actor-critic algorithms to maximize the minimum achievable rate of users. & $\surd$ & $\times$ & $\surd$ & $\times$ & $\times$ & $\times$ & $\times$ & $\surd$ & $\surd$ & $\times$  & $\surd$ \bigstrut\\
			\hline	
		\end{tabular}%
		\begin{tablenotes}
			\centering
			\item The symbol ``$\surd$" represents the article satisfies the property, while ``$\times$" represents not. 
		\end{tablenotes}
	\end{threeparttable}
	\label{tab:addlabel}%
\end{table*}%

\indent Authors in \cite{9944150} focus on imperfect IRSs, which is different from the ideal IRS-assisted UAV communications in \cite{9749020,9367288,9293155,9685217}. In fact, the enhancement of system gain by IRSs largely depends on the reliability of phase estimation and co-phase processes. To alleviate the impact of IRS imperfections, the phase estimation error of IRSs can be taken into account in the practical system design and optimization \cite{9539168}. In \cite{9944150}, authors study the joint IRS element and power allocation problem in the presence of phase estimation and compensation errors. The objective is to maximize the total reception rate of the UAV while satisfying energy constraints and minimum reception rate requirements of individual UAVs. To meet the finite phase configuration frequency of the IRS panel, TDMA technology is used, and each IRS element is only used once in each TDMA frame. A heuristic algorithm based on the estimated phase quality of IRS elements is proposed to solve the problem.

\indent \textbf{Lesson 6}: In IRS-assisted UAV communications, system designs for enhancing the system transmission rate are mainly based on the assumption that the CSI of the entire system is known. However, acquiring the CSI for all channels can evidently cause a significant overhead, and CSI may not be obtained promptly. By leveraging the spatial correlation of channels, adopting a combination of prediction and estimation to acquire the CSI can greatly reduce the system overhead.

\subsubsection{\textbf{Enhanced Communications for Latency Reduction}}

\indent In IRS-assisted UAV communications, most research has focused on the communication propagation process to reduce latency. The solutions primarily involve high-frequency communications \cite{9013626,9348040} and Finite Blocklength Coding (FBC) \cite{9741782,9912224,9781659}. There are also a few studies \cite{9789841,9804341} aiming to enhance the computational capabilities of the system.

\indent Using frequency bands such as mmWave and THz for communications is an effective method to improve transmission latency and alleviate spectrum scarcity to some extent. Authors in \cite{9013626,9348040} consider mmWave-based airborne IRS-assisted communications. In order to maintain the LoS channel for mmWave communications, authors in \cite{9013626} use an RL-based approach to simulate the radio channel, and optimize the phase shift and deployment location of the IRS. Building upon \cite{9013626}, authors in \cite{9348040} introduce the optimal precoding matrix for the BS to further enhance the system's performance. To address the difficulty of LoS channel maintenance brought by the uncertainty of mmWave channels, a distributed RL algorithm is proposed to dynamically optimize locations of the airborne IRS.

\indent The use of FBC can also reduce the latency in IRS-assisted UAV communication systems. FBC refers to appropriate encoding within a given length of data blocks, allowing the receiver to determine whether the sent data is correct within a limited time. Its feasibility in IRS-assisted UAV communication systems is demonstrated in \cite{9741782}, and the system design of FBC-based IRS-assisted UAV communications is discussed in detail in \cite{9912224,9781659}. In \cite{9781659}, authors investigate massive ultra-reliable and low-latency communications for 6G networks supported by IRS-assisted UAV communications. To reduce transmission latency and meet the error-rate bounded quality of service, authors jointly optimize the UAV's trajectory and FBC-based IRS layout. Additionally, authors solve the proposed joint optimization problem by double deep Q-network algorithms. Different from \cite{9781659}, authors in \cite{9912224} apply the FBC technology to MEC system to meet requirements on latency and error rates. Considering energy consumption of edge users and UAVs, authors jointly optimize UAV trajectories, active and passive beamforming, and analyze the effectiveness of this methodology in the context of single-user and multi-user cases.

\indent In MEC scenarios, computational latency cannot be ignored, and its related research is discussed in \cite{9789841,9804341}. Multi-access edge computing in a THz network is considered, and authors in \cite{9789841} study the downlink, uplink, and computation latency on the UAV. Joint optimization of IRS phase shifts, UAV computing resources, and THz sub-bandwidth allocation is performed to reduce the overall system latency. Hungarian algorithm is leveraged to schedule sub-bandwidths, while whale optimization algorithm is used to optimize IRS phase shifts. Different from the single UAV and IRS scenario discussed in \cite{9789841}, authors in \cite{9804341} discuss the overall cost of a collaborative MEC system with multiple UAVs and multiple IRSs, including energy consumption, latency, and maintenance costs. To minimize the total cost of the system, authors formulate a joint optimization problem based on UAV trajectories and IRS phase shifts, which is solved by a four-stage optimization algorithm based on differential evolution, clustering algorithm, and greedy algorithm.

\indent \textbf{Lesson 7}: The aforementioned studies primarily focus on communication transmission and data processing. Although they are effective in reducing latency, there is still room for improvement. On one hand, integrating the efficiency of high-frequency communication with FBC can further reduce system latency. On the other hand, rationally planning the UAV's trajectories and resource allocation based on user task requirements and servers' computational capabilities is feasible. Moreover, not all tasks necessitate a low network latency. Developing an adaptive multi-modal UAV communication method tailored to delay-sensitive tasks and communication resources is important.

\subsubsection{\textbf{Enhanced Communications for Spectrum Efficiency Improvement}}

\indent Due to the scarcity of spectrum resources, researchers' attention to spectrum efficiency is gradually increasing. Generally, CR \cite{10093979,10108047} and Symbiotic Radio (SR) \cite{9860636,9400768} are used to improve spectrum efficiency. Instead of utilizing spectrum with high frequencies, CR and SR reuse the idle spectrum to improve spectrum utilisation. Moreover, unlike multiple access technologies such as NOMA, CR and SR have an intelligent spectrum allocation strategy that avoids interference in signal transmission among multiple users.

\indent CR is composed of PUs and SUs. When SUs share the spectrum for communications, they need to ensure the service quality of PUs. Specifically, CR technology utilizes the idle portion of radio spectra, allowing SUs to occupy these idle frequency bands for communications while generating acceptable interference to PUs, thereby improving spectrum utilization ratios.

\indent The use of CR for spectrum efficiency improvement in IRS-assisted UAV communication systems is widely studied. Authors in \cite{10093979} focus on a CR-based IRS-assisted UAV communication scenario, where SUs, including vehicles and UAVs, rely on CR to access the authorized spectrum for communications. Additionally, NOMA and IRSs are employed to maintain good system performance for SUs. To analyze the performance of CR-based IRS-assisted UAV communication systems, the authors derive the expression for the outage probability of all SUs. Authors in \cite{10108047} propose a CR model based on IRS-assisted UAVs. To address the throughput degradation of SUs caused by interference from PUs, authors make the use of IRS beamforming to mitigate signal interference, and jointly optimize the UAV's trajectory and transmission power to maximize the throughput of SUs.

\indent Building upon CR, SR can ensure good spectrum efficiency while achieving low-power communications. In fact, SR can be regarded as an improved product combining CR and BackCom, where the secondary information is generated by controlling the on/off states of reflecting elements of IRSs, and transmitted through RF signals of the primary information. The outage probability and ergodic spectral efficiency of UAV communication systems assisted by SR-based IRSs are analyzed in \cite{9860636}, in which the UAV equiped with an IRS act as a relay to transmit the signal from the source node to the GU. The ground secondary node modulates and uses ambient BackCom technology to relay the RF signal from the UAV-IRS environment and send information to the ground secondary receiver. 

\indent Authors in \cite{9400768} optimize the performance of IRS-assisted UAV communications based on SR in urban environments. Multiple IRSs are deployed to sense and transmit environmental information, and decoding is achieved through channel response differences. Each reflecting element of IRSs is tuned to align the signal phase of the UAV-IRS-BS link with that of the UAV-BS link, achieving coherent signal combination at the BS side. Furthermore, favorable channel conditions for UAV-BS and UAV-IRS links can be created based on the mobility of the UAV. By jointly optimizing IRS phase shift, IRS scheduling and UAV trajectories, the BER of IRSs is maximized while meeting the minimum transmission rate requirement of UAVs.

\indent \textbf{Lesson 8}: In SR systems, the backscattering efficiency directly affects the system performance, so deployment locations and reflection design of IRSs are particularly important for the improvement of system spectral efficiency. In addition, in theory, NOMA technology does not degrade system performance significantly when the number of users increases. However, in practice, there are still difficulties in the decoding process to distinguish different users.

\subsubsection{\textbf{Enhanced Communications for Others}}

\indent Enhanced communications of IRS-assisted UAVs are also reflected in the network coverage and communication reliability. 

\indent In order to enhance network coverage, two aspects can be considered: \romannumeral1)\, The flight altitude of the UAV can be planned to achieve a balance among transmission performance, coverage ranges, and path losses. For example, authors in \cite{9804220} allow the UAV altitude to be increased when the horizontal distance between the UAV and GUs is far to strike a balance between data transfer rates and outage probabilities. \romannumeral2)\, From the perspective of IRSs, the deployment location, reflection design, and the number of IRSs can be jointly optimized to assist the UAV in improving the coverage range. For example, to meet the wireless network requirements of trains and passengers, authors in \cite{9817819} focus on IRS and UAV-assisted railway communications. High wireless network coverage can reduce frequent network switching caused by movements of high speed trains. Therefore, IRS-equipped UAVs are designed as airborne relays while IRS phase shifts and UAV trajectories are jointly optimized to extend the BS signals. Similarly to \cite{9817819}, authors in \cite{9771729} study how to effectively improve the coverage range of IRS-assisted UAV communication systems based on NOMA, mainly exploring the influence of UAV deployment density and user distribution. By analysing historical associations between UAVs and users with LoS and NLoS channels, the maximum achievable coverage probability of two users is derived and analyzed.

\indent Improving reliability of UAV communication systems can be considered from two aspects: \romannumeral1)\,Multi-path compensation. An IRS-based multi-path compensation scheme is investigated in \cite{9715145}. Specifically, at multi-antenna BSs, the signal is divided into multiple orthogonal active beams for data transmission. Subsequently, a set of IRSs reflects the orthogonal beams through different paths, and the received signals are coherently combined at the user's receiver. This approach effectively compensates for the significant multiplicative path loss induced by multipath communications. \romannumeral2)\, Multi-user interference management. Common multi-user interference management methods include beamforming, multiple access technologies, and resource allocation. Authors in \cite{8811733,9417539} focus on the signal in a specific direction to reduce the received interference from non-target users by beamforming through IRSs. Authors in \cite{9893192,9454446} study multiuser communications based on NOMA and use SIC technique to decode user received signals and eliminate interference among different users. In addition, adjusting the transmission power of each user based on the channel quality and user location information can also reduce the signal interference among multiple users \cite{9944150}.


\indent \textbf{Lesson 9}: In summary, enhancing network coverage and communication reliability requires consideration from multiple aspects, mainly including reflections of IRSs and positional design of IRSs and UAVs. Moreover, they are contradictory metrics, and their balance needs special consideration in different scenarios. Although increasing the flight altitude of UAVs and the deployment altitude of IRSs can enhance the network coverage, the practical system design usually has an altitude limitation, which is not considered in the above studies. In addition, the above studies ignore the fact that UAV flight can inevitably generate jitter, which should be solved by efficient beam tracking techniques.

\subsubsection{\textbf{ Summary for Enhanced Communications in Different Scenarios}} 
\hyperlink{table 5}{Table \uppercase\expandafter{\romannumeral5}} presents solutions to achieve enhanced communications in IRS-assisted UAV communications from the perspective of system performance metrics. However, different scenarios often have varying requirements for communication performance metrics. As a result, the current implementations of enhanced communication solutions in different application scenarios are further summarized, and potential optimization strategies are explored as follows.

In SAGIN, resource scheduling within the multi-layer network structure impacts the overall network rate. Authors in \cite{9822386} focus on relay switching in SAGIN and propose a comprehensive design and analysis of a free-space optics-based IRS-UAV relay-assisted SAGIN. This addresses the effects of weather and atmospheric conditions, and introduces a multi-rate system with a new link switching scheme to optimize network rates and energy consumption. Unlike studies on relay switching schemes \cite{9822386}, authors in \cite{10288083} focus on system beamforming design and propose an asymmetric LSTM-deep deterministic policy gradient algorithm. This algorithm simultaneously optimizes active and passive beamforming to enhance system network rate in an uplink airborne IRS-assisted hybrid free-space optics/RF-enabled SAGIN. 

\indent In V2X, communication protocols influence system resource allocation and scheduling.  Authors in \cite{10498067} propose a UAV-enhanced IRS-assisted V2X architecture and design a corresponding communication protocol to improve system transmission rates, by optimizing transmit power and IRS phase shifts. To address the lack of relay flexibility caused by fixed IRS, authors in \cite{10285074} propose a multifunctional UAV-assisted vehicular communication system equipped with IRS. In this system, UAVs act as transmit BSs for communications with ground vehicles while carrying IRS to provide relay services for V2V communications. By optimizing vehicle scheduling, UAV transmission power, IRS reflection phase, and UAV trajectories, the average bit rate for UAV-to-vehicle and V2V communications is maximized while meeting minimum communication rate requirements. Authors in \cite{10499959} propose an optimization scheme to maximize system spectral efficiency in NOMA-based multi-UAV and multi-IRS networks by optimizing UAV association, power control, and passive beamforming under the presence of imperfect SIC.

\hypertarget{Section5}{\section{Research Challenges and Open Issues}}

\indent Although IRS-assisted UAV communications can address various issues, it still faces many challenges and research opportunities in the face of the continuous development of future 6G networks.

\subsection{Research Challenges}
\indent Through the investigation of existing research, there are still some research challenges in IRS-assisted UAV communications, which are described as follows.

\subsubsection{\textbf{Anti-Pilot Contamination in IRS-Assisted UAV Communications}} 
\indent In IRS-assisted UAV communications, beamforming is an effective method to enhance system performance. Efficient beamforming requires accurate CSI, which is usually obtained through pilot signals. In practice, pilot signals may become contaminated due to noise injected by attackers or interference inter/intra-IRS, leading to channel estimation errors. Particularly, pilot contamination caused by attackers can result in ineffective transmission or even make the transmission design favorable for the attacker to launch attacks. Although authors in \cite{9344862} study pilot contamination in IRS, they do not further consider the challenges posed by UAV mobility. Specifically, in high-speed UAV scenarios, pilot sequences need frequent updates, increasing the complexity of system design. To mitigate pilot contamination, complex algorithms are needed for channel estimation, which places high demands on computational resources and real-time scenarios. In other words, considerations for pilot contamination need to be addressed from both pilot signal design and channel estimation perspectives, with attention to system energy consumption and computational resources.

\subsubsection{\textbf{Enhanced Interference Management in IRS-Assisted UAV Communications}}
\indent Currently, most research on interference management in IRS-assisted UAV communications is limited to single-user or multi-user scenarios considering multiple access technologies, and collaborative resource scheduling based on beamforming is the primary interference management method. However, interference management in IRS-assisted UAV communication scenarios remains challenging:
\begin{itemize}
	\item Due to hardware limitations, IRS can only generate one set of phase shifts within a short time interval. This means that in broadband scenarios, a single phase setting of IRS cannot effectively suppress interference across all frequencies. Although authors in \cite{9365004, 9353406} demonstrate the effectiveness of IRS in multi-user scenarios, the interference suppression effect of IRS significantly decreases as the number of users increases.
	\item In multi-cell scenarios, co-channel transmission inevitably leads to severe cross-cell interference, which becomes extremely complex and difficult to resolve under the high mobility of UAVs and rapidly changing channel conditions. Additionally, significant differences in channel characteristics among different users and cells make a single phase setting of IRS cannot achieve ideal interference management for all users. Although multiple fixed IRSs can achieve effective UAV interference management, this method inevitably increases the complexity of system design in dynamic IRS scenarios \cite{10537097} . 
\end{itemize}

 \indent It is worth noting that CoMP technology garners significant attention for its interference suppression capabilities, particularly for inter-cell interference. Authors in \cite{9279253} study the performance of IRS-assisted joint processing CoMP (a specific CoMP technology) downlink transmission in a multiple-user system, with simulation results showing a substantial increase in system throughput due to the combination of IRS and CoMP. However, how to deploy the IRS-assisted CoMP system in UAV communication scenarios has not been studied.

\subsubsection{\textbf{Design of IRS-Assisted UAV Communication Protocols}}
\indent Designing IRS communication protocols is a necessary and challenging task. Generally, control signals for IRS and relay signals are independent of each other, requiring special communication protocols to avoid interference between them. Therefore, IRS communication protocols are the foundation of IRS-assisted UAV communications, but most research focuses only on the communication process while neglecting the protocol design. Typically, IRS control information reception involves three stages: information synchronization, channel estimation, and control information transmission \cite{9453804,9198125}. For devices with signal processing capabilities, channel estimation is relatively easy to implement. However, for IRS, which lacks computational power and energy supply, the communication protocol design must prioritize low power consumption and low complexity. Additionally, IRS communication protocols need to be timely to adapt to the rapidly changing UAV communication scenarios. Timeliness mainly depends on the channel estimation process. In \cite{9453804}, authors assume perfect CSI acquisition during the channel estimation phase, reducing the practicality of the protocol. In fact, incorporating IRS with sensing capabilities could provide potential insights for IRS protocol design.

\subsection{Open Issues}
\indent With the advancement in research on IRSs and UAVs, a number of interesting topics are coming into the public eye. Therefore, this subsection provides an overview of the future field of IRS-assisted UAV communications to inspire researchers for further exploration.

\subsubsection{\textbf{IRS Reflective Unit Scheduling in IRS-Assisted UAV Communications}}
\indent Most research on IRSs focuses on system performance improvement brought by a single IRS or multiple IRSs, ignoring the fact that the IRS is composed of a finite number of reflecting units. In early research on IRSs, most assumptions are made based on the ideal IRS. In fact, both IRSs and transceivers have different degrees of hardware limitations and defects \cite{9722893}, and it is often not feasible to correct the phase shift of the IRS by ideal phase compensation techniques \cite{9395180}. Specially, the phase errors of each reflecting unit are not the same \cite{9133142}. In this case, increasing the number of IRSs may lead to a decrease in system performance, and the reason for this phenomenon is the introduction of reflecting units with large phase errors due to non-ideal phase estimation and compensation methods \cite{9816032}. Therefore, future research should be focused on how to select the reflecting units of IRSs to enhance the performance of UAV systems. This includes developing advanced algorithms for real-time phase error correction and adaptive unit selection, and exploring the trade-offs between the number of IRSs and the overall system performance.

\indent The selection of specific reflecting units of IRSs may result in idle reflecting units. How to make reasonable usage of these idle reflecting units is deserved to investigate. In fact, unused reflecting units of IRSs can be reconfigured to serve other users. First, due to the randomness of wireless network environments, certain reflecting units may perform poorly for specific users and scenarios, but not necessarily in other scenarios, providing an opportunity for differentiated services of IRS reflecting units. Second, the rational use of idle reflecting units can reduce the number of IRSs, mitigating the cost of network deployment to a certain extent. Consequently, dynamically reallocating reflecting units based on real-time network conditions, user demands, and performance metrics is necessary. Such strategies should maximize the overall network efficiency and cost-effectiveness while maintaining high service quality for diverse user requirements.

\subsubsection{\textbf{Hybrid Active and Passive IRS-Assisted UAV Communications}}
\indent Up to now, most of the designed IRS are passive in IRS-assisted UAV communications, which may not perform well in certain scenarios like IoT data collection systems due to low transmission power and the ``double fading" effect \cite{9652031}. Active IRS, equipped with power amplifiers, can amplify reflected signals to enable effective communication with low-power signals, but they introduce challenges such as increased energy consumption and amplified noise, which affect transmission quality and increase BER. A promising future direction involves the joint design of active and passive IRSs to combine the signal gain of active IRSs with the interference reduction of passive IRSs, though this approach increases system design complexity and optimization challenges, including deployment, cooperation, and balancing energy consumption with system performance. Future research should address these complexities and explore advanced algorithms for optimizing the integration and deployment of active and passive IRSs, considering hardware implementation and energy efficiency.

\subsubsection{\textbf{IRS and UAV-Assisted Maritime Communications}}
\indent Currently, maritime communications are mainly realized based on satellites. However, their reliability and timeliness are not well guaranteed due to the long distances and harsh maritime communication environment. The combination of IRSs and UAVs introduces a variety of new research topics for diverse maritime communication scenarios. First, IRSs can dynamically adjust reflected signals and be deployed on ocean platforms, buoys, and ships to optimize signal paths and reduce interference. This provides a low-cost data collection method for UAV-based marine environment monitoring systems. Second, IRS mounted on UAVs can act as flexible maritime communication relays, offering stable connections and enhanced coverage for emergency maritime communications and maritime traffic scenarios. It is worth noting that the maritime environment presents complexities such as multipath propagation \cite{9645536}, signal attenuation, dynamic changes (e.g., waves, wind, and tides), and evaporation ducting effects \cite{9460824}, necessitating novel wireless channel modeling methods.

\subsubsection{\textbf{IRS and UAV-Assisted Communication-Sensing Integration}}
\indent The current focus of most research is on using IRS as a relay to enhance transmission quality. As research into IRS progresses, researchers discover that the reflecting units of IRS can exhibit different response characteristics to incident signals from various directions \cite{9938373}. This suggests that IRS can serve as a sensing array for passive monitoring of the radio environment. The integration of IRS with UAVs may bring revolutionary changes to future integrated sensing and communication networks. Based on specific beam design, dual-function IRS deployed in dense urban areas can simultaneously provide communication and sensing services (such as positioning) to UAVs \cite{10508296}. With the support of specific transmission protocols, using the positioning and channel information sensed by IRS is expected to further improve the communication quality of UAVs \cite{9937163}. Moreover, UAVs are widely regarded as effective carriers for aerial BSs and relay communications. Thus, the issue of performance degradation caused by obstacles can be effectively resolved through UAVs. 
However, the integration of sensing and communication inevitably leads to resource competition. If resources are evenly allocated between communication and sensing without reasonable consideration, it may result in resource wastage \cite{9858656}. Therefore, balancing resources between sensing and communication and designing flexible resource allocation strategies can be a future research direction for IRS and UAV-assisted integrated sensing and communication networks.

\hypertarget{Section6}{\section{Conclusion}}

\indent In this article, we explore the research on IRS-assisted UAV communications for 6G networks. Specifically, we introduce typical application scenarios of IRS-assisted UAV communications for 6G networks, and discuss key issues faced by these applications. Then, we present prototypes from multiple perspectives and summarize key technologies of IRS-assisted UAV communications. After that, existing solutions for key issues are summarized from different perspectives, including energy-constrained communications, secure communications, and enhanced communications. Last, we highlight some open issues and future research challenges of IRS-assisted UAV communications. Through this article, we can observe advantages of IRS-assisted UAV communications in addressing issues under 6G networks. However, there are still challenges and difficulties that need to be further investigated. We hope that this article can provide useful references and insights for researchers and contribute to the advancement of IRS-assisted UAV communications.

\section{Acknowledgment}
\indent This work was supported by the Natural Science Foundation of China under Grant 62272075, Grant 62371289 and Grant 62331022, National Natural Science Foundation of Chongqing under Grant CSTB2024NSCQ-JQX0013 and Grant CSTB2024NSCQ-QCXMX0058, Science and Technology Research Program for Chongqing Municipal Education Commission KJZD-M202200601 and KJZD-K202300608,  Hong Kong RGC General Research Fund (152244/21E, 152169/22E, 152228/23E, 162161/24E), Research Impact Fund (No. R5011-23F, No. R5060-19), Collaborative Research Fund (No. C1042-23GF), Theme-based Research Scheme (T43-518/24-N), Areas of Excellence Scheme (AoE/E-601/22-R), and the InnoHK (HKGAI).

\bibliographystyle{ieeetr}
\bibliography{myref.bib}

\begin{thebibliography}{100}

\bibitem{8782879}
P.~Yang, Y.~Xiao, M.~Xiao, and S.~Li, ``{6G} wireless communications: Vision
  and potential techniques,'' {\em IEEE Network}, vol.~33, no.~4, pp.~70--75,
  2019.

\bibitem{9903905}
A.~Ihsan, W.~Chen, M.~Asif, W.~U. Khan, Q.~Wu, and J.~Li, ``Energy-efficient
  {IRS}-aided {NOMA} beamforming for {6G} wireless communications,'' {\em IEEE
  Transactions on Green Communications and Networking}, vol.~6, no.~4,
  pp.~1945--1956, 2022.

\bibitem{10239285}
T.~Q. Duong, L.~D. Nguyen, T.~T. Bui, K.~D. Pham, and G.~K. Karagiannidis,
  ``Machine learning-aided real-time optimized multibeam for {6G} integrated
  satellite-terrestrial networks: {Global} coverage for mobile services,'' {\em
  IEEE Network}, vol.~37, no.~2, pp.~86--93, 2023.

\bibitem{9681714}
Y.~Cao, S.~Xu, J.~Liu, and N.~Kato, ``Toward smart and secure {V2X}
  communication in {5G} and beyond: A {UAV}-enabled aerial intelligent
  reflecting surface solution,'' {\em IEEE Vehicular Technology Magazine},
  vol.~17, no.~1, pp.~66--73, 2022.

\bibitem{9681624}
M.~Mozaffari, X.~Lin, and S.~Hayes, ``Toward {6G} with connected sky: {UAVs}
  and beyond,'' {\em IEEE Communications Magazine}, vol.~59, no.~12,
  pp.~74--80, 2021.

\bibitem{9870557}
N.~Babu, M.~Virgili, M.~Al-jarrah, X.~Jing, E.~Alsusa, P.~Popovski, A.~Forsyth,
  C.~Masouros, and C.~B. Papadias, ``Energy-efficient trajectory design of a
  multi-{IRS} assisted portable access point,'' {\em IEEE Transactions on
  Vehicular Technology}, vol.~72, no.~1, pp.~611--622, 2023.

\bibitem{9599592}
X.~Pang, M.~Sheng, N.~Zhao, J.~Tang, D.~Niyato, and K.-K. Wong, ``When {UAV}
  meets {IRS}: Expanding air-ground networks via passive reflection,'' {\em
  IEEE Wireless Communications}, vol.~28, no.~5, pp.~164--170, 2021.

\bibitem{9326394}
Q.~Wu, S.~Zhang, B.~Zheng, C.~You, and R.~Zhang, ``Intelligent reflecting
  surface-aided wireless communications: A tutorial,'' {\em IEEE Transactions
  on Communications}, vol.~69, no.~5, pp.~3313--3351, 2021.

\bibitem{10061643}
Y.~Ge, J.~Fan, G.~Y. Li, and L.-C. Wang, ``Intelligent reflecting
  surface-enhanced {UAV} communications: {Advances}, challenges, and
  prospects,'' {\em IEEE Wireless Communications}, vol.~30, no.~6,
  pp.~119--126, 2023.

\bibitem{9453804}
X.~Cao, B.~Yang, C.~Huang, C.~Yuen, M.~D. Renzo, D.~Niyato, and Z.~Han,
  ``Reconfigurable intelligent surface-assisted aerial-terrestrial
  communications via multi-task learning,'' {\em IEEE Journal on Selected Areas
  in Communications}, vol.~39, no.~10, pp.~3035--3050, 2021.

\bibitem{9198125}
M.-M. Zhao, Q.~Wu, M.-J. Zhao, and R.~Zhang, ``Intelligent reflecting surface
  enhanced wireless networks: Two-timescale beamforming optimization,'' {\em
  IEEE Transactions on Wireless Communications}, vol.~20, no.~1, pp.~2--17,
  2021.

\bibitem{9804220}
Y.~Cai, Z.~Wei, S.~Hu, C.~Liu, D.~W.~K. Ng, and J.~Yuan, ``Resource allocation
  and {3D} trajectory design for power-efficient {IRS}-assisted {UAV}-{NOMA}
  communications,'' {\em IEEE Transactions on Wireless Communications},
  vol.~21, no.~12, pp.~10315--10334, 2022.

\bibitem{9140329}
M.~Di~Renzo, A.~Zappone, M.~Debbah, M.-S. Alouini, C.~Yuen, J.~de~Rosny, and
  S.~Tretyakov, ``Smart radio environments empowered by reconfigurable
  intelligent surfaces: {How} it works, state of research, and the road
  ahead,'' {\em IEEE Journal on Selected Areas in Communications}, vol.~38,
  no.~11, pp.~2450--2525, 2020.

\bibitem{9279253}
M.~Hua, Q.~Wu, D.~W.~K. Ng, J.~Zhao, and L.~Yang, ``Intelligent reflecting
  surface-aided joint processing coordinated multipoint transmission,'' {\em
  IEEE Transactions on Communications}, vol.~69, no.~3, pp.~1650--1665, 2021.

\bibitem{10375242}
Q.~Li, P.~Si, Y.~Zhang, J.~Wang, D.~Zhang, and F.~R. Yu, ``{UAV} altitude,
  relay selection, and user association optimization for cooperative
  relay-transmission in {UAV-IRS}-based {THz} networks,'' {\em IEEE
  Transactions on Green Communications and Networking}, vol.~8, no.~2,
  pp.~815--826, 2024.

\bibitem{zhang2018space}
L.~Zhang, X.~Q. Chen, S.~Liu, Q.~Zhang, J.~Zhao, J.~Y. Dai, G.~D. Bai, X.~Wan,
  Q.~Cheng, G.~Castaldi, {\em et~al.}, ``Space-time-coding digital
  metasurfaces,'' {\em Nature communications}, vol.~9, no.~1, p.~4334, 2018.

\bibitem{liu2022programmable}
C.~Liu, Q.~Ma, Z.~J. Luo, Q.~R. Hong, Q.~Xiao, H.~C. Zhang, L.~Miao, W.~M. Yu,
  Q.~Cheng, L.~Li, {\em et~al.}, ``A programmable diffractive deep neural
  network based on a digital-coding metasurface array,'' {\em Nature
  Electronics}, vol.~5, no.~2, pp.~113--122, 2022.

\bibitem{9386233}
A.~Mahmoud, S.~Muhaidat, P.~C. Sofotasios, I.~Abualhaol, O.~A. Dobre, and
  H.~Yanikomeroglu, ``Intelligent reflecting surfaces assisted {UAV}
  communications for {IoT} networks: {Performance} analysis,'' {\em IEEE
  Transactions on Green Communications and Networking}, vol.~5, no.~3,
  pp.~1029--1040, 2021.

\bibitem{10.1145/3479239.3485700}
K.~Heimann, B.~Sliwa, M.~Patchou, and C.~Wietfeld, ``Modeling and simulation of
  reconfigurable intelligent surfaces for hybrid aerial and ground-based
  vehicular communications,'' in {\em Proc. MSWiM '21}, p.~67–74, 2021.

\bibitem{8641424}
N.~Zhao, W.~Lu, M.~Sheng, Y.~Chen, J.~Tang, F.~R. Yu, and K.-K. Wong,
  ``{UAV}-assisted emergency networks in disasters,'' {\em IEEE Wireless
  Communications}, vol.~26, no.~1, pp.~45--51, 2019.

\bibitem{9690481}
C.~You, Z.~Kang, Y.~Zeng, and R.~Zhang, ``Enabling smart reflection in
  integrated air-ground wireless network: {IRS} meets {UAV},'' {\em IEEE
  Wireless Communications}, vol.~28, no.~6, pp.~138--144, 2021.

\bibitem{9351782}
H.~Lu, Y.~Zeng, S.~Jin, and R.~Zhang, ``Aerial intelligent reflecting surface:
  Joint placement and passive beamforming design with {3D} beam flattening,''
  {\em IEEE Transactions on Wireless Communications}, vol.~20, no.~7,
  pp.~4128--4143, 2021.

\bibitem{9722893}
B.~Zheng, C.~You, W.~Mei, and R.~Zhang, ``A survey on channel estimation and
  practical passive beamforming design for intelligent reflecting surface aided
  wireless communications,'' {\em IEEE Communications Surveys \& Tutorials},
  vol.~24, no.~2, pp.~1035--1071, 2022.

\bibitem{DAJER202287}
M.~Dajer, Z.~Ma, L.~Piazzi, N.~Prasad, X.-F. Qi, B.~Sheen, J.~Yang, and G.~Yue,
  ``Reconfigurable intelligent surface: Design the channel-a new opportunity
  for future wireless networks,'' {\em Digital Communications and Networks},
  vol.~8, no.~2, pp.~87--104, 2022.

\bibitem{SADIA2023}
H.~Sadia, A.~K. Hassan, Z.~H. Abbas, G.~Abbas, M.~Waqas, and Z.~Han,
  ``{IRS}-enabled {NOMA} communication systems: {A} network architecture primer
  with future trends and challenges,'' {\em Digital Communications and
  Networks, DOI:{https://doi.org/10.1016/j.dcan.2023.09.002}}, 2023.

\bibitem{8411465}
A.~A. Khuwaja, Y.~Chen, N.~Zhao, M.-S. Alouini, and P.~Dobbins, ``A survey of
  channel modeling for {UAV} communications,'' {\em IEEE Communications Surveys
  \& Tutorials}, vol.~20, no.~4, pp.~2804--2821, 2018.

\bibitem{10.1145/3604933}
Z.~Ning, H.~Hu, X.~Wang, L.~Guo, S.~Guo, G.~Wang, and X.~Gao, ``Mobile edge
  computing and machine learning in the {Internet} of unmanned aerial vehicles:
  {A} survey,'' {\em ACM Computing Surveys}, vol.~56, no.~1, p.~31, 2023.

\bibitem{8918497}
Y.~Zeng, Q.~Wu, and R.~Zhang, ``Accessing from the sky: A tutorial on {UAV}
  communications for {5G} and beyond,'' {\em Proceedings of the IEEE},
  vol.~107, no.~12, pp.~2327--2375, 2019.

\bibitem{8675384}
A.~Fotouhi, H.~Qiang, M.~Ding, M.~Hassan, L.~G. Giordano, A.~Garcia-Rodriguez,
  and J.~Yuan, ``Survey on {UAV} cellular communications: Practical aspects,
  standardization advancements, regulation, and security challenges,'' {\em
  IEEE Communications Surveys \& Tutorials}, vol.~21, no.~4, pp.~3417--3442,
  2019.

\bibitem{9424177}
Y.~Liu, X.~Liu, X.~Mu, T.~Hou, J.~Xu, M.~Di~Renzo, and N.~Al-Dhahir,
  ``Reconfigurable intelligent surfaces: Principles and opportunities,'' {\em
  IEEE Communications Surveys \& Tutorials}, vol.~23, no.~3, pp.~1546--1577,
  2021.

\bibitem{9122596}
S.~Gong, X.~Lu, D.~T. Hoang, D.~Niyato, L.~Shu, D.~I. Kim, and Y.-C. Liang,
  ``Toward smart wireless communications via intelligent reflecting surfaces: A
  contemporary survey,'' {\em IEEE Communications Surveys \& Tutorials},
  vol.~22, no.~4, pp.~2283--2314, 2020.

\bibitem{9806434}
W.~Song, S.~Rajak, S.~Dang, R.~Liu, J.~Li, and S.~Chinnadurai, ``Deep learning
  enabled {IRS} for 6g intelligent transportation systems: {A} comprehensive
  study,'' {\em IEEE Transactions on Intelligent Transportation Systems},
  vol.~24, no.~11, pp.~12973--12990, 2023.

\bibitem{9968053}
S.~Aboagye, A.~R. Ndjiongue, T.~M.~N. Ngatched, O.~A. Dobre, and H.~V. Poor,
  ``{RIS}-assisted visible light communication systems: {A} tutorial,'' {\em
  IEEE Communications Surveys \& Tutorials}, vol.~25, no.~1, pp.~251--288,
  2023.

\bibitem{9358097}
N.-N. Dao, Q.-V. Pham, N.~H. Tu, T.~T. Thanh, V.~N.~Q. Bao, D.~S. Lakew, and
  S.~Cho, ``Survey on aerial radio access networks: Toward a comprehensive {6G}
  access infrastructure,'' {\em IEEE Communications Surveys \& Tutorials},
  vol.~23, no.~2, pp.~1193--1225, 2021.

\bibitem{9768113}
G.~Geraci, A.~Garcia-Rodriguez, M.~M. Azari, A.~Lozano, M.~Mezzavilla,
  S.~Chatzinotas, Y.~Chen, S.~Rangan, and M.~D. Renzo, ``What will the future
  of {UAV} cellular communications be? {A} flight from {5G} to {6G},'' {\em
  IEEE Communications Surveys \& Tutorials}, vol.~24, no.~3, pp.~1304--1335,
  2022.

\bibitem{9779853}
Z.~Wei, M.~Zhu, N.~Zhang, L.~Wang, Y.~Zou, Z.~Meng, H.~Wu, and Z.~Feng,
  ``{UAV}-assisted data collection for {Internet} of things: A survey,'' {\em
  IEEE Internet of Things Journal}, vol.~9, no.~17, pp.~15460--15483, 2022.

\bibitem{7876852}
H.~Menouar, I.~Guvenc, K.~Akkaya, A.~S. Uluagac, A.~Kadri, and A.~Tuncer,
  ``{UAV}-enabled intelligent transportation systems for the smart city:
  Applications and challenges,'' {\em IEEE Communications Magazine}, vol.~55,
  no.~3, pp.~22--28, 2017.

\bibitem{10354514}
S.~K. Taskou, M.~Rasti, and E.~Hossain, ``End-to-end resource slicing for
  coexistence of {eMBB} and {URLLC} services in {5G}-advanced/{6G} networks,''
  {\em IEEE Transactions on Mobile Computing}, vol.~23, no.~7, pp.~8015--8032,
  2024.

\bibitem{9520380}
B.~Shang, Y.~Yi, and L.~Liu, ``Computing over space-air-ground integrated
  networks: Challenges and opportunities,'' {\em IEEE Network}, vol.~35, no.~4,
  pp.~302--309, 2021.

\bibitem{10288083}
M.~Wu, K.~Guo, Z.~Lin, X.~Li, K.~An, and Y.~Huang, ``Joint optimization design
  of {RIS}-assisted hybrid {FSO} {SAGINs} using deep reinforcement learning,''
  {\em IEEE Transactions on Vehicular Technology}, vol.~73, no.~3,
  pp.~3025--3040, 2024.

\bibitem{9822386}
T.~V. Nguyen, H.~D. Le, and A.~T. Pham, ``On the design of {RIS-UAV}
  relay-assisted hybrid {FSO/RF} satellite-aerial-ground integrated network,''
  {\em IEEE Transactions on Aerospace and Electronic Systems}, vol.~59, no.~2,
  pp.~757--771, 2023.

\bibitem{9771729}
C.-H. Liu, M.~A. Syed, and L.~Wei, ``Toward ubiquitous and flexible coverage of
  {UAV}-{IRS}-assisted {NOMA} networks,'' in {\em Proc. IEEE WCNC},
  pp.~1749--1754, 2022.

\bibitem{10330152}
Y.~Chen, W.~Cheng, and W.~Zhang, ``Reconfigurable intelligent surface equipped
  {UAV} in emergency wireless communications: {A} new fading–shadowing model
  and performance analysis,'' {\em IEEE Transactions on Communications},
  vol.~72, no.~3, pp.~1821--1834, 2024.

\bibitem{9144463}
Y.~Chen, Y.~Wang, J.~Zhang, and Z.~Li, ``Resource allocation for intelligent
  reflecting surface aided vehicular communications,'' {\em IEEE Transactions
  on Vehicular Technology}, vol.~69, no.~10, pp.~12321--12326, 2020.

\bibitem{9714139}
M.~A. Javed, T.~N. Nguyen, J.~Mirza, J.~Ahmed, and B.~Ali, ``Reliable
  communications for cybertwin-driven {6G} {IoVs} using intelligent reflecting
  surfaces,'' {\em IEEE Transactions on Industrial Informatics}, vol.~18,
  no.~11, pp.~7454--7462, 2022.

\bibitem{9416239}
S.~Li, B.~Duo, M.~D. Renzo, M.~Tao, and X.~Yuan, ``Robust secure {UAV}
  communications with the aid of reconfigurable intelligent surfaces,'' {\em
  IEEE Transactions on Wireless Communications}, vol.~20, no.~10,
  pp.~6402--6417, 2021.

\bibitem{9685217}
J.~Xu, X.~Kang, R.~Zhang, and Y.-C. Liang, ``Joint power and trajectory
  optimization for {IRS}-aided master-auxiliary-{UAV}-powered {IoT} networks,''
  in {\em Proc. IEEE GLOBECOM}, pp.~1--6, 2021.

\bibitem{9804495}
J.~Xu, X.~Kang, R.~Zhang, Y.-C. Liang, and S.~Sun, ``Optimization for
  master-{UAV}-powered auxiliary-aerial-{IRS}-assisted {IoT} networks: An
  option-based multi-agent hierarchical deep reinforcement learning approach,''
  {\em IEEE Internet of Things Journal}, vol.~9, no.~22, pp.~22887--22902,
  2022.

\bibitem{8629941}
N.~Chaabouni, M.~Mosbah, A.~Zemmari, C.~Sauvignac, and P.~Faruki, ``Network
  intrusion detection for {IoT} security based on learning techniques,'' {\em
  IEEE Communications Surveys \& Tutorials}, vol.~21, no.~3, pp.~2671--2701,
  2019.

\bibitem{9454446}
X.~Mu, Y.~Liu, L.~Guo, J.~Lin, and H.~V. Poor, ``Intelligent reflecting surface
  enhanced multi-{UAV} {NOMA} networks,'' {\em IEEE Journal on Selected Areas
  in Communications}, vol.~39, no.~10, pp.~3051--3066, 2021.

\bibitem{10379001}
Y.~Ma, K.~Ota, and M.~Dong, ``{QoE} optimization for virtual reality services
  in multi-{RIS}-assisted {Terahertz} wireless networks,'' {\em IEEE Journal on
  Selected Areas in Communications}, vol.~42, no.~3, pp.~538--551, 2024.

\bibitem{9580624}
Z.~Ning, H.~Chen, X.~Wang, S.~Wang, and L.~Guo, ``Blockchain-enabled electrical
  fault inspection and secure transmission in {5G} smart grids,'' {\em IEEE
  Journal of Selected Topics in Signal Processing}, vol.~16, no.~1, pp.~82--96,
  2022.

\bibitem{10108047}
Y.~Yu, X.~Liu, Z.~Liu, and T.~S. Durrani, ``Joint trajectory and resource
  optimization for {RIS} assisted {UAV} cognitive radio,'' {\em IEEE
  Transactions on Vehicular Technology}, vol.~72, no.~10, pp.~13643--13648,
  2023.

\bibitem{9894720}
S.~Zargari, A.~Hakimi, C.~Tellambura, and S.~Herath, ``User scheduling and
  trajectory optimization for energy-efficient {IRS}-{UAV} networks with
  {SWIPT},'' {\em IEEE Transactions on Vehicular Technology}, vol.~72, no.~2,
  pp.~1815--1830, 2023.

\bibitem{9756208}
Y.~Liu, F.~Han, and S.~Zhao, ``Flexible and reliable multiuser {SWIPT} {IoT}
  network enhanced by {UAV}-mounted intelligent reflecting surface,'' {\em IEEE
  Transactions on Reliability}, vol.~71, no.~2, pp.~1092--1103, 2022.

\bibitem{9743298}
J.~Yu, X.~Liu, Y.~Gao, C.~Zhang, and W.~Zhang, ``Deep learning for channel
  tracking in {IRS}-assisted {UAV} communication systems,'' {\em IEEE
  Transactions on Wireless Communications}, vol.~21, no.~9, pp.~7711--7722,
  2022.

\bibitem{10066841}
J.~Wang, S.~Xu, S.~Han, and L.~Xiao, ``{UAV}-powered multi-user intelligent
  reflecting surface backscatter communication,'' {\em IEEE Transactions on
  Vehicular Technology}, vol.~72, no.~8, pp.~10251--10262, 2023.

\bibitem{10139787}
Z.~Hou, Y.~Huang, J.~Chen, G.~Li, X.~Guan, Y.~Xu, R.~Chen, and Y.~Xu, ``Joint
  {IRS} selection and passive beamforming in multiple {IRS-UAV}-enhanced
  anti-jamming {D2D} communication networks,'' {\em IEEE Internet of Things
  Journal}, vol.~10, no.~22, pp.~19558--19569, 2023.

\bibitem{9696283}
L.~Wang, Y.~Chen, P.~Wang, and Z.~Yan, ``Security threats and countermeasures
  of unmanned aerial vehicle communications,'' {\em IEEE Communications
  Standards Magazine}, vol.~5, no.~4, pp.~41--47, 2021.

\bibitem{9940551}
T.~Cheng, B.~Wang, K.~Cao, R.~Dong, and D.~Diao, ``{IRS}-enabled secure {G2A}
  communications for {UAV} system with aerial eavesdropping,'' {\em IEEE
  Systems Journal}, vol.~17, no.~3, pp.~3670--3681, 2023.

\bibitem{10115020}
W.~Zhai, L.~Liu, Y.~Ding, S.~Sun, and Y.~Gu, ``Etd: {An} efficient time delay
  attack detection framework for {UAV} networks,'' {\em IEEE Transactions on
  Information Forensics and Security}, vol.~18, pp.~2913--2928, 2023.

\bibitem{10261240}
Z.~Yu, Z.~Wang, J.~Yu, D.~Liu, H.~Song, and Z.~Li, ``Cybersecurity of unmanned
  aerial vehicles: {A} survey,'' {\em IEEE Aerospace and Electronic Systems
  Magazine, DOI:{10.1109/MAES.2023.3318226}}, pp.~1--25, 2023.

\bibitem{10121733}
F.~Naeem, M.~Ali, G.~Kaddoum, C.~Huang, and C.~Yuen, ``Security and privacy for
  reconfigurable intelligent surface in {6G}: {A} review of prospective
  applications and challenges,'' {\em IEEE Open Journal of the Communications
  Society}, vol.~4, pp.~1196--1217, 2023.

\bibitem{9771762}
H.~Zhao, J.~Hao, and Y.~Guo, ``Joint trajectory and beamforming design for
  {IRS}-assisted anti-jamming {UAV} communication,'' in {\em Proc. IEEE WCNC},
  pp.~369--374, 2022.

\bibitem{9538830}
W.~Wang, H.~Tian, and W.~Ni, ``Secrecy performance analysis of {IRS}-aided
  {UAV} relay system,'' {\em IEEE Wireless Communications Letters}, vol.~10,
  no.~12, pp.~2693--2697, 2021.

\bibitem{9527176}
X.~Fang, Z.~Du, X.~Yin, L.~Liu, X.~Sha, and H.~Zhang, ``Toward physical layer
  security and efficiency for {SAGIN}: A {WFRFT}-based parallel complex-valued
  spectrum spreading approach,'' {\em IEEE Transactions on Intelligent
  Transportation Systems}, vol.~23, no.~3, pp.~2819--2829, 2022.

\bibitem{8869705}
W.~Saad, M.~Bennis, and M.~Chen, ``A vision of {6G} wireless systems:
  {Applications}, trends, technologies, and open research problems,'' {\em IEEE
  Network}, vol.~34, no.~3, pp.~134--142, 2020.

\bibitem{9749020}
X.~Zhang, H.~Zhang, W.~Du, K.~Long, and A.~Nallanathan, ``{IRS} empowered {UAV}
  wireless communication with resource allocation, reflecting design and
  trajectory optimization,'' {\em IEEE Transactions on Wireless
  Communications}, vol.~21, no.~10, pp.~7867--7880, 2022.

\bibitem{9912224}
X.~Zhang, J.~Wang, and H.~V. Poor, ``Joint beamforming and trajectory
  optimizations for statistical delay and error-rate bounded {QoS} over
  {MIMO}-{UAV}/{IRS}-based {6G} mobile edge computing networks using {FBC},''
  in {\em Proc. IEEE ICDCS}, pp.~983--993, 2022.

\bibitem{9745104}
M.~Tatar~Mamaghani and Y.~Hong, ``Aerial intelligent reflecting surface-enabled
  terahertz covert communications in beyond-{5G} {Internet} of things,'' {\em
  IEEE Internet of Things Journal}, vol.~9, no.~19, pp.~19012--19033, 2022.

\bibitem{9789841}
Y.~M. Park, S.~S. Hassan, Y.~K. Tun, Z.~Han, and C.~S. Hong, ``Joint resources
  and phase-shift optimization of {MEC}-enabled {UAV} in {IRS}-assisted {6G}
  {THz} networks,'' in {\em Proc. IEEE/IFIP NOMS}, pp.~1--7, 2022.

\bibitem{9367288}
Y.~Pan, K.~Wang, C.~Pan, H.~Zhu, and J.~Wang, ``{UAV}-assisted and intelligent
  reflecting surfaces-supported terahertz communications,'' {\em IEEE Wireless
  Communications Letters}, vol.~10, no.~6, pp.~1256--1260, 2021.

\bibitem{9528924}
G.~Sun, X.~Tao, N.~Li, and J.~Xu, ``Intelligent reflecting surface and {UAV}
  assisted secrecy communication in millimeter-wave networks,'' {\em IEEE
  Transactions on Vehicular Technology}, vol.~70, no.~11, pp.~11949--11961,
  2021.

\bibitem{9893192}
Y.~Li, H.~Zhang, K.~Long, and A.~Nallanathan, ``Exploring sum rate maximization
  in {UAV}-based multi-{IRS} networks: {IRS} association, {UAV} altitude, and
  phase shift design,'' {\em IEEE Transactions on Communications}, vol.~70,
  no.~11, pp.~7764--7774, 2022.

\bibitem{10.1007/s10846-021-01383-5}
T.~Baca, M.~Petrlik, M.~Vrba, V.~Spurny, R.~Penicka, D.~Hert, and M.~Saska,
  ``The mrs uav system: Pushing the frontiers of reproducible research,
  real-world deployment, and education with autonomous unmanned aerial
  vehicles,'' {\em Journal of Intelligent Robotic Systems}, vol.~102, no.~1,
  2021.

\bibitem{9836083}
D.~Hert, T.~Baca, P.~Petracek, V.~Kratky, V.~Spurny, M.~Petrlik, M.~Vrba,
  D.~Zaitlik, P.~Stoudek, V.~Walter, P.~Stepan, J.~Horyna, V.~Pritzl,
  G.~Silano, D.~Bonilla~Licea, P.~Stibinger, R.~Penicka, T.~Nascimento, and
  M.~Saska, ``{MRS} modular {UAV} hardware platforms for supporting research in
  real-world outdoor and indoor environments,'' in {\em Proc. ICUAS},
  pp.~1264--1273, 2022.

\bibitem{10596064}
E.~Basar, G.~C. Alexandropoulos, Y.~Liu, Q.~Wu, S.~Jin, C.~Yuen, O.~A. Dobre,
  and R.~Schober, ``Reconfigurable intelligent surfaces for {6G}: {Emerging}
  hardware architectures, applications, and open challenges,'' {\em IEEE
  Vehicular Technology Magazine, DOI:{10.1109/MVT.2024.3415570}}, pp.~2--22,
  2024.

\bibitem{9794781}
B.~K.~S. Lima, A.~S. De~Sena, R.~Dinis, D.~Benevides Da~Costa, M.~Beko,
  R.~Oliveira, and M.~Debbah, ``Aerial intelligent reflecting surfaces in
  {MIMO-NOMA} networks: {Fundamentals}, potential achievements, and
  challenges,'' {\em IEEE Open Journal of the Communications Society}, vol.~3,
  pp.~1007--1024, 2022.

\bibitem{9852389}
B.~Sihlbom, M.~I. Poulakis, and M.~Di~Renzo, ``Reconfigurable intelligent
  surfaces: {Performance} assessment through a system-level simulator,'' {\em
  IEEE Wireless Communications}, vol.~30, no.~4, pp.~98--106, 2023.

\bibitem{park2020devising}
S.~Park, W.~G. La, W.~Lee, and H.~Kim, ``Devising a distributed co-simulator
  for a multi-{UAV} network,'' {\em Sensors}, vol.~20, no.~21, p.~6196, 2020.

\bibitem{9893879}
G.~Grieco, G.~Iacovelli, P.~Boccadoro, and L.~A. Grieco, ``Internet of drones
  simulator: {Design}, implementation, and performance evaluation,'' {\em IEEE
  Internet of Things Journal}, vol.~10, no.~2, pp.~1476--1498, 2023.

\bibitem{10356747}
G.~Grieco, G.~Iacovelli, D.~Pugliese, D.~Striccoli, and L.~A. Grieco, ``A
  system-level simulation module for multi-{UAV} {IRS}-assisted
  communications,'' {\em IEEE Transactions on Vehicular Technology}, vol.~73,
  no.~5, pp.~6740--6751, 2024.

\bibitem{10296049}
D.~Diao, B.~Wang, K.~Cao, B.~Zheng, R.~Dong, T.~Cheng, and J.~Chen,
  ``Reflecting elements analysis for secure and energy-efficient {UAV-RIS}
  system with phase errors,'' {\em IEEE Wireless Communications Letters},
  vol.~13, no.~2, pp.~293--297, 2024.

\bibitem{10049533}
P.~Saxena and Y.~H. Chung, ``Analysis of jamming effects in {IRS} assisted
  {UAV} dual-hop {FSO} communication systems,'' {\em IEEE Transactions on
  Vehicular Technology}, vol.~72, no.~7, pp.~8956--8971, 2023.

\bibitem{9395180}
M.~Al-Jarrah, E.~Alsusa, A.~Al-Dweik, and D.~K.~C. So, ``Capacity analysis of
  {IRS}-based {UAV} communications with imperfect phase compensation,'' {\em
  IEEE Wireless Communications Letters}, vol.~10, no.~7, pp.~1479--1483, 2021.

\bibitem{9293155}
Z.~Wei, Y.~Cai, Z.~Sun, D.~W.~K. Ng, J.~Yuan, M.~Zhou, and L.~Sun, ``Sum-rate
  maximization for {IRS}-assisted {UAV} {OFDMA} communication systems,'' {\em
  IEEE Transactions on Wireless Communications}, vol.~20, no.~4,
  pp.~2530--2550, 2021.

\bibitem{9804341}
M.~Asim, M.~ELAffendi, and A.~A.~A. El-Latif, ``Multi-{IRS} and
  multi-{UAV}-assisted {MEC} system for {5G}/{6G} networks: Efficient joint
  trajectory optimization and passive beamforming framework,'' {\em IEEE
  Transactions on Intelligent Transportation Systems}, vol.~24, no.~4,
  pp.~4553--4564, 2023.

\bibitem{9771971}
F.~Wang and X.~Zhang, ``{IRS/UAV}-based edge-computing/traffic-offloading over
  {RF}-powered {6G} mobile wireless networks,'' in {\em Proc. IEEE WCNC},
  pp.~1272--1277, 2022.

\bibitem{9785612}
X.~Song, Y.~Zhao, Z.~Wu, Z.~Yang, and J.~Tang, ``Joint trajectory and
  communication design for {IRS}-assisted {UAV} networks,'' {\em IEEE Wireless
  Communications Letters}, vol.~11, no.~7, pp.~1538--1542, 2022.

\bibitem{9377648}
R.~Long, Y.-C. Liang, Y.~Pei, and E.~G. Larsson, ``Active reconfigurable
  intelligent surface-aided wireless communications,'' {\em IEEE Transactions
  on Wireless Communications}, vol.~20, no.~8, pp.~4962--4975, 2021.

\bibitem{10214219}
Y.~Ge, J.~Fan, and J.~Zhang, ``Active reconfigurable intelligent surface
  enhanced secure and energy-efficient communication of jittering {UAV},'' {\em
  IEEE Internet of Things Journal}, vol.~10, no.~24, pp.~22386--22400, 2023.

\bibitem{10472415}
A.~Huang, X.~Mu, L.~Guo, and G.~Zhu, ``Hybrid active-passive {RIS} transmitter
  enabled energy-efficient multi-user communications,'' {\em IEEE Transactions
  on Wireless Communications, DOI:{10.1109/TWC.2024.3373900}}, pp.~1--1, 2024.

\bibitem{10198901}
S.~Zhang, H.~Gao, Y.~Su, J.~Cheng, and M.~Jo, ``Intelligent mixed
  reflecting/relaying surface-aided secure wireless communications,'' {\em IEEE
  Transactions on Vehicular Technology}, vol.~73, no.~1, pp.~532--543, 2024.

\bibitem{9771718}
S.~Lin, M.~Wen, and F.~Chen, ``Cascaded channel estimation using full duplex
  for {IRS}-aided multiuser communications,'' in {\em Proc. IEEE WCNC},
  pp.~375--380, 2022.

\bibitem{9559873}
Y.~Wei, M.-M. Zhao, M.-J. Zhao, and Y.~Cai, ``Channel estimation for
  {IRS}-aided multiuser communications with reduced error propagation,'' {\em
  IEEE Transactions on Wireless Communications}, vol.~21, no.~4,
  pp.~2725--2741, 2022.

\bibitem{9967965}
H.~Dong, C.~Ji, L.~Zhou, J.~Dai, and Z.~Ye, ``Sparse channel estimation with
  surface clustering for {IRS}-assisted {OFDM} systems,'' {\em IEEE
  Transactions on Communications}, vol.~71, no.~2, pp.~1083--1095, 2023.

\bibitem{9521836}
Z.~Chen, J.~Tang, X.~Y. Zhang, D.~K.~C. So, S.~Jin, and K.-K. Wong, ``Hybrid
  evolutionary-based sparse channel estimation for {IRS}-assisted {mmWave}
  {MIMO} systems,'' {\em IEEE Transactions on Wireless Communications},
  vol.~21, no.~3, pp.~1586--1601, 2022.

\bibitem{9513592}
J.~Zhao, J.~Liu, F.~Gao, W.~Jia, and W.~Zhang, ``Gridless compressed sensing
  based channel estimation for {UAV} wideband communications with beam
  squint,'' {\em IEEE Transactions on Vehicular Technology}, vol.~70, no.~10,
  pp.~10265--10277, 2021.

\bibitem{9373363}
B.~Zheng, C.~You, and R.~Zhang, ``Efficient channel estimation for double-{IRS}
  aided multi-user {MIMO} system,'' {\em IEEE Transactions on Communications},
  vol.~69, no.~6, pp.~3818--3832, 2021.

\bibitem{9603291}
X.~Guan, Q.~Wu, and R.~Zhang, ``Anchor-assisted channel estimation for
  intelligent reflecting surface aided multiuser communication,'' {\em IEEE
  Transactions on Wireless Communications}, vol.~21, no.~6, pp.~3764--3778,
  2022.

\bibitem{8847452}
Q.~Bai, J.~Wang, Y.~Zhang, and J.~Song, ``Deep learning-based channel
  estimation algorithm over time selective fading channels,'' {\em IEEE
  Transactions on Cognitive Communications and Networking}, vol.~6, no.~1,
  pp.~125--134, 2020.

\bibitem{9505267}
C.~Liu, X.~Liu, D.~W.~K. Ng, and J.~Yuan, ``Deep residual learning for channel
  estimation in intelligent reflecting surface-assisted multi-user
  communications,'' {\em IEEE Transactions on Wireless Communications},
  vol.~21, no.~2, pp.~898--912, 2022.

\bibitem{9622178}
Z.~Chen, J.~Tang, X.~Y. Zhang, Q.~Wu, Y.~Wang, D.~K.~C. So, S.~Jin, and K.-K.
  Wong, ``Offset learning based channel estimation for intelligent reflecting
  surface-assisted indoor communication,'' {\em IEEE Journal of Selected Topics
  in Signal Processing}, vol.~16, no.~1, pp.~41--55, 2022.

\bibitem{9745538}
X.~Zheng, P.~Wang, J.~Fang, and H.~Li, ``Compressed channel estimation for
  {IRS}-assisted millimeter wave {OFDM} systems: A low-rank tensor
  decomposition-based approach,'' {\em IEEE Wireless Communications Letters},
  vol.~11, no.~6, pp.~1258--1262, 2022.

\bibitem{665}
B.~Van~Veen and K.~Buckley, ``Beamforming: {A} versatile approach to spatial
  filtering,'' {\em IEEE ASSP Magazine}, vol.~5, no.~2, pp.~4--24, 1988.

\bibitem{10075500}
Y.~Song, S.~Xu, G.~Sun, and B.~Ai, ``Weighted sum-rate maximization in
  multi-{IRS}-aided multi-cell {mmWave} communication systems for suppressing
  {ICI},'' {\em IEEE Transactions on Vehicular Technology}, vol.~72, no.~8,
  pp.~10234--10250, 2023.

\bibitem{9810528}
Z.~Ji, W.~Yang, X.~Guan, X.~Zhao, G.~Li, and Q.~Wu, ``Trajectory and transmit
  power optimization for {IRS}-assisted {UAV} communication under malicious
  jamming,'' {\em IEEE Transactions on Vehicular Technology}, vol.~71, no.~10,
  pp.~11262--11266, 2022.

\bibitem{9656117}
X.~Pang, N.~Zhao, J.~Tang, C.~Wu, D.~Niyato, and K.-K. Wong, ``{IRS}-assisted
  secure {UAV} transmission via joint trajectory and beamforming design,'' {\em
  IEEE Transactions on Communications}, vol.~70, no.~2, pp.~1140--1152, 2022.

\bibitem{9325920}
B.~Ning, Z.~Chen, W.~Chen, Y.~Du, and J.~Fang, ``Terahertz multi-user massive
  {MIMO} with intelligent reflecting surface: Beam training and hybrid
  beamforming,'' {\em IEEE Transactions on Vehicular Technology}, vol.~70,
  no.~2, pp.~1376--1393, 2021.

\bibitem{9348156}
T.~Jiang, H.~V. Cheng, and W.~Yu, ``Learning to beamform for intelligent
  reflecting surface with implicit channel estimate,'' in {\em Proc. IEEE
  GLOBECOM}, pp.~1--6, 2020.

\bibitem{9479733}
K.~Yu, X.~Yu, and J.~Cai, ``{UAV}s assisted intelligent reflecting surfaces
  {SWIPT} system with statistical {CSI},'' {\em IEEE Journal of Selected Topics
  in Signal Processing}, vol.~15, no.~5, pp.~1095--1109, 2021.

\bibitem{9374101}
H.-L. Song and Y.-C. Ko, ``Robust and low complexity beam tracking with
  monopulse signal for {UAV} communications,'' {\em IEEE Transactions on
  Vehicular Technology}, vol.~70, no.~4, pp.~3505--3513, 2021.

\bibitem{9869296}
D.~Jang, H.-L. Song, Y.-C. Ko, and H.~J. Kim, ``Distributed beam tracking for
  vehicular communications via {UAV}-assisted cellular network,'' {\em IEEE
  Transactions on Vehicular Technology}, vol.~72, no.~1, pp.~589--600, 2023.

\bibitem{9390407}
H.-L. Chiang, K.-C. Chen, W.~Rave, M.~Khalili~Marandi, and G.~Fettweis,
  ``Machine-learning beam tracking and weight optimization for {mmWave}
  multi-{UAV} links,'' {\em IEEE Transactions on Wireless Communications},
  vol.~20, no.~8, pp.~5481--5494, 2021.

\bibitem{10120938}
Q.~Deng, X.~Chen, X.~Liang, F.~Shu, J.~Du, G.~Yu, and J.~Wang, ``Adaptive beam
  alignment and optimization for {IRS}-aided high-speed {UAV} communications,''
  {\em IEEE Transactions on Green Communications and Networking}, vol.~7,
  no.~3, pp.~1583 -- 1595, 2023.

\bibitem{9813590}
L.~Yan, X.~Fang, Y.~Fang, L.~Hao, Q.~Xue, and C.~Xu, ``{KF-LSTM} based beam
  tracking for {UAV}-assisted {mmWave} {HSR} wireless networks,'' {\em IEEE
  Transactions on Vehicular Technology}, vol.~71, no.~10, pp.~10796--10807,
  2022.

\bibitem{10288199}
R.~Ye, Y.~Peng, F.~Al-Hazemi, and R.~Boutaba, ``A robust cooperative jamming
  scheme for secure {UAV} communication via intelligent reflecting surface,''
  {\em IEEE Transactions on Communications}, vol.~72, no.~2, pp.~1005--1019,
  2024.

\bibitem{9417539}
Z.~Mohamed and S.~Aïssa, ``Resource allocation for energy-efficient cellular
  communications via aerial {IRS},'' in {\em Proc. IEEE WCNC}, pp.~1--6, 2021.

\bibitem{9944150}
S.~Jangsher, M.~Al-Jarrah, A.~Al-Dweik, E.~Alsusa, and P.-Y. Kong, ``Energy
  constrained sum-rate maximization in {IRS}-assisted {UAV} networks with
  imperfect channel information,'' {\em IEEE Transactions on Aerospace and
  Electronic Systems}, vol.~59, no.~3, pp.~2898--2908, 2022.

\bibitem{9400768}
M.~Hua, L.~Yang, Q.~Wu, C.~Pan, C.~Li, and A.~L. Swindlehurst, ``{UAV}-assisted
  intelligent reflecting surface symbiotic radio system,'' {\em IEEE
  Transactions on Wireless Communications}, vol.~20, no.~9, pp.~5769--5785,
  2021.

\bibitem{9625737}
Z.~Ning, Y.~Yang, X.~Wang, L.~Guo, X.~Gao, S.~Guo, and G.~Wang, ``Dynamic
  computation offloading and server deployment for {UAV}-enabled multi-access
  edge computing,'' {\em IEEE Transactions on Mobile Computing}, vol.~22,
  no.~5, pp.~2628--2644, 2023.

\bibitem{10285605}
Y.~Liao, J.~Liu, X.~Chen, Y.~Han, Q.~Ai, and G.-M. Muntean, ``Energy
  minimization of inland waterway {USVs} for {IRS}-assisted hybrid
  {UAV}-terrestrial {MEC} network,'' {\em IEEE Transactions on Vehicular
  Technology}, vol.~73, no.~3, pp.~4121--4135, 2024.

\bibitem{10003080}
Q.~Pan, J.~Wu, A.~K. Bashir, J.~Li, S.~Vashisht, and R.~Nawaz, ``Blockchain and
  {AI} enabled configurable reflection resource allocation for {IRS}-aided
  coexisting drone-terrestrial networks,'' {\em IEEE Wireless Communications},
  vol.~29, no.~6, pp.~46--54, 2022.

\bibitem{10088448}
M.~Misbah, Z.~Kaleem, W.~Khalid, C.~Yuen, and A.~Jamalipour, ``Phase and {3D}
  placement optimization for rate enhancement in {RIS}-assisted {UAV}
  networks,'' {\em IEEE Wireless Communications Letters}, vol.~12, no.~7,
  pp.~1135 -- 1138, 2023.

\bibitem{9552547}
X.~Wang, Z.~Ning, S.~Guo, M.~Wen, L.~Guo, and H.~V. Poor, ``Dynamic {UAV}
  deployment for differentiated services: {A} multi-agent imitation learning
  based approach,'' {\em IEEE Transactions on Mobile Computing}, vol.~22,
  no.~4, pp.~2131--2146, 2023.

\bibitem{9802633}
L.~Dong, Z.~Liu, F.~Jiang, and K.~Wang, ``Joint optimization of deployment and
  trajectory in {UAV} and {IRS}-assisted {IoT} data collection system,'' {\em
  IEEE Internet of Things Journal}, vol.~9, no.~21, pp.~21583--21593, 2022.

\bibitem{9817819}
Y.~M. Park, Y.~K. Tun, Z.~Han, and C.~S. Hong, ``Trajectory optimization and
  phase-shift design in {IRS}-assisted {UAV} network for smart railway,'' {\em
  IEEE Transactions on Vehicular Technology}, vol.~71, no.~10,
  pp.~11317--11321, 2022.

\bibitem{10044705}
S.~Han, J.~Wang, L.~Xiao, and C.~Li, ``Broadcast secrecy rate maximization in
  {UAV}-empowered {IRS} backscatter communications,'' {\em IEEE Transactions on
  Wireless Communications}, vol.~22, no.~10, pp.~6445--6458, 2023.

\bibitem{9771577}
C.~Mei, Y.~Fang, and L.~Qiu, ``Dual based optimization method for {IRS}-aided
  {UAV}-enabled {SWIPT} system,'' in {\em Proc. IEEE WCNC}, pp.~890--895, 2022.

\bibitem{9849020}
Z.~Li, W.~Chen, H.~Cao, H.~Tang, K.~Wang, and J.~Li, ``Joint communication and
  trajectory design for intelligent reflecting surface empowered {UAV} {SWIPT}
  networks,'' {\em IEEE Transactions on Vehicular Technology}, vol.~71, no.~12,
  pp.~12840--12855, 2022.

\bibitem{10075533}
A.~B.~M. Adam, X.~Wan, M.~A.~M. Elhassan, M.~S.~A. Muthanna, A.~Muthanna,
  N.~Kumar, and M.~Guizani, ``Intelligent and robust {UAV}-aided multiuser
  {RIS} communication technique with jittering {UAV} and imperfect hardware
  constraints,'' {\em IEEE Transactions on Vehicular Technology}, vol.~72,
  no.~8, pp.~10737--10753, 2023.

\bibitem{9906843}
Z.~Mohamed and S.~Aissa, ``Energy-efficient joint broadcast-unicast
  communications via dual-polarized aerial {RIS},'' {\em IEEE Transactions on
  Wireless Communications}, vol.~22, no.~3, pp.~2113--2126, 2023.

\bibitem{9866052}
Y.~Su, X.~Pang, S.~Chen, X.~Jiang, N.~Zhao, and F.~R. Yu, ``Spectrum and energy
  efficiency optimization in {IRS}-assisted {UAV} networks,'' {\em IEEE
  Transactions on Communications}, vol.~70, no.~10, pp.~6489--6502, 2022.

\bibitem{9526285}
A.~Khalili, E.~M. Monfared, S.~Zargari, M.~R. Javan, N.~M. Yamchi, and E.~A.
  Jorswieck, ``Resource management for transmit power minimization in
  {UAV}-assisted {RIS} {HetNets} supported by dual connectivity,'' {\em IEEE
  Transactions on Wireless Communications}, vol.~21, no.~3, pp.~1806--1822,
  2022.

\bibitem{9789440}
S.~Basharat, S.~A. Hassan, A.~Mahmood, Z.~Ding, and M.~Gidlund,
  ``Reconfigurable intelligent surface-assisted backscatter communication: A
  new frontier for enabling {6G} {IoT} networks,'' {\em IEEE Wireless
  Communications}, vol.~29, no.~6, pp.~96--103, 2022.

\bibitem{8368232}
N.~Van~Huynh, D.~T. Hoang, X.~Lu, D.~Niyato, P.~Wang, and D.~I. Kim, ``Ambient
  backscatter communications: A contemporary survey,'' {\em IEEE Communications
  Surveys \& Tutorials}, vol.~20, no.~4, pp.~2889--2922, 2018.

\bibitem{8476597}
B.~Clerckx, R.~Zhang, R.~Schober, D.~W.~K. Ng, D.~I. Kim, and H.~V. Poor,
  ``Fundamentals of wireless information and power transfer: From {RF} energy
  harvester models to signal and system designs,'' {\em IEEE Journal on
  Selected Areas in Communications}, vol.~37, no.~1, pp.~4--33, 2019.

\bibitem{9531372}
Z.~Li, W.~Chen, Q.~Wu, K.~Wang, and J.~Li, ``Joint beamforming design and power
  splitting optimization in {IRS}-assisted {SWIPT} {NOMA} networks,'' {\em IEEE
  Transactions on Wireless Communications}, vol.~21, no.~3, pp.~2019--2033,
  2022.

\bibitem{9453799}
H.~D. Tuan, A.~A. Nasir, A.~V. Savkin, H.~V. Poor, and E.~Dutkiewicz,
  ``{MPC}-based {UAV} navigation for simultaneous solar-energy harvesting and
  two-way communications,'' {\em IEEE Journal on Selected Areas in
  Communications}, vol.~39, no.~11, pp.~3459--3474, 2021.

\bibitem{10555361}
X.~Song, Y.~Wang, S.~Xu, R.~Zhang, Y.~Zhang, and Z.~Xie, ``{SD-Jaya} based
  multi-objective optimization algorithm for {IRS}-aided air-to-ground task
  offloading in charging electric vehicle networks,'' {\em IEEE Transactions on
  Industry Applications, DOI{10.1109/TIA.2024.3413048}}, pp.~1--13, 2024.

\bibitem{10017780}
F.~O. Olowononi, D.~B. Rawat, C.~A. Kamhoua, and B.~M. Sadler, ``Deep
  reinforcement learning for deception in {IRS}-assisted {UAV}
  communications,'' in {\em Proc. IEEE MILCOM}, pp.~763--768, 2022.

\bibitem{10452297}
H.~Yang, K.~Lin, L.~Xiao, Y.~Zhao, Z.~Xiong, and Z.~Han, ``Energy harvesting
  {UAV-RIS}-assisted maritime communications based on deep reinforcement
  learning against jamming,'' {\em IEEE Transactions on Wireless
  Communications, DOI:{10.1109/TWC.2024.3367034}}, pp.~1--1, 2024.

\bibitem{10070838}
W.~Wei, X.~Pang, J.~Tang, N.~Zhao, X.~Wang, and A.~Nallanathan, ``Secure
  transmission design for aerial {IRS} assisted wireless networks,'' {\em IEEE
  Transactions on Communications}, vol.~71, no.~6, pp.~3528--3540, 2023.

\bibitem{9434412}
X.~Guo, Y.~Chen, and Y.~Wang, ``Learning-based robust and secure transmission
  for reconfigurable intelligent surface aided millimeter wave {UAV}
  communications,'' {\em IEEE Wireless Communications Letters}, vol.~10, no.~8,
  pp.~1795--1799, 2021.

\bibitem{10111039}
Y.~Qian, C.~Yang, Z.~Mei, X.~Zhou, L.~Shi, and J.~Li, ``On joint optimization
  of trajectory and phase shift for {IRS-UAV} assisted covert communication
  systems,'' {\em IEEE Transactions on Vehicular Technology}, vol.~72, no.~10,
  pp.~12873--12883, 2023.

\bibitem{9943536}
C.~Wang, X.~Chen, J.~An, Z.~Xiong, C.~Xing, N.~Zhao, and D.~Niyato, ``Covert
  communication assisted by {UAV-IRS},'' {\em IEEE Transactions on
  Communications}, vol.~71, no.~1, pp.~357--369, 2023.

\bibitem{10271264}
X.~Chen, Z.~Chang, M.~Liu, N.~Zhao, T.~Hämäläinen, and D.~Niyato,
  ``{UAV-IRS} assisted covert communication: {Introducing} uncertainty via
  phase shifting,'' {\em IEEE Wireless Communications Letters}, vol.~13, no.~1,
  pp.~103--107, 2024.

\bibitem{9454372}
Y.~Wu, W.~Yang, X.~Guan, and Q.~Wu, ``{UAV}-enabled relay communication under
  malicious jamming: Joint trajectory and transmit power optimization,'' {\em
  IEEE Transactions on Vehicular Technology}, vol.~70, no.~8, pp.~8275--8279,
  2021.

\bibitem{9200570}
Y.~Wu, W.~Yang, X.~Guan, and Q.~Wu, ``Energy-efficient trajectory design for
  {UAV}-enabled communication under malicious jamming,'' {\em IEEE Wireless
  Communications Letters}, vol.~10, no.~2, pp.~206--210, 2021.

\bibitem{9552611}
R.~Dong, B.~Wang, K.~Cao, and T.~Cheng, ``Securing transmission for {UAV}
  swarm-enabled communication network,'' {\em IEEE Systems Journal}, vol.~16,
  no.~4, pp.~5200--5211, 2022.

\bibitem{9374975}
M.-M. Zhao, Q.~Wu, M.-J. Zhao, and R.~Zhang, ``Exploiting amplitude control in
  intelligent reflecting surface aided wireless communication with imperfect
  {CSI},'' {\em IEEE Transactions on Communications}, vol.~69, no.~6,
  pp.~4216--4231, 2021.

\bibitem{9382022}
X.~Jiang, X.~Chen, J.~Tang, N.~Zhao, X.~Y. Zhang, D.~Niyato, and K.-K. Wong,
  ``Covert communication in {UAV}-assisted air-ground networks,'' {\em IEEE
  Wireless Communications}, vol.~28, no.~4, pp.~190--197, 2021.

\bibitem{10499205}
J.~Li, G.~Chen, T.~Zhang, W.~Feng, W.~Jiang, T.~Q. Quek, and R.~Tafazolli,
  ``{UAV-RIS}-aided space-air-ground integrated network: {Interference}
  alignment design and {DoF} analysis,'' {\em IEEE Transactions on Wireless
  Communications, DOI:{10.1109/TWC.2024.3384257}}, pp.~1--1, 2024.

\bibitem{10089834}
T.~M. Hoang, C.~Xu, A.~Vahid, H.~D. Tuan, T.~Q. Duong, and L.~Hanzo,
  ``Secrecy-rate optimization of double {RIS}-aided space–ground networks,''
  {\em IEEE Internet of Things Journal}, vol.~10, no.~15, pp.~13221--13234,
  2023.

\bibitem{9760044}
A.~Bansal, N.~Agrawal, and K.~Singh, ``Rate-splitting multiple access for
  {UAV}-based {RIS}-enabled interference-limited vehicular communication
  system,'' {\em IEEE Transactions on Intelligent Vehicles}, vol.~8, no.~1,
  pp.~936--948, 2023.

\bibitem{10638476}
H.~Zhang and J.~Liu, ``Dynamic aerial reconfigurable intelligent surface aided
  multi-cell multi-user communications,'' {\em IEEE Transactions on Wireless
  Communications, DOI:{10.1109/TWC.2024.3441093}}, pp.~1--1, 2024.

\bibitem{10411934}
J.~Chen, K.~Zhai, Z.~Wang, Y.~Liu, J.~Jia, and X.~Wang, ``{CoMP} and
  {RIS}-assisted multicast transmission in a multi-{UAV} communication
  system,'' {\em IEEE Transactions on Communications}, vol.~72, no.~6,
  pp.~3602--3617, 2024.

\bibitem{9013433}
Y.~Liu, L.~Zhang, B.~Yang, W.~Guo, and M.~A. Imran, ``Programmable wireless
  channel for multi-user {MIMO} transmission using meta-surface,'' in {\em
  Proc. GLOBECOM}, pp.~1--6, 2019.

\bibitem{9013626}
Q.~Zhang, W.~Saad, and M.~Bennis, ``Reflections in the sky: Millimeter wave
  communication with {UAV}-carried intelligent reflectors,'' in {\em Proc. IEEE
  GLOBECOM}, pp.~1--6, 2019.

\bibitem{9348040}
Q.~Zhang, W.~Saad, and M.~Bennis, ``Distributional reinforcement learning for
  mmwave communications with intelligent reflectors on a {UAV},'' in {\em Proc.
  IEEE GLOBECOM}, pp.~1--6, 2020.

\bibitem{9781659}
X.~Zhang, J.~Wang, and H.~V. Poor, ``Joint optimization of {IRS} and
  {UAV}-trajectory: For supporting statistical delay and error-rate bounded
  {QoS} over {mURLLC}-driven {6G} mobile wireless networks using {FBC},'' {\em
  IEEE Vehicular Technology Magazine}, vol.~17, no.~2, pp.~55--63, 2022.

\bibitem{9539168}
M.~Al-Jarrah, A.~Al-Dweik, E.~Alsusa, Y.~Iraqi, and M.-S. Alouini, ``On the
  performance of {IRS}-assisted multi-layer {UAV} communications with imperfect
  phase compensation,'' {\em IEEE Transactions on Communications}, vol.~69,
  no.~12, pp.~8551--8568, 2021.

\bibitem{9741782}
N.~Agrawal, A.~Bansal, K.~Singh, C.-P. Li, and S.~Mumtaz, ``Finite block length
  analysis of {RIS}-assisted {UAV}-based multiuser {IoT} communication system
  with non-linear {EH},'' {\em IEEE Transactions on Communications}, vol.~70,
  no.~5, pp.~3542--3557, 2022.

\bibitem{10093979}
K.~Guo, R.~Liu, M.~Alazab, R.~H. Jhaveri, X.~Li, and M.~Zhu,
  ``{STAR-RIS}-empowered cognitive non-terrestrial vehicle network with
  {NOMA},'' {\em IEEE Transactions on Intelligent Vehicles}, vol.~8, no.~6,
  pp.~3735--3749, 2023.

\bibitem{9860636}
S.~Solanki, S.~Gautam, V.~Singh, S.~K. Sharma, and S.~Chatzinotas, ``Symbiotic
  radio based spectrum sharing in cooperative {UAV}-{IRS} wireless networks,''
  in {\em Proc. IEEE VTC}, pp.~1--5, 2022.

\bibitem{9715145}
W.~Mei and R.~Zhang, ``Intelligent reflecting surface for multi-path beam
  routing with active/passive beam splitting and combining,'' {\em IEEE
  Communications Letters}, vol.~26, no.~5, pp.~1165--1169, 2022.

\bibitem{8811733}
Q.~Wu and R.~Zhang, ``Intelligent reflecting surface enhanced wireless network
  via joint active and passive beamforming,'' {\em IEEE Transactions on
  Wireless Communications}, vol.~18, no.~11, pp.~5394--5409, 2019.

\bibitem{10498067}
Y.~Mao, X.~Yang, L.~Wang, D.~Wang, O.~Alfarraj, K.~Yu, S.~Mumtaz, and
  F.~Richard~Yu, ``A high-capacity {MAC} protocol for {UAV}-enhanced
  {RIS}-assisted {V2X} architecture in {3-D} {IoT} traffic,'' {\em IEEE
  Internet of Things Journal}, vol.~11, no.~13, pp.~23711--23726, 2024.

\bibitem{10285074}
P.~Wang, D.~Li, Y.~Zhang, and X.~Chen, ``{UAV}-assisted {Vehicular}
  communication system optimization with aerial base station and intelligent
  reflecting surface,'' {\em IEEE Transactions on Intelligent Vehicles,
  DOI:{10.1109/TIV.2023.3324385}}, pp.~1--12, 2023.

\bibitem{10499959}
N.~Sehito, Y.~Shouyi, H.~M. Alshahrani, M.~Alamgeer, A.~K. Dutta, S.~Alsubai,
  L.~Nkenyereye, and R.~K. Dhanaraj, ``Optimizing user association, power
  control and beamforming for {6G} multi-{IRS} multi-{UAV} {NOMA}
  communications in smart cities,'' {\em IEEE Transactions on Consumer
  Electronics, DOI:{10.1109/TCE.2024.3388596}}, pp.~1--1, 2024.

\bibitem{9344862}
M.~Jung, W.~Saad, and G.~Kong, ``Performance analysis of active large
  intelligent surfaces {(LISs)}: {Uplink} spectral efficiency and pilot
  training,'' {\em IEEE Transactions on Communications}, vol.~69, no.~5,
  pp.~3379--3394, 2021.

\bibitem{9365004}
X.~Mu, Y.~Liu, L.~Guo, J.~Lin, and N.~Al-Dhahir, ``Capacity and optimal
  resource allocation for {IRS}-assisted multi-user communication systems,''
  {\em IEEE Transactions on Communications}, vol.~69, no.~6, pp.~3771--3786,
  2021.

\bibitem{9353406}
W.~Ni, X.~Liu, Y.~Liu, H.~Tian, and Y.~Chen, ``Resource allocation for
  multi-cell {IRS}-aided {NOMA} networks,'' {\em IEEE Transactions on Wireless
  Communications}, vol.~20, no.~7, pp.~4253--4268, 2021.

\bibitem{10537097}
Z.~Ning, H.~Hu, X.~Wang, Q.~Wu, C.~Yuen, F.~R. Yu, and Y.~Zhang, ``Joint user
  association, interference cancellation and power control for multi-{IRS}
  assisted {UAV} communications,'' {\em IEEE Transactions on Wireless
  Communications, DOI:{10.1109/TWC.2024.3401152}}, pp.~1--1, 2024.

\bibitem{9133142}
C.~You, B.~Zheng, and R.~Zhang, ``Channel estimation and passive beamforming
  for intelligent reflecting surface: Discrete phase shift and progressive
  refinement,'' {\em IEEE Journal on Selected Areas in Communications},
  vol.~38, no.~11, pp.~2604--2620, 2020.

\bibitem{9816032}
S.~Jangsher, M.~Al-Jarrah, A.~Al-Dweik, E.~Alsusa, and M.-S. Alouini, ``{BER}
  reduction using partial-elements selection in {IRS}-{UAV} communications with
  imperfect phase compensation,'' {\em IEEE Transactions on Aerospace and
  Electronic Systems}, vol.~59, no.~1, pp.~623--633, 2023.

\bibitem{9652031}
L.~Dong, H.-M. Wang, and J.~Bai, ``Active reconfigurable intelligent surface
  aided secure transmission,'' {\em IEEE Transactions on Vehicular Technology},
  vol.~71, no.~2, pp.~2181--2186, 2022.

\bibitem{9645536}
Y.~He, C.-X. Wang, H.~Chang, J.~Huang, J.~Sun, W.~Zhang, and E.-H.~M. Aggoune,
  ``A novel {3D} non-stationary maritime wireless channel model,'' {\em IEEE
  Transactions on Communications}, vol.~70, no.~3, pp.~2102--2116, 2022.

\bibitem{9460824}
Y.~Liu, C.-X. Wang, H.~Chang, Y.~He, and J.~Bian, ``A novel non-stationary {6G}
  {UAV} channel model for maritime communications,'' {\em IEEE Journal on
  Selected Areas in Communications}, vol.~39, no.~10, pp.~2992--3005, 2021.

\bibitem{9938373}
K.~Meng, Q.~Wu, R.~Schober, and W.~Chen, ``Intelligent reflecting surface
  enabled multi-target sensing,'' {\em IEEE Transactions on Communications},
  vol.~70, no.~12, pp.~8313--8330, 2022.

\bibitem{10508296}
J.~Zhang, J.~Xu, W.~Lu, N.~Zhao, X.~Wang, and D.~Niyato, ``Secure transmission
  for {IRS}-aided {UAV-ISAC} networks,'' {\em IEEE Transactions on Wireless
  Communications, DOI:{10.1109/TWC.2024.3390169}}, pp.~1--1, 2024.

\bibitem{9937163}
Z.~Yu, X.~Hu, C.~Liu, M.~Peng, and C.~Zhong, ``Location sensing and beamforming
  design for {IRS}-enabled multi-user {ISAC} systems,'' {\em IEEE Transactions
  on Signal Processing}, vol.~70, pp.~5178--5193, 2022.

\bibitem{9858656}
K.~Meng, Q.~Wu, S.~Ma, W.~Chen, K.~Wang, and J.~Li, ``Throughput maximization
  for {UAV}-enabled integrated periodic sensing and communication,'' {\em IEEE
  Transactions on Wireless Communications}, vol.~22, no.~1, pp.~671--687, 2023.

\end{thebibliography}

\begin{IEEEbiography}[{\includegraphics[width=1in,height=1.25in,clip,keepaspectratio]{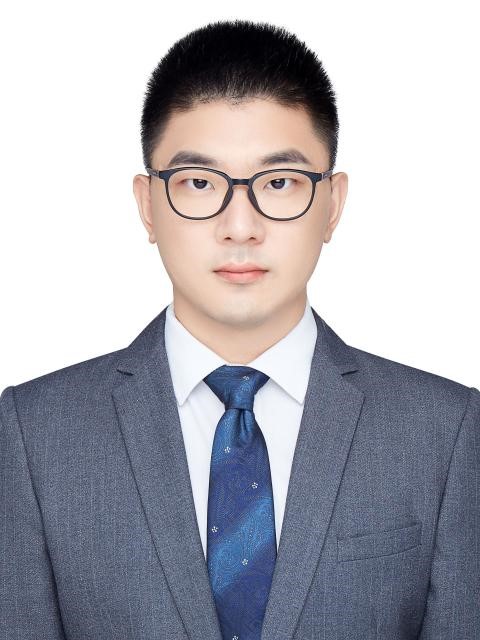}}]{Zhaolong Ning}
	(M'14-SM'18) received the Ph.D. degree from Northeastern University, China in 2014. He was a Research Fellow at Kyushu University from 2013 to 2014, Japan. Currently, he is a full professor with the School of Communications and Information Engineering, the Chongqing University of Posts and Telecommunications, Chongqing, China. His research interests include mobile edge computing, 6G networks, machine learning, and resource management. He has published over 150 scientific papers in international journals and conferences. Dr. Ning serves as an associate editor or guest editor of several journals, such as IEEE Transactions on Industrial Informatics, IEEE Transactions on Social Computational Systems, The Computer Journal and so on. He is a Highly Cited Researcher (Web of Science) since 2020.
\end{IEEEbiography}

\begin{IEEEbiography}[{\includegraphics[width=1in,height=1.25in,clip,keepaspectratio]{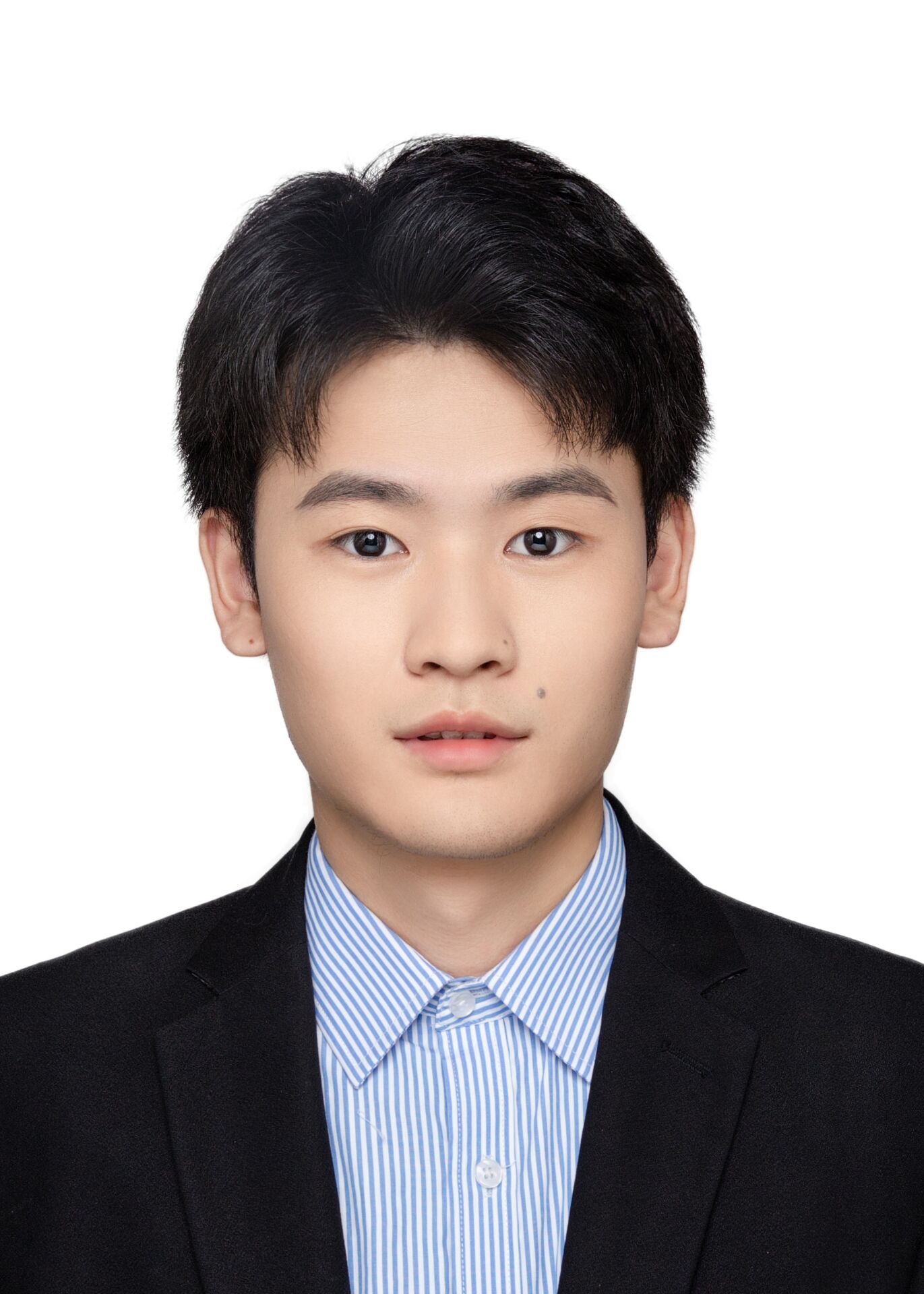}}]{Tengfeng Li}
	received the B.E. degree in Communication Engineering from the Chongqing University of Posts and Telecommunications, Chongqing, China, in 2022. He is currently working toward the M.E. degree with the School of Communications and Information Engineering, Chongqing University of Posts and Telecommunications, Chongqing, China. His research interests are intelligent reflecting surface, unmanned aerial vehicle communications and resource allocation.
\end{IEEEbiography}

\begin{IEEEbiography}[{\includegraphics[width=1in,height=1.25in,clip,keepaspectratio]{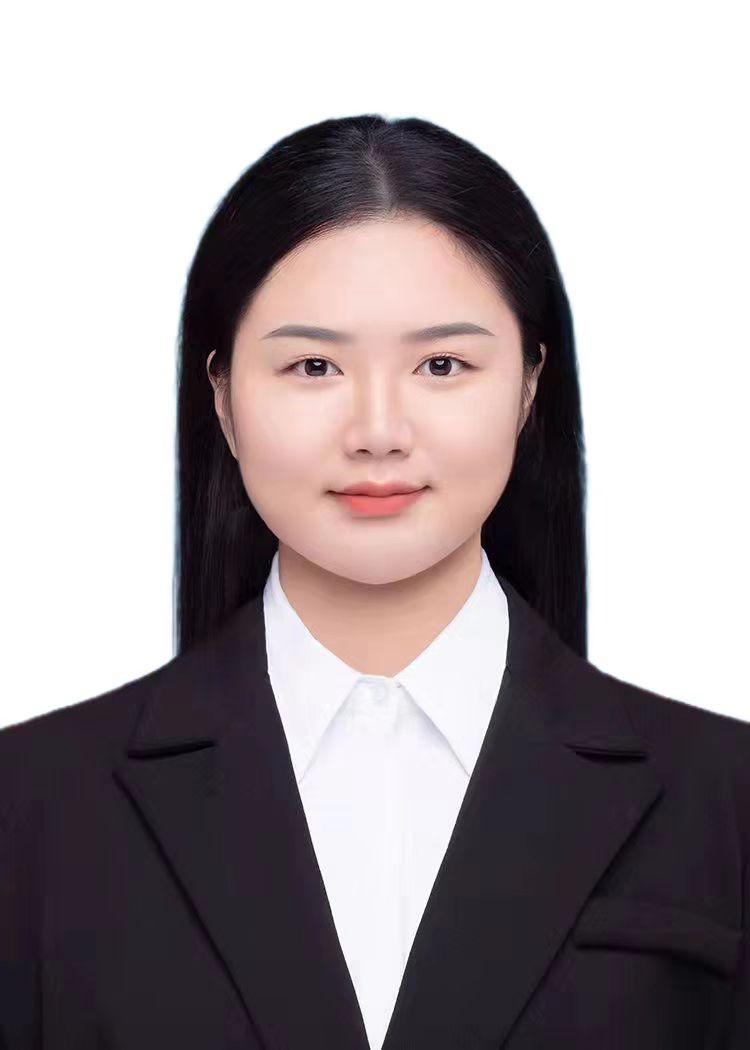}}]{Yu Wu}
	is a lecture with the School of Cyber Security and Information Law, Chongqing University of Posts and Telecommunications, China. She received the Ph.D. degree in computer science from Chongqing University in 2022. Her research interests include storage security and mobile distributed clusters.
\end{IEEEbiography}

\begin{IEEEbiography}[{\includegraphics[width=1in,height=1.25in,clip,keepaspectratio]{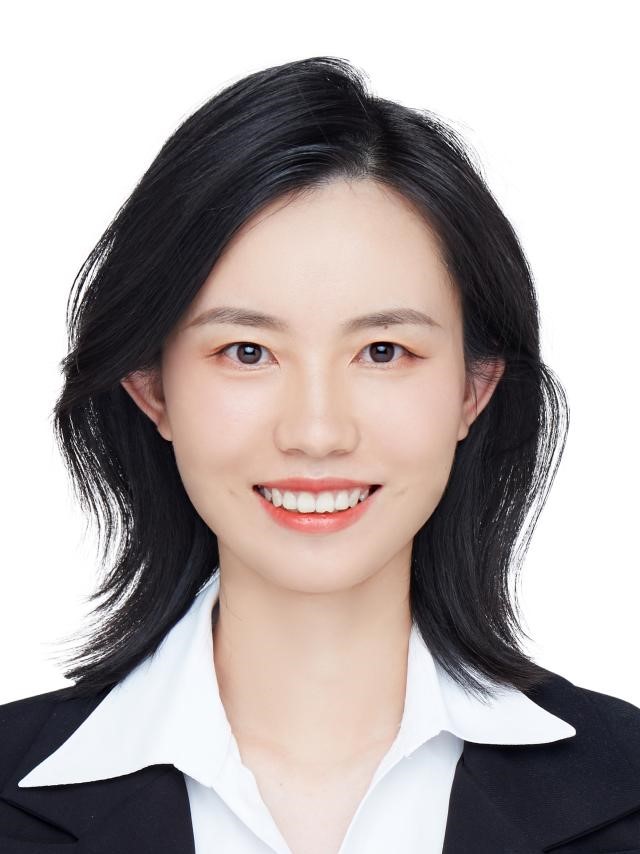}}]{Xiaojie Wang}
	(M'19-SM'22) received the PhD degree from Dalian University of Technology, Dalian, China, in 2019. After that, she was a postdoctor in the Hong Kong Polytechnic University. Currently, she is a full professor with the School of Communications and Information Engineering, the Chongqing University of Posts and Telecommunications, Chongqing, China. Her research interests are wireless networks, mobile edge computing and machine learning. She has published over 60 scientific papers in international journals and conferences, such as IEEE TMC, IEEE JSAC, IEEE TPDS and IEEE COMST. She is a Highly Cited Researcher (Web of Science) since 2023.
\end{IEEEbiography}

\begin{IEEEbiography}[{\includegraphics[width=1in,height=1.25in,clip,keepaspectratio]{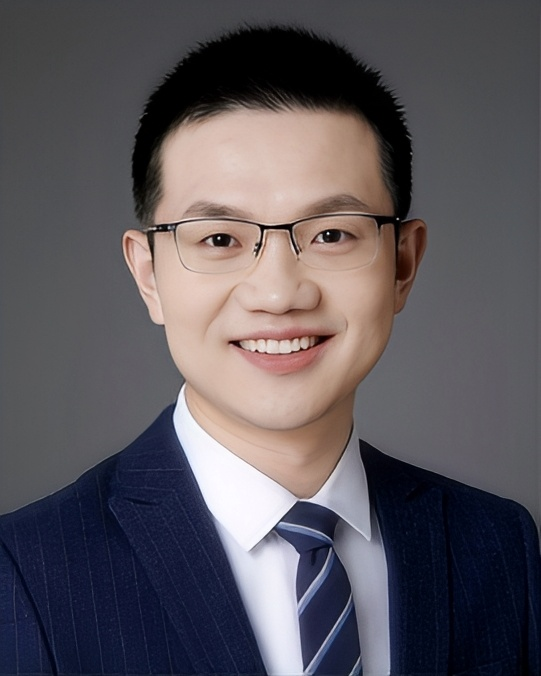}}]{Qingqing Wu}
	(S'13-M'16-SM'21) is an Associate Professor with Shanghai Jiao Tong University. His current research interest includes intelligent reflecting surface (IRS), unmanned aerial vehicle (UAV) communications, and MIMO transceiver design. He has coauthored more than 100 IEEE journal papers with 40+ ESI highly cited papers, which have received more than 35,000 Google citations. He has been listed as the Clarivate ESI Highly Cited Researcher since 2021.\\
	He was the recipient of the IEEE Communications Society Fred Ellersick Prize in 2023, Best Tutorial Paper Award in 2023,  and Young Author Best Paper Award in 2021 and 2024. He serves as an Associate/Senior/Area Editor for IEEE Transactions on Wireless Communications, IEEE Transactions on Communications, IEEE Communications Letters, IEEE Wireless Communications Letters. He is the Lead Guest Editor for IEEE Journal on Selected Areas in Communications. He is the workshop co-chair for IEEE ICC 2019-2023 and IEEE GLOBECOM 2020.  He is the Founding Chair of  IEEE Communications Society Young Professional committee in Asia Pacific Region and Chair of IEEE VTS Drone Committee.
\end{IEEEbiography}

\begin{IEEEbiography}[{\includegraphics[width=1in,height=1.25in,clip,keepaspectratio]{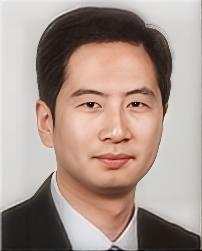}}]{Fei Richard Yu}
	(M'04-SM'08-F'18) received the PhD degree in electrical engineering from the University of British Columbia (UBC). His research interests include machine learning and embodied AI, autonomous systems, big data, and blockchain. He has been named in the Clarivate’s list of ``Highly Cited Researchers" in computer science and Standford’s Top 2\% Most Highly Cited Scientist. He received several Best Paper Awards from some first-tier conferences, Carleton Research Achievement Awards in 2012 and 2021, and the Ontario Early Researcher Award in 2011. He is a Board Member the IEEE VTS and the Editor-in-Chief for IEEE VTS Mobile World newsletter. He is a Member of the Academia Europaea (MAE), a Fellow of the IEEE, Canadian Academy of Engineering (CAE), Engineering Institute of Canada (EIC), and IET. He is a Distinguished Lecturer of the IEEE.
\end{IEEEbiography}

\begin{IEEEbiography}[{\includegraphics[width=1in,height=1.25in,clip,keepaspectratio]{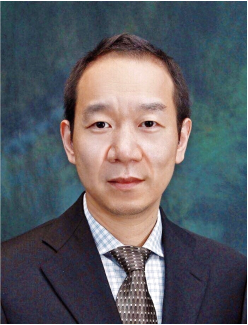}}]{Song Guo}
	(M'02-SM'11-F'20) is a Chair Professor in the Department of Computer Science and Engineering at the Hong Kong University of Science and Technology. Prof. Guo made fundamental and pioneering contributions to the development of edge AI and cloud-edge computing. He has published many papers in top venues with wide impact in these areas and been consistently recognized as a Clarivate Highly Cited Researcher. He is the recipient of Edward J. McCluskey Technical Achievement Award (IEEE Computer Society) in 2024, First Prize in Natural Science (China Electronics Society) in 2023, Gold Medal of Geneva Inventions Expo in 2024 \& 2023, and over a dozen Best Paper Awards from IEEE/ACM conferences, journals, and technical committees. Prof. Guo is the Editor-in-Chief of IEEE Transactions on Cloud Computing and founding Editor-in-Chief of IEEE Open Journal of the Computer Society. He was an IEEE Communications Society (ComSoc) Distinguished Lecturer, a member of IEEE ComSoc Board of Governors, and the Chair of IEEE ComSoc Space and Satellite Communications Technical Committee. He has served as a member of IEEE Fellow Evaluation Committees for both Compuer and Communications Societies, and a chair of organizing and technical committees of many international conferences. Prof. Guo is a Fellow of the Canadian Academy of Engineering, a Member of Academia Europaea, and a Fellow of the IEEE.
\end{IEEEbiography}

\end{document}